\documentclass{jfm}
\usepackage[export]{adjustbox}
\usepackage{subcaption}
\usepackage{siunitx}
\usepackage{xcolor}
\usepackage[section]{placeins}
\usepackage{float}  
\usepackage{physics} 
\usepackage{tikz,tkz-euclide}
\usetikzlibrary{shapes}
\usetikzlibrary{positioning}
\usepackage[normalem]{ulem}
\usetikzlibrary{calc,patterns,angles,quotes,arrows}
\usepackage{bm}  
\usepackage{longtable}

\begin{document}

\newtheorem{lemma}{Lemma}
\newtheorem{corollary}{Corollary}

\shorttitle{Controlling secondary flows in Taylor-Couette flow} 
\shortauthor{V. Jeganathan, K. Alba and R. Ostilla-M\'onico} 

\title{Controlling secondary flows in Taylor-Couette flow using stress-free boundary conditions }

\author
 {
 Vignesh Jeganathan\aff{1}
 \and
 Kamran Alba\aff{2}
  \and
  Rodolfo Ostilla-M\'onico \aff{1}  
  \corresp{\email{rostilla@central.uh.edu}}
 }

\affiliation
{
\aff{1}
Department of Mechanical Engineering, University of Houston, Houston, TX 77004, USA
\aff{2}
Department of Engineering Technology, University of Houston, Houston, TX 77004, USA
}

\maketitle

\begin{abstract}
Taylor-Couette (TC) flow, the flow between two independently rotating and co-axial cylinders is commonly used as a canonical model for shear flows. Unlike plane Couette, pinned secondary flows can be found in TC flow. These are known as Taylor rolls and drastically affect the flow behaviour. We study the possibility of modifying these secondary structures using patterns of stress-free and no-slip boundary conditions on the inner cylinder. For this, we perform direct numerical simulations of narrow-gap TC flow with pure inner cylinder rotation at four different shear Reynolds numbers up to $Re_s=3\times 10^4$.
We find that one-dimensional azimuthal patterns do not have a significant effect on the flow topology, and that the resulting torque is a large fraction ($\sim80-90\%$) of torque in the fully no-slip case. One-dimensional axial patterns decrease the torque more, and for certain pattern frequency disrupt the rolls by interfering with the existing Reynolds stresses that generate secondary structures. For $Re\geq 10^4$, this disruption leads to a smaller torque than what would be expected from simple boundary layer effects and the resulting effective slip length and slip velocity. We find that two-dimensional checkerboard patterns have similar behaviour to azimuthal patterns and do not affect the flow or the torque substantially, but two-dimensional spiral inhomogeneities can move around the pinned secondary flows as they induce persistent axial velocities. We quantify the roll's movement for various angles and the widths of the spiral pattern, and find a non-monotonic behaviour as a function of pattern angle and pattern frequency.
\end{abstract}

\section{Introduction}
\label{sec:introduction}

Taylor-Couette (TC) flow \citep{gro16}, the flow between two concentric and independently rotating cylinders has been used as a basic model for shear flows for many decades \citep{donnelly1991taylor}. The most particular and prominent feature of TC flow is its secondary large-scale flow structures, called Taylor rolls. In their laminar state, Taylor rolls arise due to centrifugal effects \citep{tay23}. They are axisymmetric, and fill the entire gap between the two cylinders. As the Reynolds number increases, instabilities build on top of the rolls. They first develop azimuthal waves, entering the wavy Taylor vortex flow regime. Further increases in the Reynolds number first cause the onset of temporal modulation in the roll behaviour, and finally the transition to turbulence of the rolls \citep{andereck1986flow}. Their fate when Reynolds number becomes infinitely large depends on the curvature and the way the cylinders rotate. \cite{lathrop1992turbulent} found that as the Reynolds number increases beyond $Re_s\sim\mathcal{O}(10^5)$ the Taylor rolls disappeared for pure inner cylinder rotation. Howevefr, \cite{huisman2014multiple} showed that at Reynolds numbers of $Re_s\sim \mathcal{O}(10^6)$, Taylor rolls persist in time for certain cases of cylinder counter-rotation, which corresponded to the cases with largest torque. This was explained in \cite{sacco2019dynamics}, which showed that for high Reynolds numbers, the presence or absence of rolls was determined primarily by the amount of solid-body rotation in a reference frame where both cylinders rotated with equal but opposite speeds. In the fully turbulent regime, only moderate amounts of anti-cyclonic rotation acting as a Coriolis force would generate rolls, which persisted for the largest Reynolds number achieved in simulations.

Taylor rolls are an important flow phenomenon because when present, they account for a large fraction of the convective transport of angular velocity across the gap \citep{brauckmann2013direct,ostilla2016near}. 
If we are able to somehow affect or control the behavior of Taylor rolls, we can modify the torque and other flow statistics in a TC or TC-like system. The simplest method would be to change the geometry of the system to modify or eliminate the curvature and the mean rotation. However, this is often not possible in existing engineering-technology apparatuses such as centrifugal mixers or bio-reactors. 
Other flow control mechanisms are limited by the fact that turbulent Taylor rolls seem to be a robust feature in the parameter space where they occur. They survive the early stages of turbulence decay \citep{ostilla2017life}, and are still present in the highly turbulent regime even when large axisymmetric grooves are placed on the cylinders   \citep{zhu2016direct}.

In this manuscript, we explore a different way of modifying these rolls. Namely, we will study the possibility of affecting the rolls by inducing a separate secondary flow that could
interfere destructively with them. To create this flow, we take inspiration from recent work which has shown that patterns of heterogeneous roughness induce swirling motions in the regions between high- and low-momentum flow pathways \citep{nugroho2013large,barros2014observations,willingham2014turbulent,anderson2015numerical}. The same kind of induced secondary flows can also be generated by using heterogeneous stress-free boundaries in the place of roughness \citep{turk2014turbulent}. These methods work because turbulent secondary flows can be generated and sustained due to spanwise gradients in the Reynolds-stress components. These gradients cause an imbalance between production and dissipation of turbulent kinetic energy that necessitates secondary advective velocities to balance. The induced flows are known as ``Prandtl secondary flows of the second kind''. 

\cite{bakhuis2019controlling} have shown that this is a plausible mechanism for controlling rolls in Taylor-Couette. They combined experiments and simulations that introduced an alternating pattern of spanwise roughness in the inner cylinder of a TC system. A significant effect on the global flow properties and the local flow structures was reported at Reynolds numbers of $\mathcal{O}(10^6)$, and for certain distributions of roughness a secondary flow was induced. However, this method of affecting secondary flows will come at the price of increased drag which means energy losses in real world applications.

We also draw inspiration from recent work in Rayleigh-Be{\'n}ard (RB) convection, the flow between two parallel plates heated from below and cooled from above. RB convection is in close mathematical analogy to TC flow: TC flow can be understood as a convective flow driven by the shear between the cylinders, where the angular velocity is transported from one cylinder to the other \citep{eckhardt2007torque}. Furthermore, large-scale convective cells are present in RB convection which are very similar to Taylor rolls. It has been shown that these cells can be altered through the use of boundary heterogeneities. In particular, numerical \citep{ripesi2014natural,bakhuis2018mixed} and experimental \citep{wang2017thermal} studies analyzed modifications of the canonical RB flow problem that used alternating conducting and insulating surfaces in the plates to selectively restrict heat transfer. These studies reported changes in both the large-scale flow structure and the overall heat transfer rate which depended on the size of the patterns. 

Even if this analogy becomes weaker at higher Reynolds numbers, and for low-curvature system, we take inspiration from it and choose to generate Reynolds stress imbalances by reducing the local drag through the use of hydrophobic, or stress-free surfaces. We choose to focus on ideal stress-free surfaces, where there is no shear at the wall, rather than introduce real boundary conditions where there is a finite slip, such as those in the TC experiment of \cite{srinivasan2015sustainable}. This provides an ideal model for what can be achieved in real-world circumstances, and allows us to conduct a wide parameter space study from which experiments can take inspiration as well as elucidating the basic physical mechanisms that underlie the process of Taylor roll control.

Aside from affecting secondary structures, the major impact of introducing stress-free boundary conditions in wall-bounded flows is drag reduction. This has been a very active research area, both from the point of view of manufacturing hydrophobic surfaces that can generate slip \citep{wat99,ou2004laminar}, as well as to finding the optimal pattern geometries to apply on surfaces. For laminar flows, the drag reduction from using patterned stress-free surfaces has been studied both theoretically  \citep{philip1972flows,philip1972integral,lauga2003effective}, and numerically for both structured patterns \citep{cheng2009microchannel} and random patterns \citep{samaha2011modeling}. For turbulent flows, \cite{turk2014turbulent} and \cite{jelly2014turbulence} studied the drag reduction obtained from stress-free spanwise patterns in channel flow finding drag reductions of around $20\%$. Meanwhile, \cite{watanabe2017drag} studied striped patterns at an oblique angle to the flow, finding the maximum drag reduction for spanwise patterns. We wish to highlight that a pure focus on drag reduction, such as what is explored for example in \cite{lauga2003effective} is not the main interest of our study here. By patterning a surface with stress-free boundaries we naturally expect to see drag reduction. Similarly to the previous studies, we expect the pattern frequency and direction to affect the amount of drag reduction seen. However, we wish to focus on the relationship between how much drag reduction we see, and how the secondary structures are affected to better understand how these structures affect the torque. 

We note that Taylor-Couette flow with stress-free patterned surfaces was already studied by \cite{naim2019turbulent}, who performed numerical simulations of TC at Reynolds numbers in the order $Re_s\sim \mathcal{O}(10^3)$ and used one- and two-dimensional stress-free patterns on the cylinders. The patterns studied in that manuscript are similar to ours (cf. figure \ref{fig:fullschema}), and we will not describe them here. \cite{naim2019turbulent} report a maximum drag reduction of 34\% at $Re_s=5000$. They also report substantial modification of the large-scale structures. However, their Reynolds number is in the range $4000\le Re_s \le 5000$, which corresponds to a small region of parameter space.

In this manuscript, we will examine the effect of boundary heterogeneity on the flow for Reynolds number in the range of $Re_s \in (722,30000)$, which explores regions of parameter space from steady Taylor vortices to the fully turbulent regime. We will study the role of pattern geometry using several pattern shapes. To keep the parameter space manageable, we focus on the resulting flow organization and torque in TC with pure inner cylinder rotation, and will only apply stress-free conditions on the inner cylinder.

\section{Numerical method}
\label{sec:numerical_method}

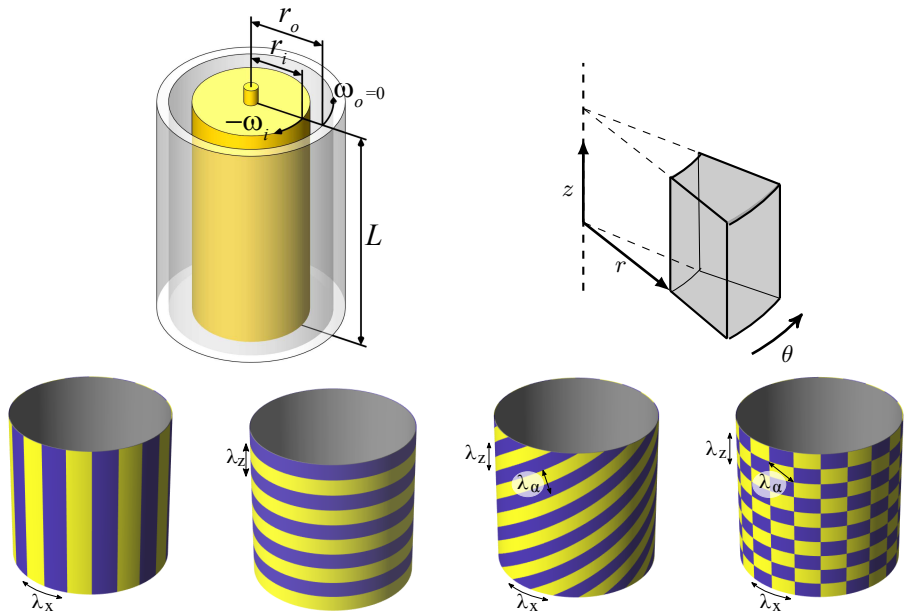
\begin{figure}
\begin{center}

\includegraphics[width=0.23\textwidth]{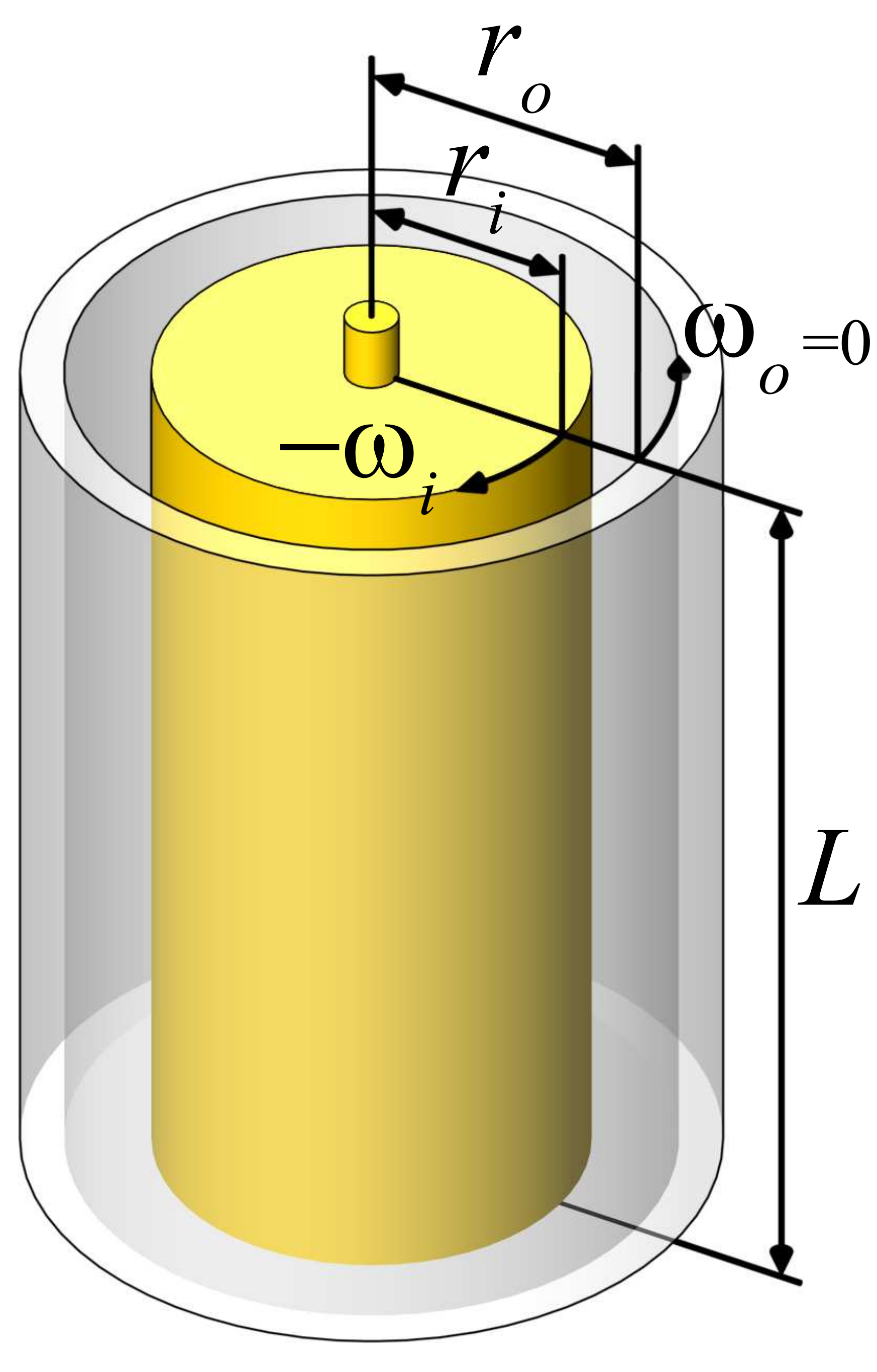}
\hspace{2cm}
    \begin{tikzpicture}
      \coordinate (O) at (0,0);
      \coordinate (Ox) at (-3,-3);
      \coordinate (Oy) at (4.243,0);
      \coordinate (Oz) at (0, 3);
      \coordinate (vz) at (0, 2);
      \coordinate (Phi) at (2.7,-0.6) ;
      \coordinate (r) at (0.5,0.5) ;
      \coordinate (z) at (-0.2,1.5) ;
      \coordinate (A1) at (0, 2.4);
      \coordinate (A2) at (0, 0.9);
      \coordinate (D) at (1.52, 1.82);
      \coordinate (P) at (1.9,0.9);
      \coordinate (Q) at (1.9,-0.63);
      \coordinate (S) at (1.15, 1.5);
      \coordinate (R) at (1.15, 0);
      \coordinate (C) at (1.54, 0.26);
      \coordinate (A) at (2.6, 1.4);
      \coordinate (B) at (2.6, -0.1);
      \coordinate (D) at (1.52, 1.82);
      \coordinate (Q) at (1.9,-0.63);
 
      \node[below] at (Phi) {$\theta$};
      \node[below] at (r) {$r$};
      \node[below] at (z) {$z$};

      \draw[-latex, line width=1] (A2)-- (vz);
      \draw[thick,dashed] (O)-- (Oz);       
      \draw[thin, dashed] (A1)--(S);
      \draw[thin, dashed] (A2)--(R);
      \draw[-latex,line width=1] (A2)--(R);
      \draw[thick] (P)--(Q);
      \draw[thin, dashed] (A1)--(D); 
      \draw[thin, dashed] (A2)--(C); 
      \draw[thick] ($(0, 2.4) + (310:1.8cm and 1.2cm)$(P) arc
         (310:330:1.8cm and 1.2cm);
      \draw[ultra thin] ($(0, 0.9) + (310:1.8cm and 1.2cm)$(P)arc
         (310:330:1.8cm and 1.2cm);
      \draw[thick] ($(0, 0.9) + (310:3cm and 2cm)$(P) arc
         (310:330:3cm and 2cm);
      \draw[thick] ($(0, 2.4) + (310:3cm and 2cm)$(P) arc
         (310:330:3cm and 2cm);
       \draw[->,>=stealth',line width=1] ($(0.3, 0.7) + (310:3cm and 2cm)$(P) arc
         (310:330:3cm and 2cm);
      \draw[thick] (A) --(B);
      \draw[thick] (S) --(R);
      \draw[ultra thin] (D) --(C);
      \draw[thick] (R) --(Q);
      \draw[thick] (S) --(P);
      \draw[thick] (D) --(A);
      \draw[ultra thin] (C) --(B);

      \filldraw[opacity=0.2]
          (D)--(A) arc (325:306:3cm and 2.2cm)--(S)
           arc (305:325:1.8cm and 1.2cm)--cycle;
      \filldraw[opacity=0.2]
          (P) arc (306:325:3cm and 2.2cm)--(B)
           arc (325:306:3.0cm and 2.2cm)--cycle;
       \filldraw[opacity=0.2]
         (P)--(Q)--(R)--(S)--cycle;

    \end{tikzpicture}
\end{center}
\includegraphics[width=0.23\textwidth]{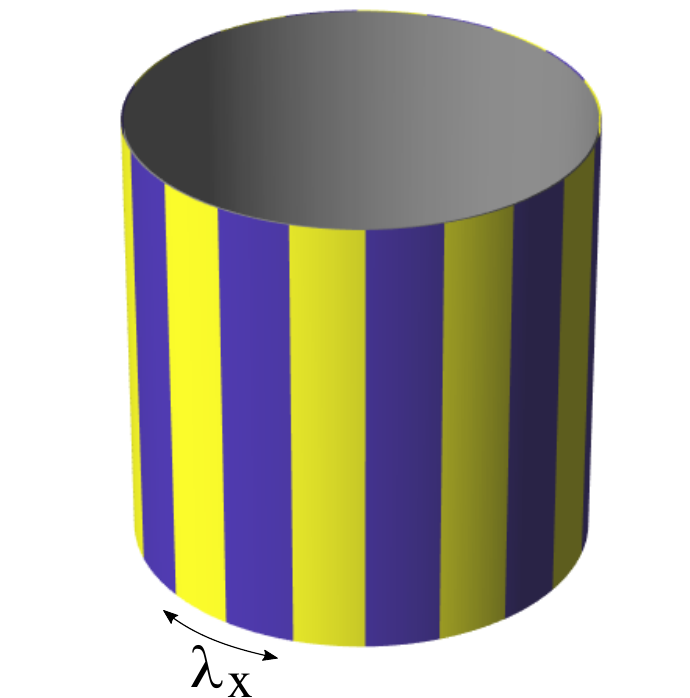}
\includegraphics[width=0.23\textwidth]{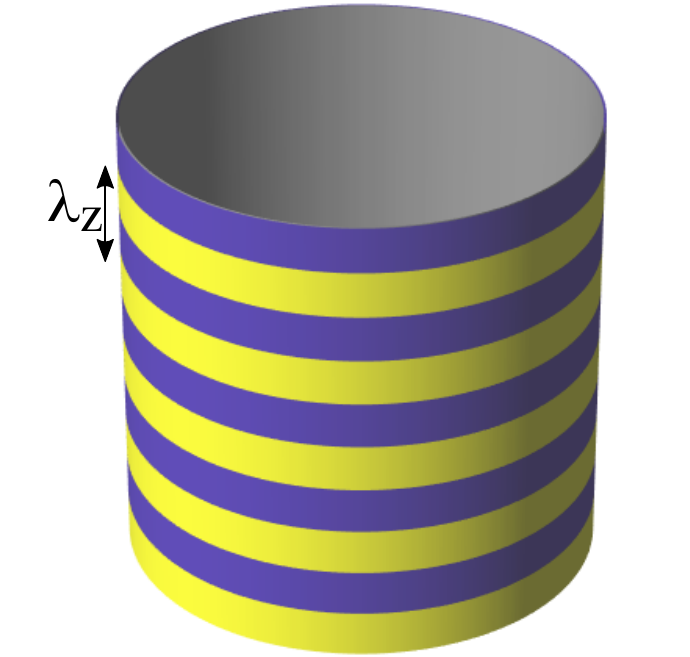}
\includegraphics[width=0.23\textwidth]{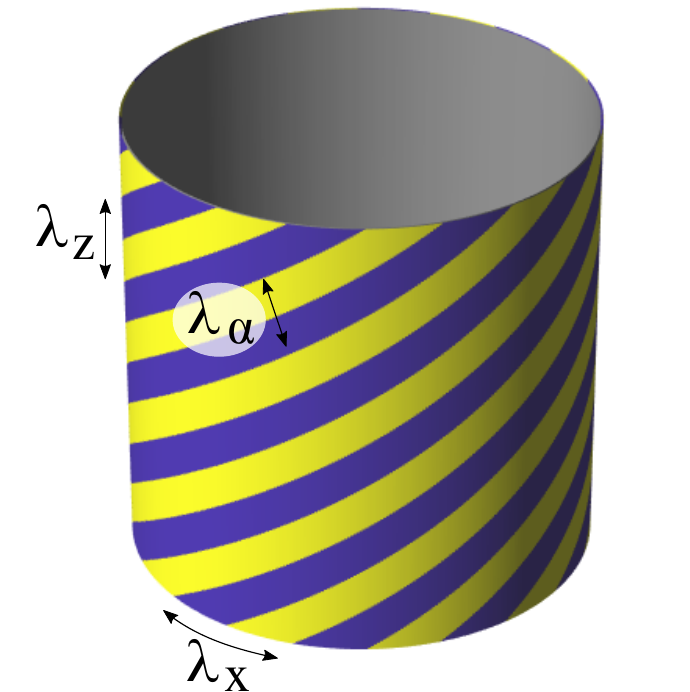}
\includegraphics[width=0.23\textwidth]{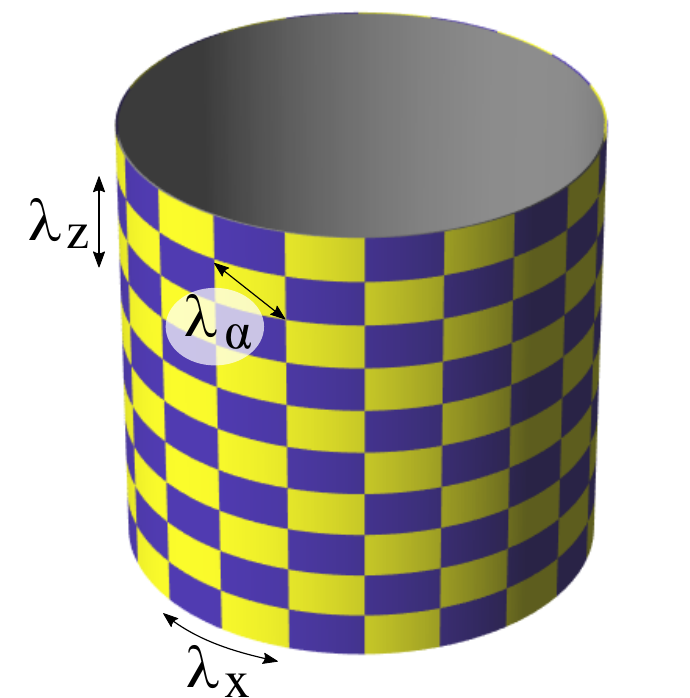}

\centering
\caption{Top left: Schematic of a Taylor-Couette geometry. Top right: Schematic of the computational domain and the coordinate system. Bottom row, from left to right: alternating free-slip and no-slip patterns over the rotating inner-cylinder, axially-oriented pattern which results in azimuthal (streamwise) inhomogeneity, azimuthally-oriented pattern which results in axial (spanwise) inhomogeneity, spiral pattern and checkerboard pattern.}
    \label{fig:fullschema}
\end{figure}

We directly simulate Taylor-Couette flow by solving the incompressible Navier-Stokes equations in cylindrical coordinates in a rotating frame described below:

\begin{equation}
    \pdv{\textbf{u}}{t} + \textbf{u}\cdot \nabla{\textbf{u}} + 2\Omega_{rf} \times \textbf{u} = -\nabla{p} + \nu \nabla^2{\textbf{u}},
\end{equation}
\begin{equation}
    \div{\textbf{u}} = 0,
\end{equation}

\noindent where $\textbf{u}$ is the velocity, $\Omega_{rf}$ the angular velocity of the rotating frame, $p$ the pressure and $t$ the time. We use a second-order energy-conserving central finite-difference scheme for the spatial discretization. Time is advanced using a third-order Runge-Kutta for the explicit terms and a second-order Crank-Nicolson scheme for the implicit treatment of the wall-normal viscous terms. The complete algorithm is described in \cite{verzicco1996finite} and \cite{van2015pencil}. We use the open-source code AFiD, which has been parallelized using MPI directives and has been heavily validated for Taylor-Couette flow \citep{ostilla2014exploring}.

Taylor-Couette flow has two non-dimensional geometrical parameters which define the system: the radius ratio $\eta=r_i/r_o$, where $r_i$ and $r_o$ are the radius of inner and outer cylinders respectively, and the aspect ratio $\Gamma=L/d$, where $d=r_o-r_i$ is the gap between both cylinders, and $L$ is the vertical height (or axial periodic length) of the system. In our case, we fix $\eta=0.909$, a small gap, and $\Gamma=2.33$ with axially periodic boundary conditions, which corresponds to allowing a single pair of rolls of wavelength $\lambda_{TR}/d=2.33$.  While we artifically fix $\lambda_{TR}$, in real systems the roll axial wavelength adjusts to the flow and to the endwall locations (if nearby), For the purpose of this research, we ignore this effect, and justification in Appendix \ref{sec:appa}, where we show results for a different roll wavelength $\lambda_{TR}/d=3$, as well as for two pairs of rolls, to ensure that the results are as independent as possible from domain size effects. We find that while the quantitative values of the results change (and less than $10\%$), the qualitative features of the results presented below are robust. 

We simulate TC flow in the rotating frame discussed by \cite{dubrulle2005stability}, where the inner and outer cylinders velocities are set to $\pm \frac{1}{2}U$, with $U$ being the characteristic velocity. With this, we have two control parameters: the shear Reynolds number $Re_s=Ud/\nu$, where $\nu$ is the fluid kinematic viscosity, and the non-dimensional Coriolis parameter $R_\Omega=2d\Omega/U$. For pure inner cylinder rotation at $u_i$ inner cylinder velocity, we obtain $R_\Omega=1-\eta$ and $U=2u_i/(1+\eta)$. A schematic of the system is shown in figure \ref{fig:fullschema}, where the azimuthal ($\theta$), radial ($r$) and axial ($z$) coordinates are indicated. For convenience, we define the non-dimensional radial coordinate $\tilde{r}=(r-r_i)/d$, the non-dimensional axial coordinate $\tilde{z}=z/d$, and the non-dimensional streamwise coordinate $\tilde{x}=r\theta/d$. Quantities are non-dimensionalized using the inner cylinder velocity $u_i$, and the characteristic length $d$ unless stated otherwise. From here on, any quantity will be represented non-dimensionally unless stated otherwise.

We set the order of rotational symmetry in the azimuthal direction to $n_{sym}=20$ to reduce computational costs. This results in a streamwise periodic length of $2\pi$ half-gaps in the mid-gap, such that $\tilde{x}$ is in the $(0,\pi)$ range. This is enough to generate accurate statistics when no patterns are present \citep{ostilla2015effects,sacco2019dynamics}. To further prove that the results are independent of $n_{sym}$, additional simulations are provided in Appendix \ref{sec:appa}. Depending on the Reynolds number, we vary the number of discretization points $N_i$ of the simulation grid. In the current study, we simulate $Re_s=7.22\times10^2$ (steady Taylor vortex regime) with a resolution of $N_\theta \times N_r \times N_z =32 \cross 64 \cross 64$, $Re_s=2.28\times 10^3$ (modulated Taylor vortex regime) with $128 \cross 256 \cross 128$, $Re_s=10^4$ (turbulent Taylor vortex regime) with $192 \cross 384 \cross 256$ and $Re_s=3\times10^4$ (turbulent Taylor vortex regime) with $384 \cross 512 \cross 512$. This covers everything from steady Taylor Vortex flow to fully turbulent flow in the ultimate regime. A table containing all simulated cases and the resulting torque is provided in Appendix \ref{sec:appb}.

The adequacy of the resolution is checked by ensuring the energy input from the cylinders matches the viscous dissipation to within $2\%$ \citep{ostilla2014exploring}. The simulations are started from initial conditions of white noise, and are ran for 400 large-eddy turnover times (defined as $d/U$) to collect statistics after the transient has passed.  An additional criterion for temporal convergence of the statistics is that the angular velocity transport, defined as $J_\omega(r)=r^3(\langle u_r u_\theta/r \rangle_{\theta,z,t} - \nu\partial_r\langle u_\theta/r \rangle_{\theta,z,t})$ \citep{eckhardt2007torque} must be constant to within $2\%$, where $\langle ... \rangle_{x_i}$ denotes an averaging operator over variable $x_i$. What this expresses is that the time-average of the convective and viscous parts of the torque are approximately constant at every radial coordinate. This provides a more stringent criterion than just looking at the convergence of torque, as we use a second-order statistic $(\langle u_r u_\theta/r\rangle_{\theta,z,t})$ to ensure converge. The torque $T$ is then calculated as the radial average of $J(r)$.

In a classical TC problem, the cylinders have a homogeneous no-slip boundary condition, where the velocity of the fluid at the wall is simply the cylinder velocity. The flow is maintained by a torque applied at the cylinders $T$, which is usually non-dimensionalized in the form of Nusselt number $Nu=T/T_{pa}$, where $T_{pa}$ is the torque for the purely azimuthal flow solution. In the current study, we alternate no-slip and free-slip boundary conditions. Free-slip boundary conditions are mathematically expressed by the combination of no penetration ($u_r=0$), and vanishing normal derivatives of the two velocity components tangential to the wall ($\partial_r u_\theta=\partial_r u_z=0$). We impose the stress-free boundary condition by setting the shear $\tau$ originating from the wall to zero at the first grid point. This is done through modifying the viscous term, which for a staggered velocity is first approximated using a finite difference of shears ($[\tau^+-\tau^-]/\Delta$), and these shears are then approximated using a finite difference of velocities. The finite difference approach of the code rapidly allows us to change between no-slip and stress free conditions by setting $\tau^-$ to be either zero, or to be the velocity gradient between the current velocity and the wall. More details on the code operation are provided in \cite{van2015pencil}. 

There are infinitely many geometric patterns of no-slip and free-slip conditions possible. In this study, we reduce the geometrical parameter space to only consider the four patterns shown in figure \ref{fig:fullschema}. We further reduce the parameter space by equally partitioning the patterned surfaces between no-slip and free-slip boundary conditions. If the repeating pattern is one-dimensional infinite stripes, it suffices to state the orientation of the stripes, and the spatial frequency to determine it. In particular, the spatial frequency $f_j$ is the number of repeating patterns per unit gap-width, and it is the inverse of the spatial period, or wavelength $\lambda_j$ of the pattern. In the case of alternating free-slip and no-slip boundary conditions in the azimuthal (streamwise) direction, the pattern wavelength is given by $\lambda_x = \hat{L}_x/f=2\pi r_i/(dn_{sym}f)$, while for stripes that alternate in the axial (spanwise) direction (figure \ref{fig:ax1}) the pattern wavelength is $\lambda_z=\Gamma/f$. For the inclined/spiral patterns, there are two ways to calculate the effective wavelength $\lambda_\alpha$ of the pattern, as it extends in both stream and spanwise directions. It is given by either $\lambda_x \sin(\alpha)$, or $\lambda_z\cos(\alpha)$, where $\alpha$ is the angle of inclination of pattern with respect to the streamwise direction. These formulas become invalid in the limits of $\alpha=0$ and $\alpha=\pi/2$, respectively. A visual can be seen in figure \ref{fig:istr}.

The checkerboard patterns can be quantified through two wavelengths: one in the axial direction and one in the azimuthal direction. When these are equal, we obtain a checkerboard pattern composed of repeating squares, and otherwise we obtain repeating rectangles. In the limit of one of the two wavelengths becoming infinity, we obtain either pure axial or azimuthal variations. Further visuals on how $\lambda_\alpha$ or an equivalent orientation of a checkerboard pattern can be defined are provided in figure \ref{fig:cbstr}.

\section{Pure axial (spanwise) and azimuthal (streamwise) inhomogeneities}
\label{sec:results1}

\begin{figure}
\centering
\begin{subfigure}{0.32\textwidth}
\begin{center}
  \caption{} 
  \adjincludegraphics[width=\textwidth,trim={{0.14\width} {.05\width} {0.05\width} {0.04\width}},clip]{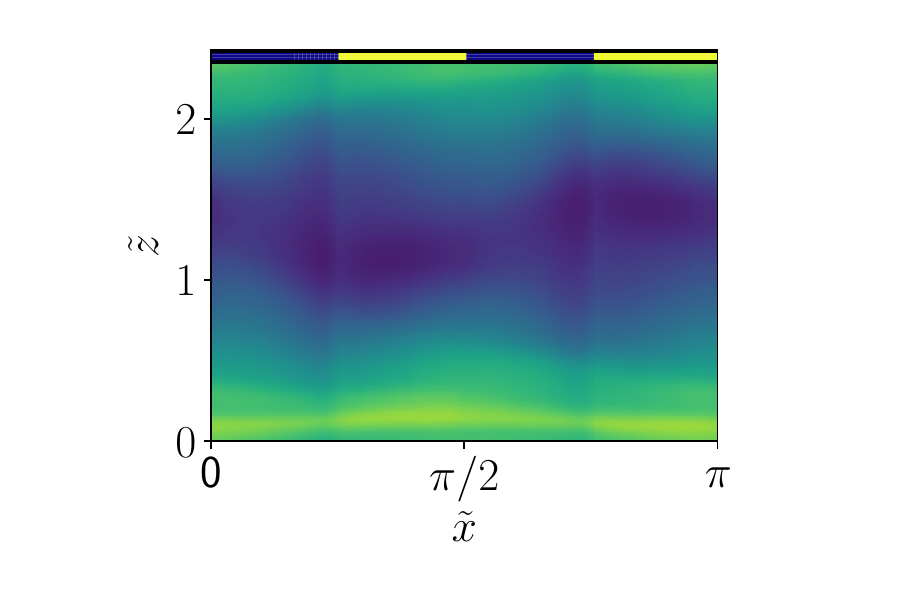}
  \label{fig:ax1_2280}
\end{center}
\end{subfigure}
\begin{subfigure}{0.32\textwidth}
\begin{center}
  \caption{} 
  \adjincludegraphics[width=\textwidth,trim={{0.14\width} {.05\width} {0.05\width} {0.04\width}},clip]{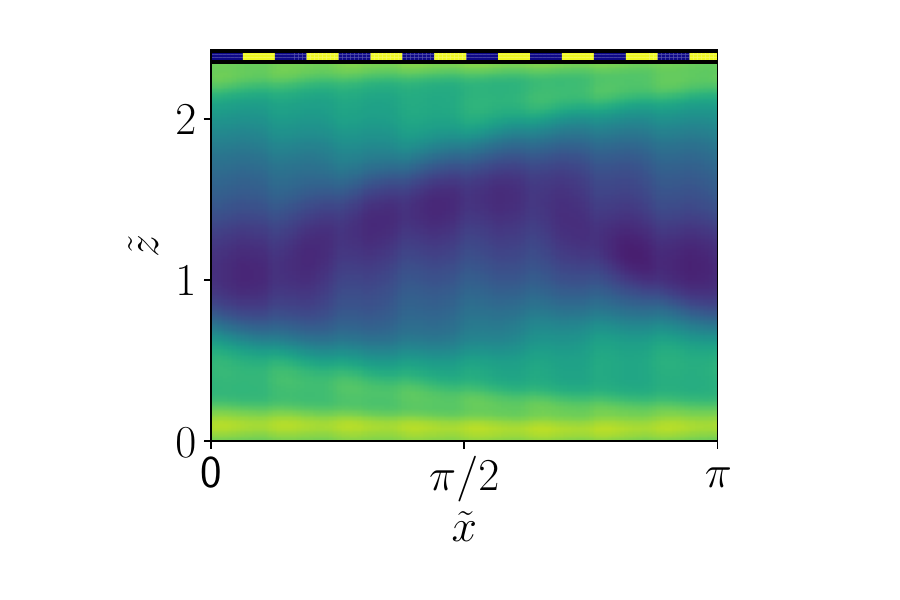}
    \label{fig:ax2_2280}
\end{center}
\end{subfigure}
\begin{subfigure}{0.32\textwidth}
\begin{center}
  \caption{} 
  \adjincludegraphics[width=\textwidth,trim={{0.1\width} {.05\width} {0.05\width} {0.04\width}},clip]{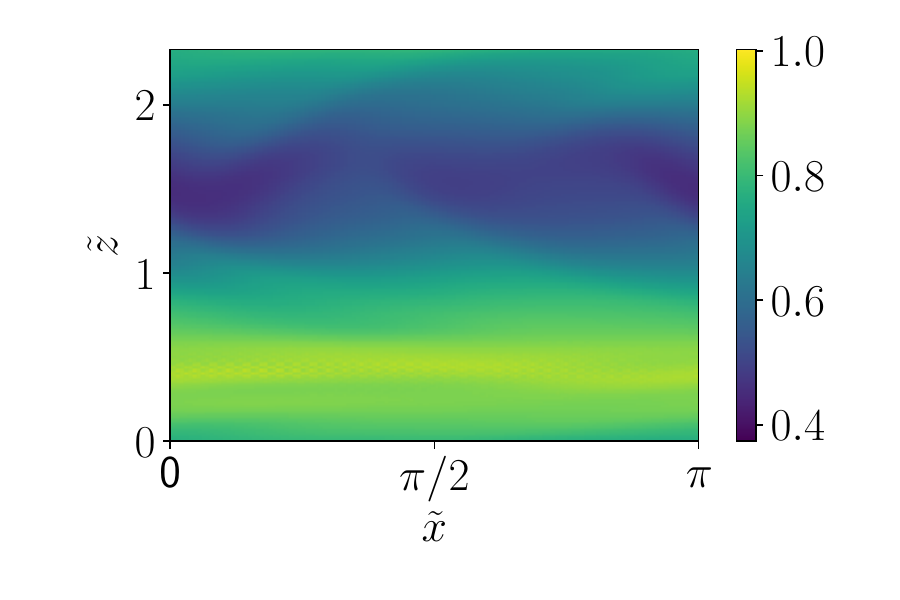}
    \label{fig:ax3_2280}
\end{center}
\end{subfigure}
\begin{subfigure}{0.32\textwidth}
\begin{center}
  \caption{} 
  \adjincludegraphics[width=\textwidth,trim={{0.14\width} {.05\width} {0.05\width} {0.04\width}},clip]{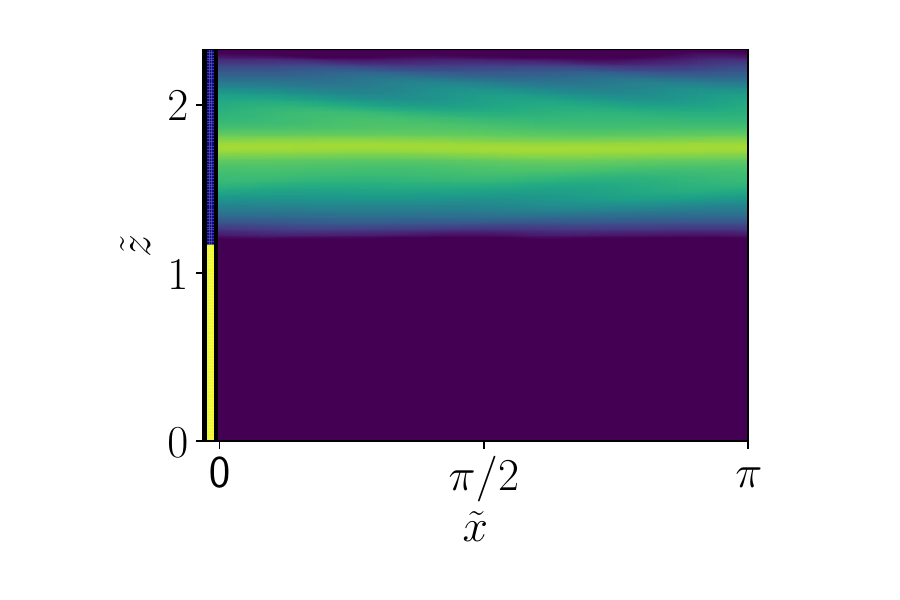}
    \label{fig:az1_2280}
\end{center}
\end{subfigure}
\begin{subfigure}{0.32\textwidth}
\begin{center}
  \caption{} 
  \adjincludegraphics[width=\textwidth,trim={{0.14\width} {.05\width} {0.05\width} {0.04\width}},clip]{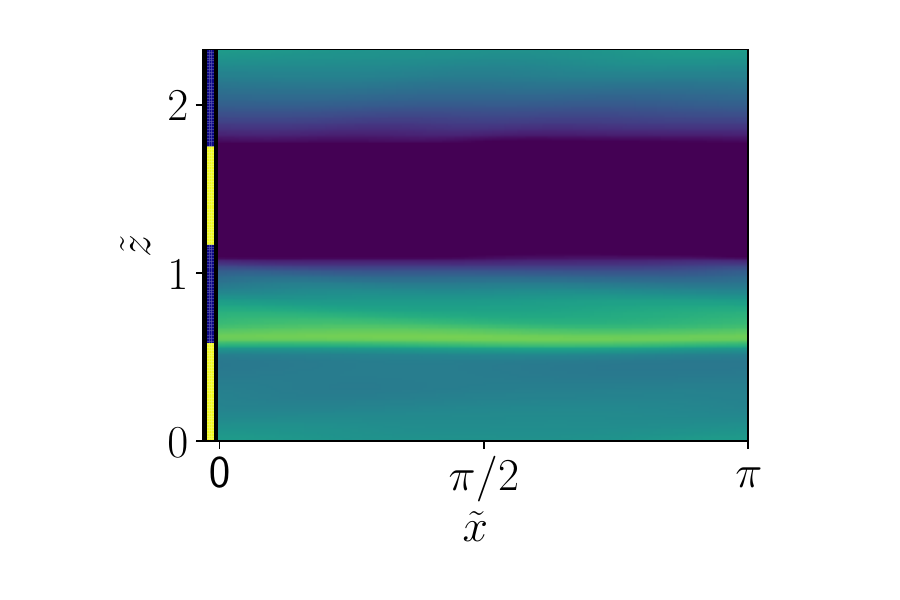}
    \label{fig:az2_2280}
\end{center}
\end{subfigure}
\begin{subfigure}{0.32\textwidth}
\begin{center}
  \caption{} 
  \adjincludegraphics[width=\textwidth,trim={{0.14\width} {.05\width} {0.05\width} {0.04\width}},clip]{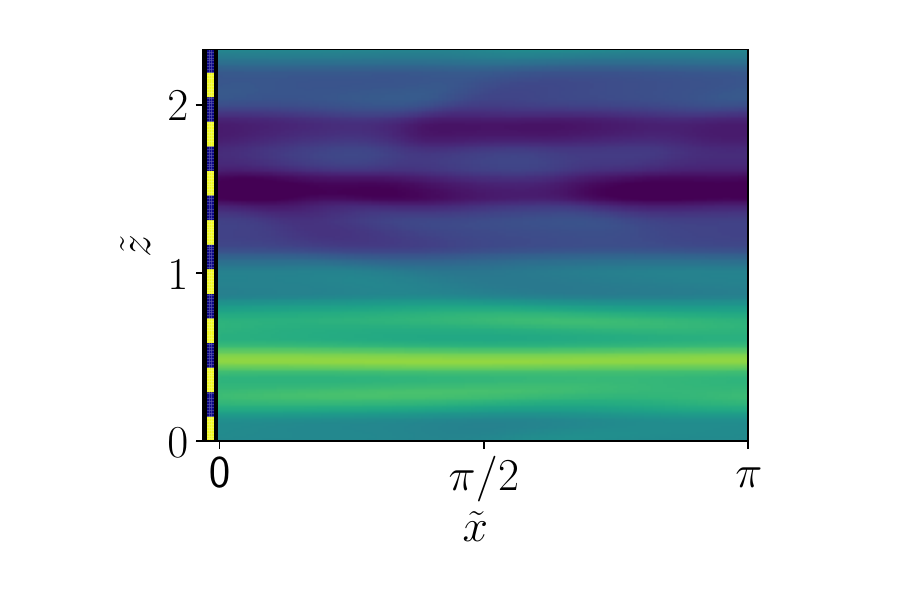}
    \label{fig:az3_2280}
\end{center}
\end{subfigure}
\centering
\caption{ Near-wall instantaneous azimuthal velocity for (\textit{a}) and (\textit{b}): azimuthal inhomogeneity with $f_x=0.32$ and $f_x=2.55$ respectively; (\textit{c}): no inhomogeneity; and (\textit{d}),(\textit{e}) and (\textit{f}): axial inhomogeneities with $f_z=0.43$, $f_z=0.86$ and $f_z=3.43$ respectively. All results are for $Re_s=2.28 \times 10^3$. Yellow stripes correspond to stress-free zones and blue zones correspond to no-slip zones.}
  \label{fig:inst_2280}
\end{figure}

\begin{figure}
\centering
\begin{subfigure}{0.32\textwidth}
\begin{center}
  \caption{} 
  \adjincludegraphics[width=\textwidth,trim={{0.14\width} {.05\width} {0.05\width} {0.04\width}},clip]{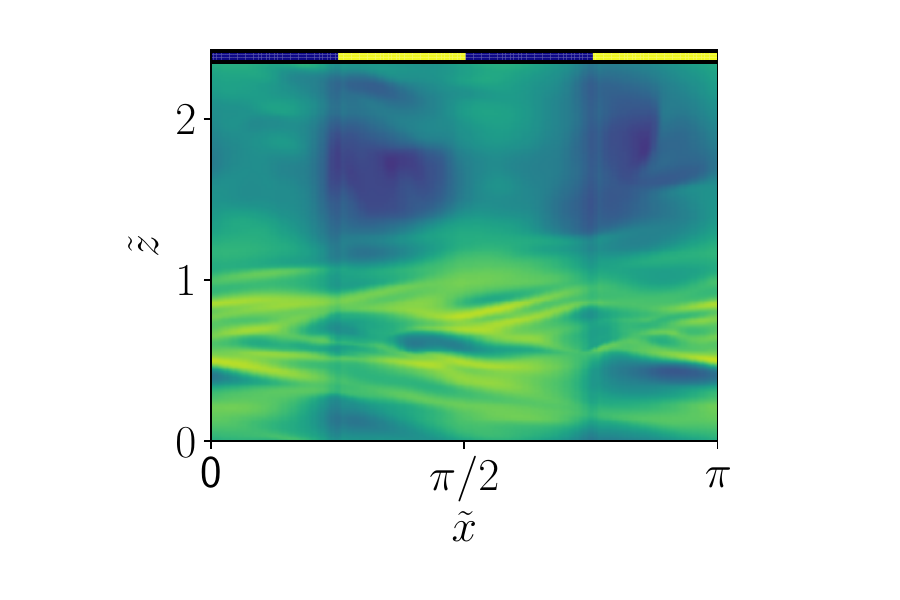}
  \label{fig:ax1}
\end{center}
\end{subfigure}
\begin{subfigure}{0.32\textwidth}
\begin{center}
  \caption{} 
  \adjincludegraphics[width=\textwidth,trim={{0.14\width} {.05\width} {0.05\width} {0.04\width}},clip]{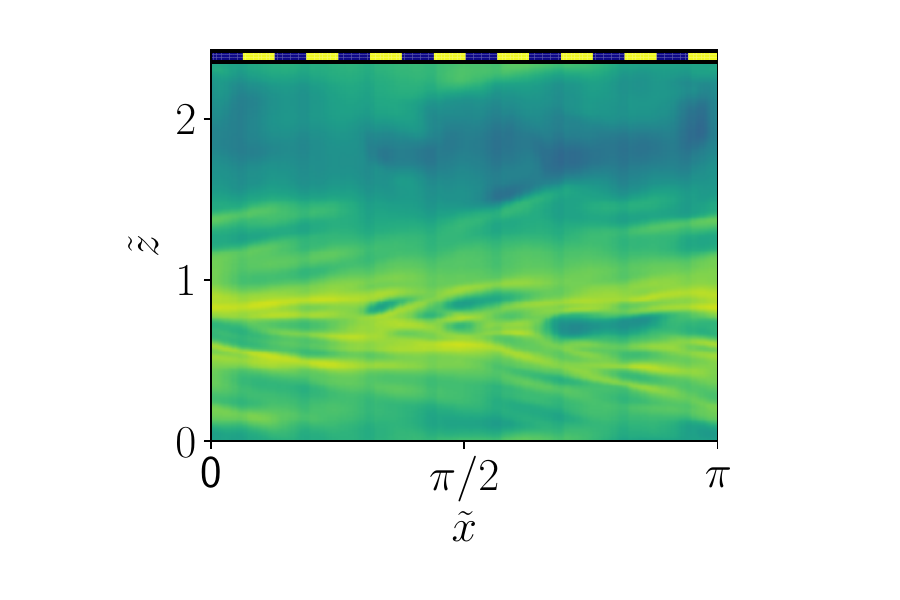}
    \label{fig:ax2}
\end{center}
\end{subfigure}
\begin{subfigure}{0.32\textwidth}
\begin{center}
  \caption{} 
  \adjincludegraphics[width=\textwidth,trim={{0.1\width} {.05\width} {0.05\width} {0.04\width}},clip]{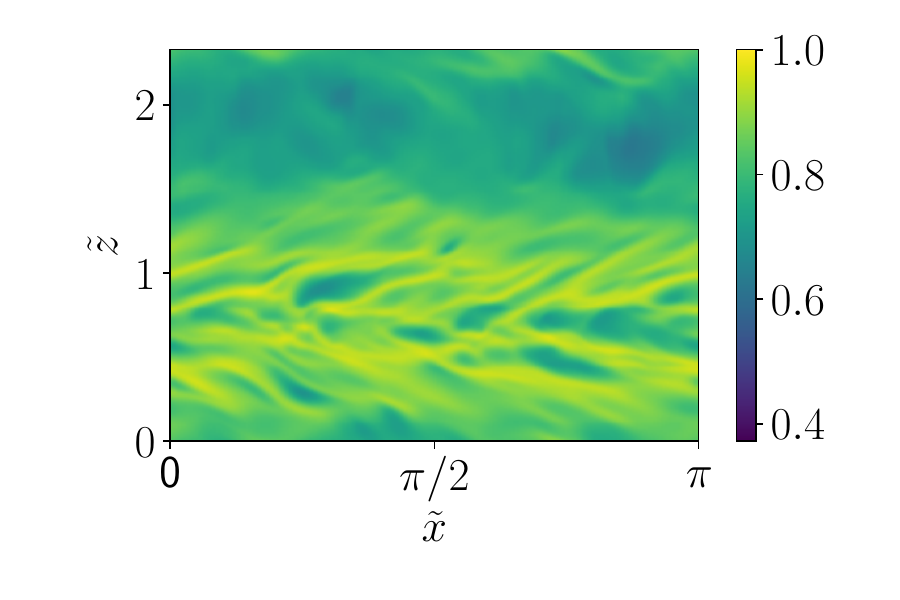}
    \label{fig:ax3}
\end{center}
\end{subfigure}
\begin{subfigure}{0.32\textwidth}
\begin{center}
  \caption{} 
  \adjincludegraphics[width=\textwidth,trim={{0.14\width} {.05\width} {0.05\width} {0.04\width}},clip]{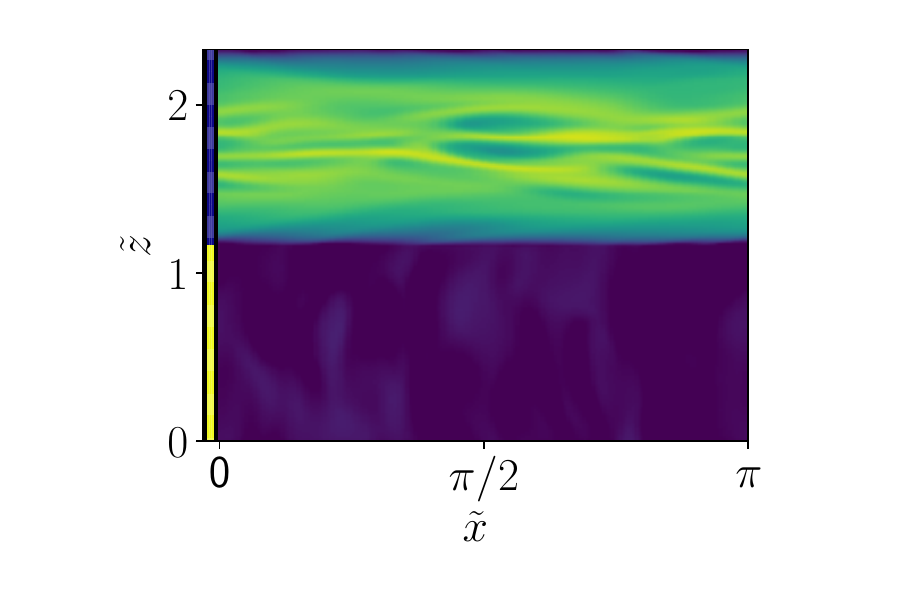}
    \label{fig:az1}
\end{center}
\end{subfigure}
\begin{subfigure}{0.32\textwidth}
\begin{center}
  \caption{} 
  \adjincludegraphics[width=\textwidth,trim={{0.14\width} {.05\width} {0.05\width} {0.04\width}},clip]{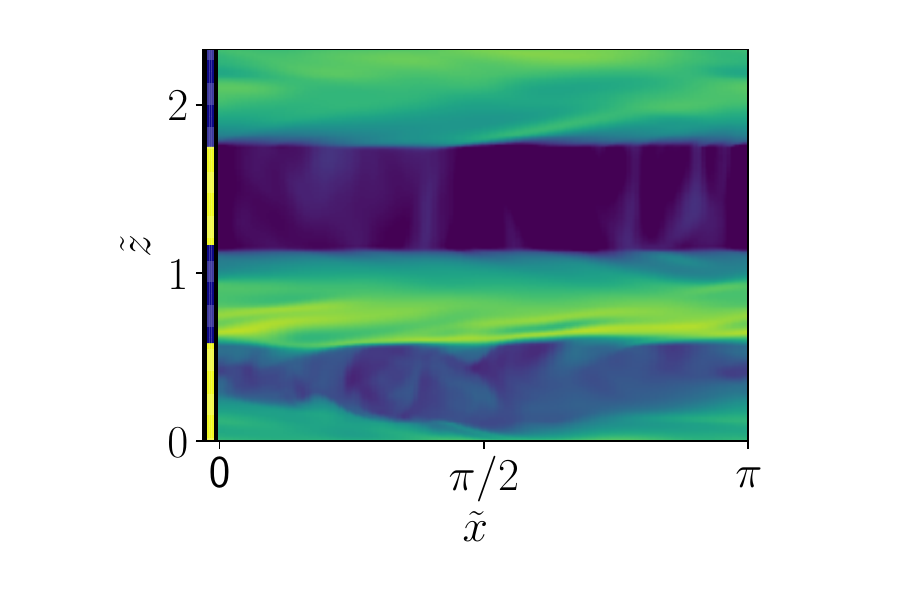}
    \label{fig:az2}
\end{center}
\end{subfigure}
\begin{subfigure}{0.32\textwidth}
\begin{center}
  \caption{} 
  \adjincludegraphics[width=\textwidth,trim={{0.14\width} {.05\width} {0.05\width} {0.04\width}},clip]{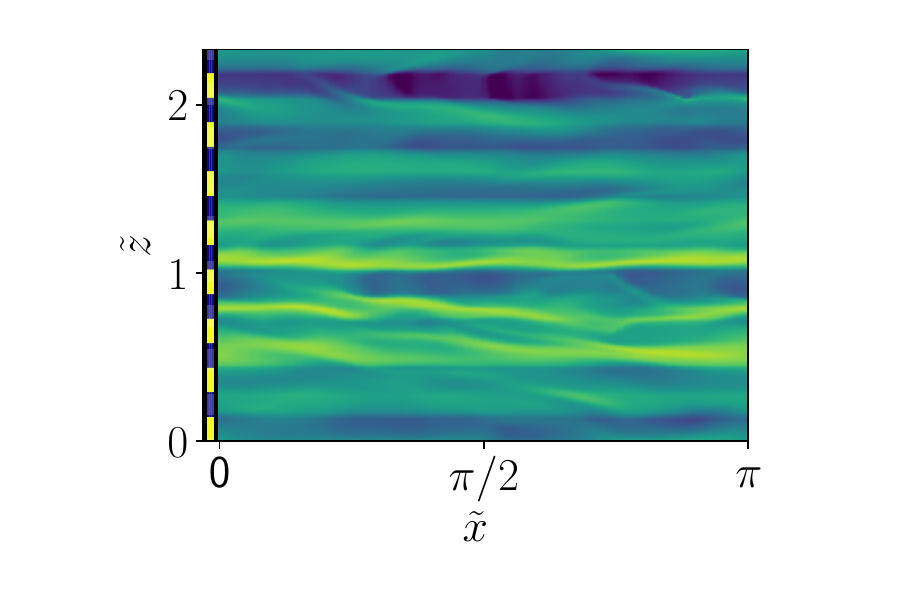}
    \label{fig:az3}
\end{center}
\end{subfigure}
\centering
\caption{ Near-wall instantaneous azimuthal velocity for (\textit{a}) and (\textit{b}): azimuthal inhomogeneity with $f_x=0.32$ and $f_x=2.55$ respectively; (\textit{c}): no inhomogeneity; and (\textit{d}),(\textit{e}) and (\textit{f}): axial inhomogeneities with $f_z=0.43$, $f_z=0.86$ and $f_z=3.43$ respectively. All results are for $Re_s=10^4$. Yellow stripes correspond to stress-free zones and blue zones correspond to no-slip zones.}
  \label{fig:azt}
\end{figure}

The first patterns we discuss are azimuthal and axial patterns, which can be understood as purely streamwise and purely spanwise inhomogeneous boundary conditions respectively (bottom first and bottom second from left in Figure \ref{fig:fullschema}). The case of axial (spanwise) variations is particularly interesting, because it is expected to generate the sort of Reynolds stress imbalances that induce additional secondary flows. In addition, axial patterns are ``frozen'' from the point of view of cylinder and because they are invariant in time, they are easier to analyze.

We begin by showing the instantaneous velocity near the inner cylinder at $\tilde{r}\approx0.05$ for the intermediate Reynolds number case $Re_s=2.28\times10^3$, in the modulated Taylor vortex regime. We first focus on the homogeneous case, where we can see strong axial variation, due to the presence of the Taylor roll, as well as azimuthal patterns caused by the wavyness of the roll. This azimuthal waviness is disrupted in Figures \ref{fig:ax1_2280}-\ref{fig:ax2_2280} by the insertion of the azimuthal inhomogeneities. It is also disrupted by the insertion of axial inhomogeneties. Furthermore, the axial stratification changes character when axial patterns are introduced. There is not much evidence of turbulence, and we expect the Reynolds stresses to be weak in this case. We highlight that this Reynolds number is most similar to those which have been explored by \cite{naim2019turbulent}. However, due to the small radius ratio they considered, the inner cylinder is more active in producing turbulent streaks which explains the qualitative differences between the cases in \cite{naim2019turbulent} and our simulations.

In the turbulent Taylor vortex regime, the picture is somewhat similar to what was analyzed above. In figure \ref{fig:azt}, we show the instantaneous azimuthal velocities near the inner cylinder for $Re_s=10^4$ at $\tilde{r}\approx0.015$ for various patterns. This value of $\tilde{r}$ roughly corresponds to $y^+=13-15$ in wall-units, depending on the case, where $y^+=(r-r_i)/\delta_\nu$, $\delta_\nu$ is the viscous unit $\delta_\nu=\nu/u_\tau$, and $u_\tau$ is the frictional velocity. Near-wall turbulent streaks can be seen due to the higher Reynolds number. For the case of azimuthal variations, the influence of the pattern on the flow is minor for the lowest frequencies. The web of streaks is somewhat disrupted, but the effect is small due to the inherent elongation of streaks in the streamwise direction which bridges the gap across patterns effectively. Aside from weak signatures of the pattern, these cases do not significantly deviate from the homogeneous case shown in the top-right panel. This is consistent with the fact that azimuthal inhomogeneities will not generate additional spanwise gradients of Reynolds stresses which are known to induce secondary flows. Thus, as first approximation, we do not expect them to affect the Taylor rolls much.

Axial inhomogenities have a larger effect on the flow structure near the walls for the lower frequencies. Parts of the flow are never forced, while other parts are always forced. The azimuthal elongation of streaks cannot adequately redistribute momentum in the axial direction, and larger changes in velocity than those seen in the homogeneous case  can be observed. For the lowest frequency, the roll structure is somewhat disrupted. The pattern wavelength coincides with the roll wavelength, so the roll ``places'' itself above the no-slip boundary, and this mitigates the overall effect the inhomogeneities have on the flow. However, for $f_z=0.86$ (Fig.~\ref{fig:az2}), the pattern wavelength is double the roll wavelength, and this significantly interferes with the roll formation. As the frequency is further increased, the destructive interference is weakened. For the highest frequency simulated ($f_z=3.43$, Fig.~\ref{fig:az3}) the axial variation is less strong even if the signature of the pattern is much more apparent than for azimuthal variations.

\begin{figure}
\centering
\begin{subfigure}{0.49\textwidth}
\begin{center}
  \caption{} 
  \adjincludegraphics[width=\textwidth,trim={{0\width} {.05\width} {0\width} {0\width}},clip]{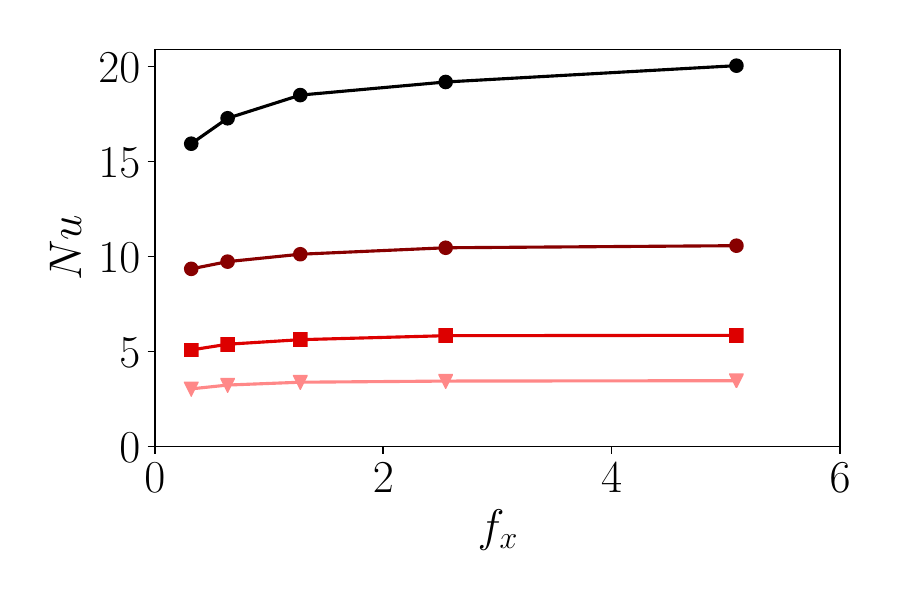}
\label{fig:axtn}
\end{center}
\end{subfigure}
\begin{subfigure}{0.49\textwidth}
\caption{} 
\begin{center}
  \adjincludegraphics[width=\textwidth,trim={{0\width} {.05\width} {0\width} {0\width}},clip]{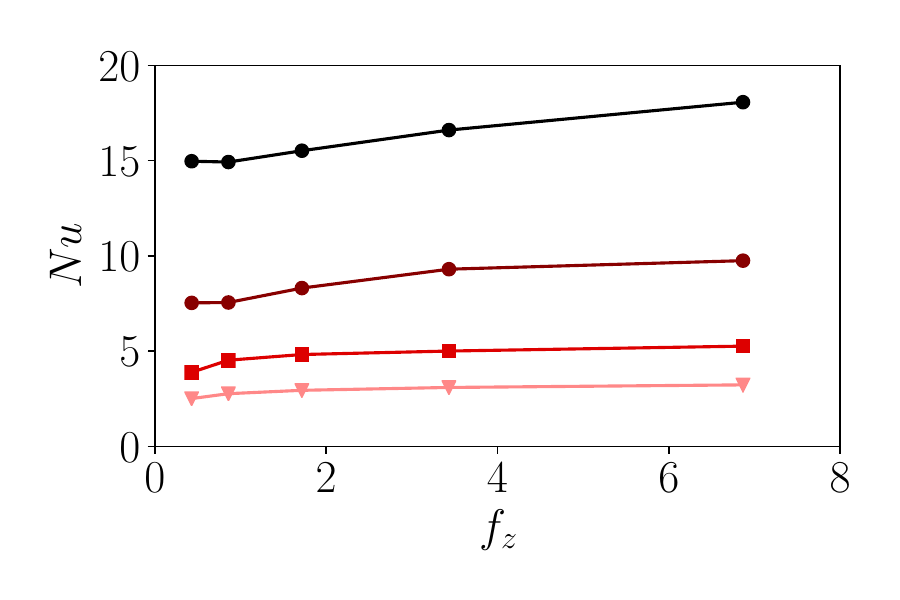}
\end{center}
\label{fig:aztnn}
\end{subfigure}\\
\begin{subfigure}{0.49\textwidth}
\begin{center}
  \caption{} 
  \adjincludegraphics[width=\textwidth,trim={{0\width} {.05\width} {0\width} {0\width}},clip]{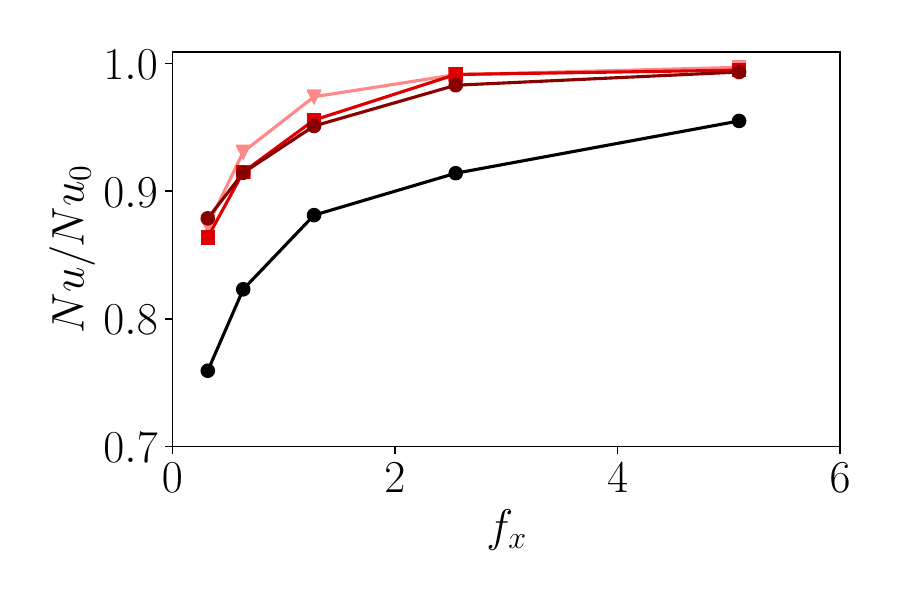}
\label{fig:axtn2}
\end{center}
\end{subfigure}
\begin{subfigure}{0.49\textwidth}
\caption{} 
\begin{center}
  \adjincludegraphics[width=\textwidth,trim={{0\width} {.05\width} {0\width} {0\width}},clip]{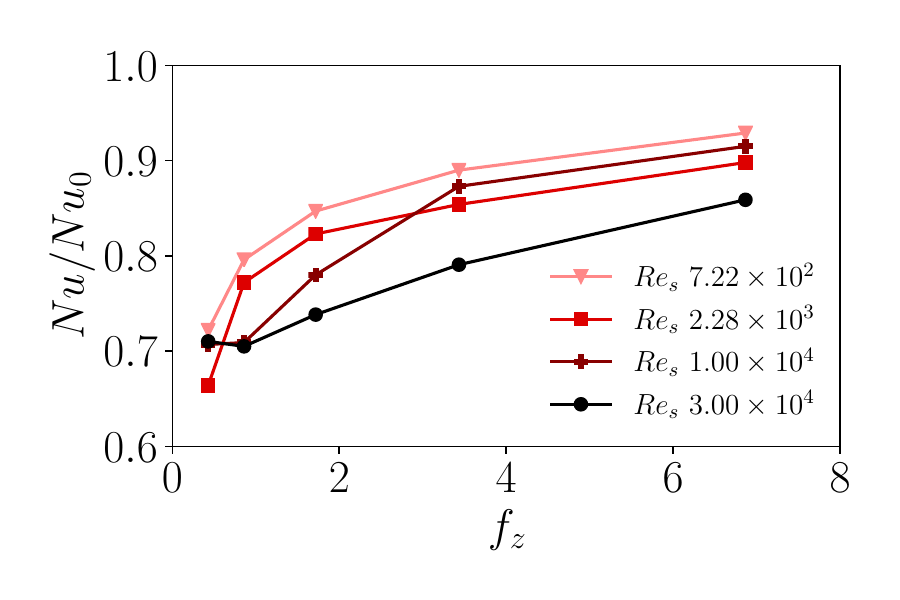}
\end{center}
\label{fig:aztn}
\end{subfigure}
\centering
\caption{(\textit{a,b}) Non-dimensionalized ($Nu$) and (\textit{c,d}) normalized torque ($Nu/Nu_0$, with $Nu_0$ the non-dimensionalized torque for homogeneous cylinders) at various wavelengths and Reynolds number. (\textit{a,c}): Azimuthal inhomogeneities as in figure \ref{fig:fullschema}c. (\textit{b,d}): Axial inhomogeneities as in figure \ref{fig:fullschema}d. }
\label{fig:torquesaznax}
\end{figure}

We continue by showing the non-dimensional torque $Nu$ for various Reynolds number and stripe wavelengths in the top panels of figure \ref{fig:torquesaznax}. Our first observation is that the $Nu(k)$ curve is monotonic for almost all cases except for the large $Re$ cases of Figure \ref{fig:torquesaznax}(b,d) to which we will return later to. Smaller pattern sizes result in higher torques even if the area ratio between no-slip and stress-free regions is kept constant at 50\% each, consistent with the existing literature \citep{watanabe2017drag,naim2019turbulent}. These results are emphasised in the bottom plots, which show the Nusselt number normalized by that obtained when the walls are fully no-slip ($Nu/Nu_0$). In the case of azimuthal variations, the patterns have a small effect: except for $Re_s=3\times10^4$, the torque drops down only to $85\%$ of the homogeneous value for the smallest frequencies (largest patterns) simulated. For the axial (spanwise) inhomogeneities the relative drops in torque is much larger ($\sim30\%$). This finding is consistent with \cite{watanabe2017drag} and \cite{naim2019turbulent}, who obtained the largest drop in friction for stripes aligned in the axial direction. The magnitudes of drag reduction achieved are also in the same order as those in these previous studies, in particular, with the $32\%$ drag reduction seen for axial patterns in \cite{naim2019turbulent}.

Increasing the frequency increases the torque, and for the smaller $Re_s$ cases with azimuthal inhomogeneities, the asymptotic value of $Nu/Nu_0 \to 1$ is reached in practice for $f_x>3$ even if only $50\%$ of the cylinder surface is providing the torque. This means that the local shear at these parts will be almost double than the one present for the homogeneous case. For $Re_s=3\times10^4$ the results also tend to an asymptote but at a slower rate. Our results are consistent with the numerical Rayleigh-B\'enard studies by \cite{bakhuis2018mixed}, which found a similar effect on the Nusselt number: with a $50\%$ area coverage of conducting surface, the full heat transfer of the homogeneous case was recovered for the smallest patterns of alternating insulating and conducting surfaces.

\begin{figure}
\begin{center}
\begin{subfigure}{.2\textwidth}
  \centering
    \caption{} 
    \adjincludegraphics[width=1\textwidth, trim={{.3\width} {.05\width} {.38\width} {.04\width}},clip]{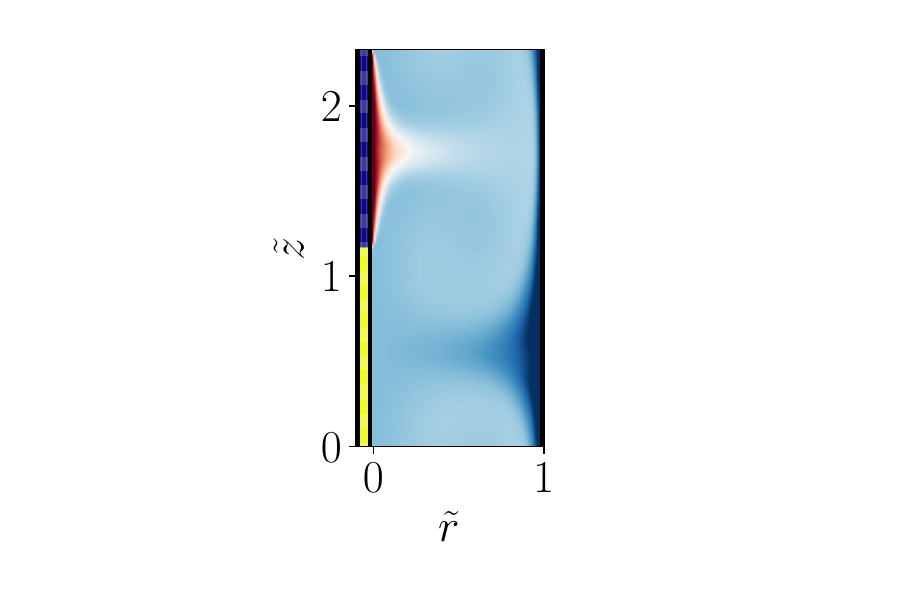}
  \label{fig:azv1}
\end{subfigure}
\begin{subfigure}{.2\textwidth}
  \centering
    \caption{} 
    \adjincludegraphics[width=1\textwidth, trim={{.3\width} {.05\width} {.38\width} {.04\width}},clip]{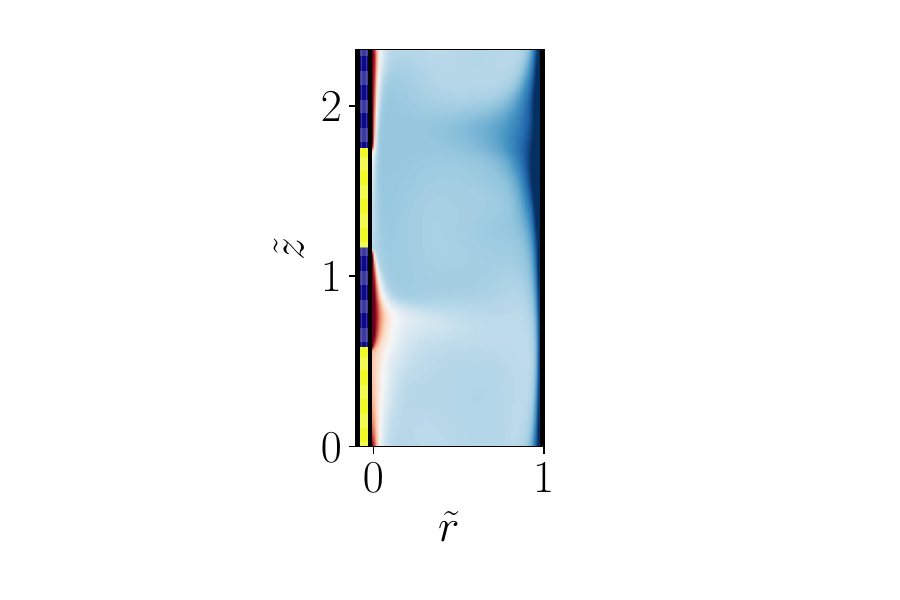}  
  \label{fig:azv2}
\end{subfigure}
\begin{subfigure}{.2\textwidth}
  \centering
  \caption{} 
    \adjincludegraphics[width=1\textwidth, trim={{.3\width} {.05\width} {.38\width} {.04\width}},clip]{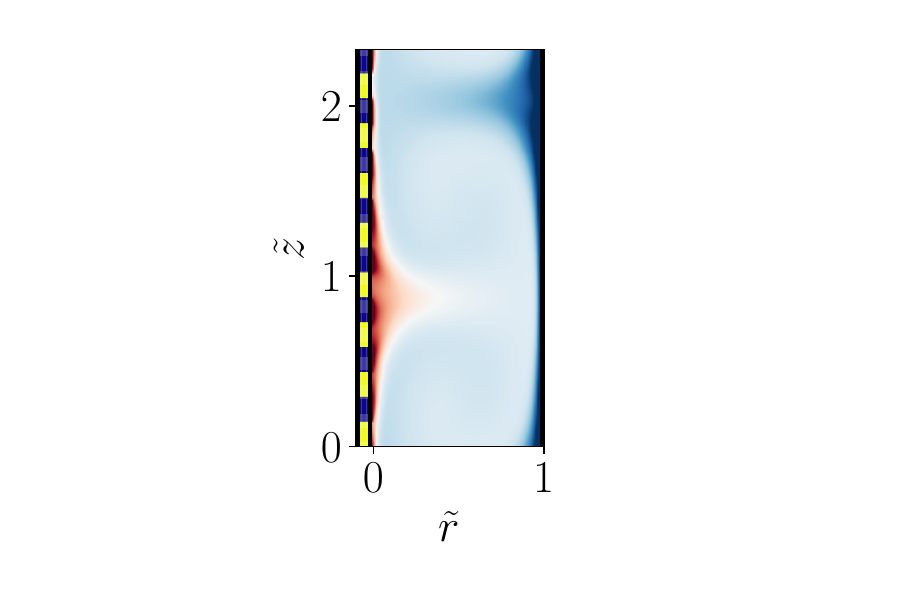}  
  \label{fig:azv3}
\end{subfigure}
\begin{subfigure}{.3\textwidth}
  \centering
  \caption{}
    \adjincludegraphics[width=1\textwidth, trim={{.47\width} {.05\width} {.05\width} {0.04\width}},clip]{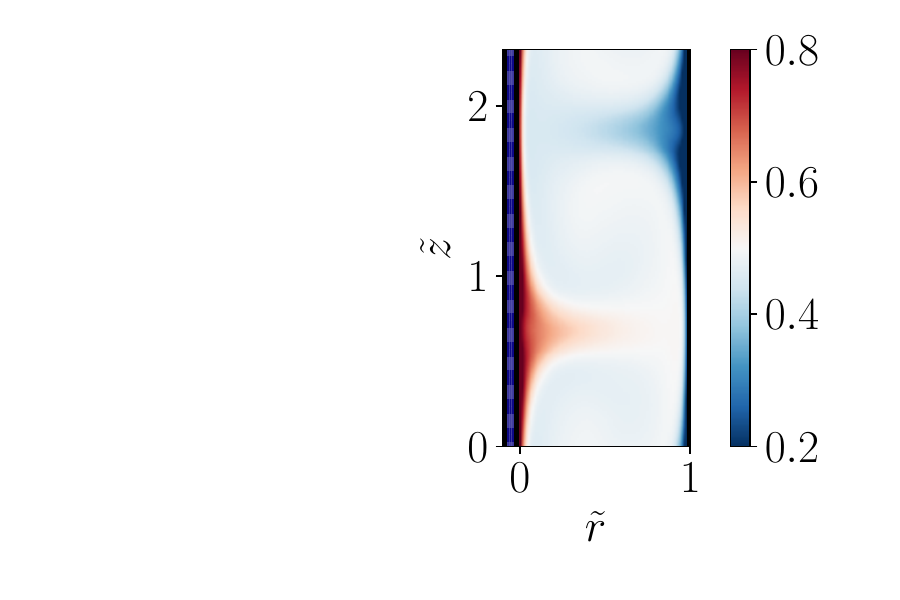}  
  \label{fig:azv4}
\end{subfigure}
\begin{subfigure}{0.49\textwidth}
\caption{} 
\begin{center}
  \adjincludegraphics[width=\textwidth]{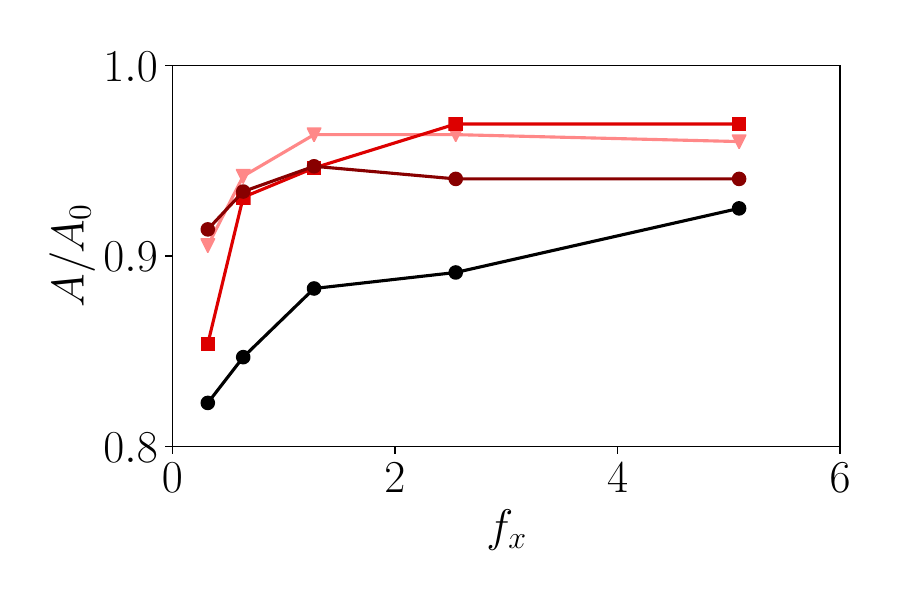}
\end{center}
\label{fig:rollamp90}
\end{subfigure}
\begin{subfigure}{0.49\textwidth}
\caption{} 
\begin{center}
  \adjincludegraphics[width=\textwidth]{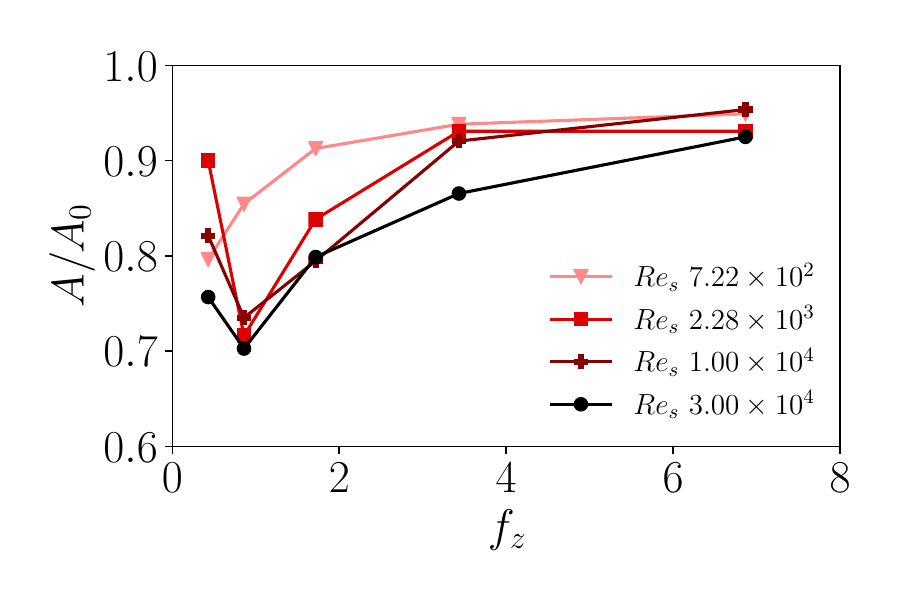}
\end{center}
\label{fig:rollamp0}
\end{subfigure}
\caption{Azimuthally and temporally averaged azimuthal velocity for $\eta=0.909$, $Re_s =10^4$ and axial inhomogeneities at several frequencies (\textit{a}) $f_z=0.43$, (\textit{b}) $f_z=0.86$, (\textit{c}) $f_z=3.43$ and (\textit{d}) No inhomogeneity. Yellow stripes correspond to stress-free zones and blue zones correspond to no-slip zones. (\textit{e,f}) Roll amplitude normalized by the roll amplitude for the homogeneous case for azimuthal and axial variations as a function of $Re_s$ and pattern frequency.}
  \label{fig:azv}
\end{center}
\end{figure}

To explore the non-monotonicity of the $Nu(f_z)$ curve at higher Reynolds numbers (Figure \ref{fig:aztnn}) we turn to the secondary flows. As mentioned in the introduction, we can expect that non-homogeneity patterns in the boundary layers generate secondary flows that interfere with the existing Taylor rolls.  
To corroborate  this, we show the temporally and azimuthally averaged azimuthal velocities $\langle u_\theta \rangle_{\theta,t}$ for $Re_s=10^4$ and several values of $f_z$ in figure \ref{fig:azv} alongside the homogeneous case. Here it is even more apparent that the pattern frequency $f_z=0.86$ has a strong impact on the rolls, because it imposes an inhomogeneity that has a wavelength of one-half that of the inhomogeneity generated by a roll pair, and the Prandtl flow of the secondary type induced by the pattern has a destructive interference with the roll. Decreasing $f_z$ to one stripe pair per roll pair matches the imposed axial inhomogeneity with the natural one from the Taylor roll pair. The roll pair will then locate itself such that the ``ejection'', or outflow region, where the outgoing streams are generated will reside on top of the no-slip region. Conversely, increasing the frequency of the imposed axial inhomogeneity, results in the associated secondary flow being increasingly absorbed inside of the boundary layer while the roll remains unaffected. 

We can quantify this disruption using the roll amplitude. We calculate this taking the temporally- and azimuthally averaged radial velocity ($\langle u_r \rangle_{t,\theta}$), and computing the magnitude of the first Fourier mode in the axial direction at the mid-gap $\tilde{r}=0.5$. In Figure \ref{fig:azv}(e,f) we show this amplitude $A$ normalized by the amplitude for the homogeneous case ($A_0$) as a function of Reynolds number and pattern frequency for both azimuthal and axial patterns. For the latter, we can see very clearly that for the three higher $Re_s$, the amplitude shows a large dip for $f_z=0.86$. For the low-$Re$ case in the steady Taylor vortex regime there are no Reynolds stresses, as well as for all cases with azimuthal variations where there is no axial/spanwise imbalance, the behaviour of $A/A_0$ is monotonic, i.e. the roll amplitude increases as the pattern gets smaller.

\begin{figure}
\centering
\begin{subfigure}{0.49\textwidth}
\begin{center}
  \caption{} 
  \adjincludegraphics[width=\textwidth,trim={{0\width} {.05\width} {0\width} {0\width}},clip]{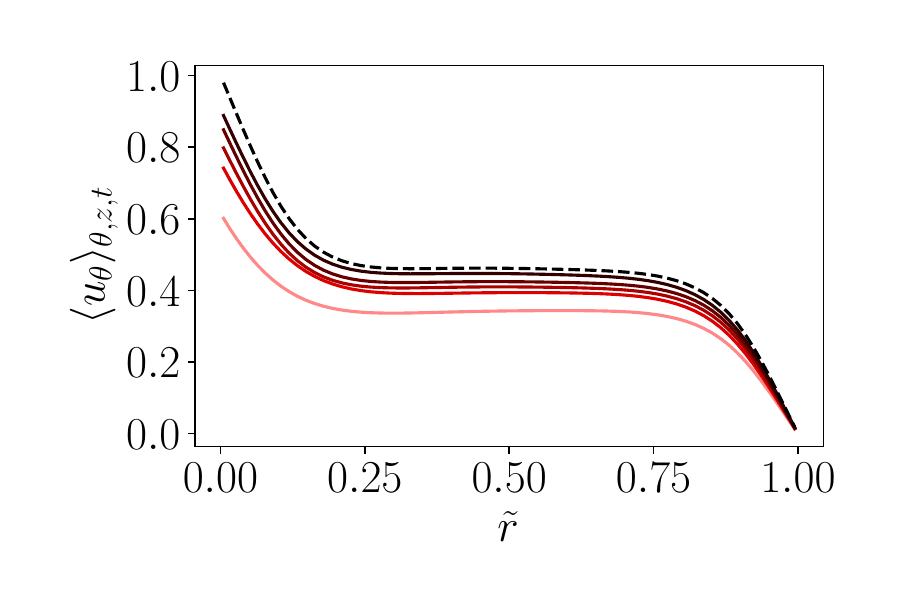}
\label{fig:vthetaRe7e2}
\end{center}
\end{subfigure}
\begin{subfigure}{0.49\textwidth}
\begin{center}
  \caption{} 
  \adjincludegraphics[width=\textwidth,trim={{0\width} {.05\width} {0\width} {0\width}},clip]{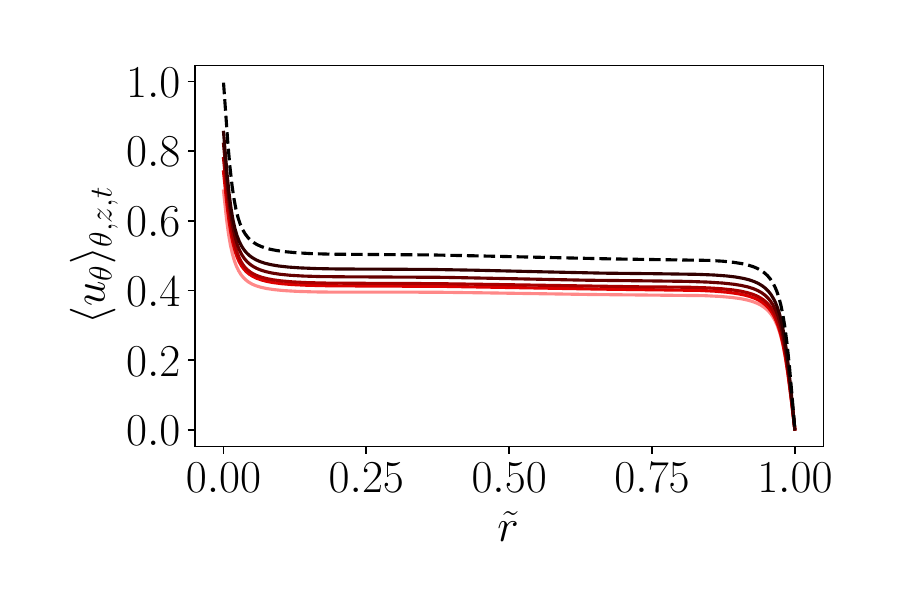}
\label{fig:vthetaRe3e4}
\end{center}
\end{subfigure}
\begin{subfigure}{0.49\textwidth}
\begin{center}
  \caption{} 
  \adjincludegraphics[width=\textwidth,trim={{0\width} {.05\width} {0\width} {0\width}},clip]{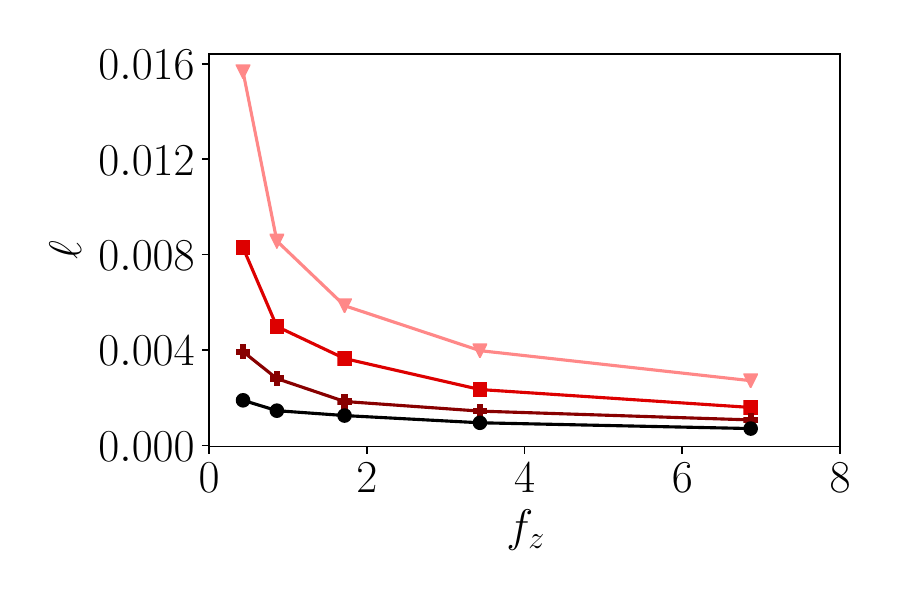}
\label{fig:slipl}
\end{center}
\end{subfigure}
\begin{subfigure}{0.49\textwidth}
\caption{} 
\begin{center}
  \adjincludegraphics[width=\textwidth,trim={{0\width} {.05\width} {0\width} {0\width}},clip]{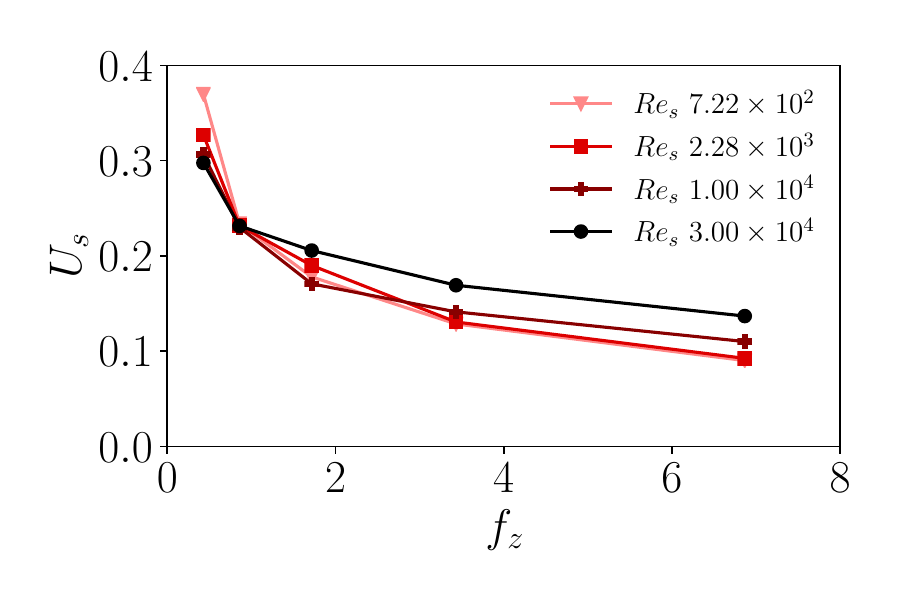}
\end{center}
\label{fig:slipv}
\end{subfigure}
\centering
\caption{Mean azimuthal velocity along the radial gap for various pattern frequencies $f_z$ for (\textit{a}) $Re_s=722$ and (\textit{b}) $Re_s=3 \times 10^4$; Darker colors signify higher values of $f_z$, while the dashed black line is the homogeneous case. (\textit{c}) Effective slip length; and (\textit{d}) effective slip velocity for all simulated cases of axial inhomogeneities. }
\label{fig:slips}
\end{figure}

\begin{figure}
\centering
\begin{subfigure}{0.49\textwidth}
\begin{center}
  \caption{} 
  \adjincludegraphics[width=\textwidth,trim={{0\width} {.05\width} {0\width} {0\width}},clip]{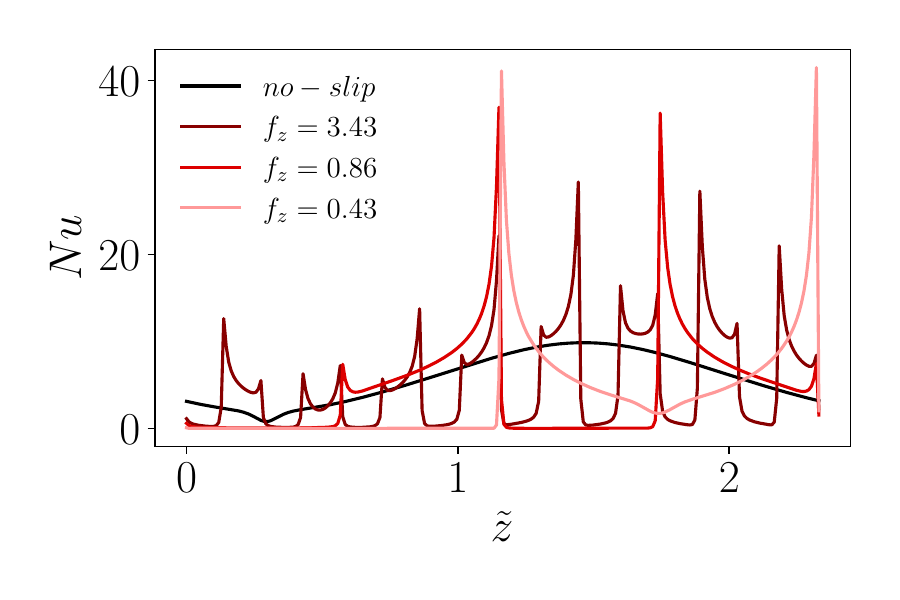}
\label{fig:Nu_axial_2e3}
\end{center}
\end{subfigure}
\begin{subfigure}{0.49\textwidth}
\caption{} 
\begin{center}
  \adjincludegraphics[width=\textwidth,trim={{0\width} {.05\width} {0\width} {0\width}},clip]{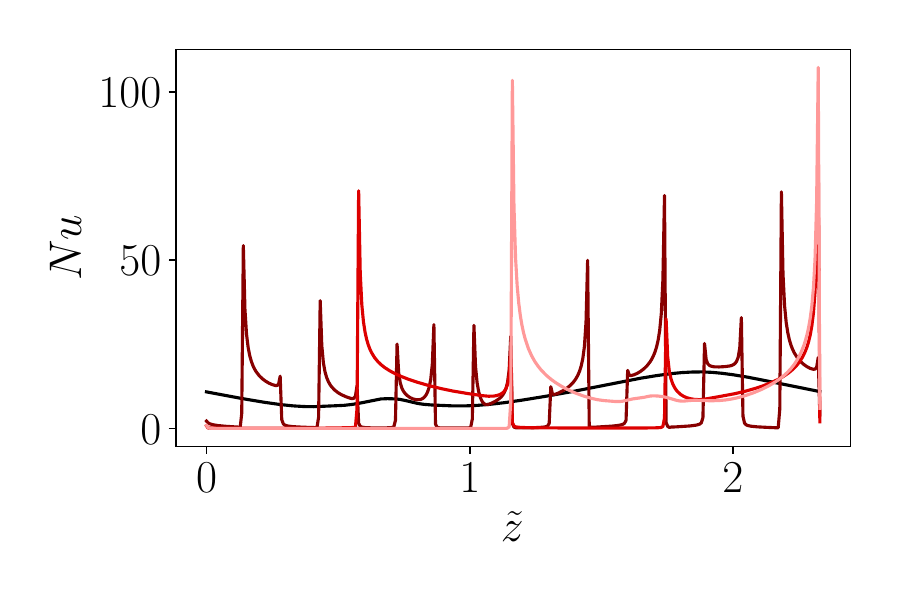}
\end{center}
\label{fig:Nu_ax_1e4}
\end{subfigure}
\centering
\caption{Temporally- and azimuthally averaged Nusselt number at the inner cylinder along the axial direction for various axial inhomogeneities $f_z$ at (\textit{a}) $Re_s=2.28\times 10^3$ and (\textit{b}) $Re_s=1 \times 10^4$.}
\label{fig:Nu_axial}
\end{figure}

The disruption of the roll can be distinguished from a pure boundary layer effect by examining how the axial patterns affect the behaviour of the temporally, axially and azimuthally averaged azimuthal velocity $\langle u_\theta(r_i) \rangle_{\theta,t,z}$. Figures \ref{fig:vthetaRe7e2}-\ref{fig:vthetaRe3e4} show this for the lowest and highest $Re$ simulated.
The special behaviour of the $f_z=0.86$ seen previously vanishes, and the curves are ordered according to $f_z$ for both low and high Reynolds numbers. We can further examine this by looking at how the effective slip in the inner cylinder behaves. The effective slip velocity $U_s$ is defined as the deficit, or slip, between the inner cylinder velocity and $\langle u_\theta(r_i) \rangle_{\theta,t,z}$. The effective slip length $\ell$ is then calculated using the averaged azimuthal velocity gradient at the inner cylinder as $\ell=U_s/\langle \partial_r u_\theta(r_i) \rangle_{\theta,t,z}$. These quantities are shown in Figures \ref{fig:slipl} and \ref{fig:slipv} for the azimuthal variations as a function of Reynolds number and pattern frequency. Remarkably, no apparent change is observed at $f_z=0.86$ for the effective slip properties. All the curves behave monotonically, with the effective slip length, and the effective slip velocity decreasing with increasing pattern frequency. To highlight this, we show the local torque, non-dimensionalized as a Nusselt number along the axial direction at the inner cylinder for various axial inhomogeneities for $Re_s=2.28 \times 10^3$ and $Re_s=1 \times 10^4$ in the figure \ref{fig:Nu_axial}. We can first see how the presence of the roll causes even the fully no-slip case to have axial inhomogeneities, and that boundary inhomogeneities interact with this existing inhomogeneity. Furthermore, the inhomogeneities cause the average value of the torque to increase, especially at the interface between regions. The local peaks are higher for smaller pattern frequencies as the slip velocity is higher. These results highlight the crucial difference between Taylor-Couette and the other shear flows previously studied: the disruption coming from the boundary is able to interact with large-scale components to multiply its effects.

To further visualize what changes are taking place due to the axial (spanwise) pattern, we turn to the Reynolds stress. In rectangular coordinates, we can write a relatively simple equation for a time and streamwise-averaged streamwise vorticity $\Omega_x$ \citep{einstein1958}, 

\begin{equation}
\label{eq:einsteinli}
    \overline{v} \displaystyle\frac{\partial \Omega_x }{\partial y} +  \overline{w} \displaystyle\frac{\partial \Omega_x }{\partial z} = \left (  \displaystyle\frac{\partial^2 }{\partial y^2} -  \displaystyle\frac{\partial^2 }{\partial z^2} \right)(-\langle \overline{v'w'} \rangle) +  \displaystyle\frac{\partial^2 }{\partial y \partial z}( \overline{v^{\prime 2}}  - \overline{w^{\prime 2}}) + \nu \left (  \displaystyle\frac{\partial^2 \Omega_x }{\partial z^2} + \displaystyle\frac{\partial^2 \Omega_x  }{\partial z^2} \right ),
\end{equation}

\noindent where $\overline{\phi}$ denotes the temporal and streamwise averaging operator, $\phi^\prime$ are fluctuations around that mean, and $\overline{\omega}_x$ is simply $\Omega_x$, and represents the secondary flow. 

This equation is derived by performing a streamwise- and temporal Reynolds-averaging operation on the vorticity advection-diffusion equation:

\begin{equation}
  \displaystyle\frac{D\bm{\omega}}{Dt} = -\bm{\omega}\cdot\nabla\textbf{u} + \nu\nabla^2\bm{\omega},
\end{equation}

\noindent where \bm{$\omega$} is the flow vorticity and $D/Dt$ denotes a convective derivative. By using the definition of vorticity, the continuity equation and the properties of cross-derivatives, the seventeen terms originating from Reynolds stresses can be reduced to the four, to which the viscid terms are added \citep{einstein1958}.

For Taylor-Couette flow, this equation becomes even more complicated because of cylindrical coordinates. The  reduction to equation \ref{eq:einsteinli} would add many curvature terms and complicate the interpretation. Basing ourselves on the analysis by \cite{bra15}, we can estimate the magnitude of curvature terms to be of the order of $R_C=(1-\eta)/\sqrt{\eta}$, which is $0.09$ in our narrow-gap and low curvature case. Hence, we keep the rectangular coordinates as they appear to us to be more useful to gain some insights, and for now neglect the error introduced due to curvature. This essentially means taking $\Omega_x$, i.e. the Reynolds-averaged streamwise vorticity to be equivalent to $\Omega_\theta$, setting $v=u_r$ and $w=u_z$, and using $\langle \phi \rangle_{\theta,t}$ for the $\overline{\phi}$ operator.

\begin{figure}
    \adjincludegraphics[width=0.49\textwidth, trim={{.05\width} {.15\width} {.0\width} {.2\width}},clip]{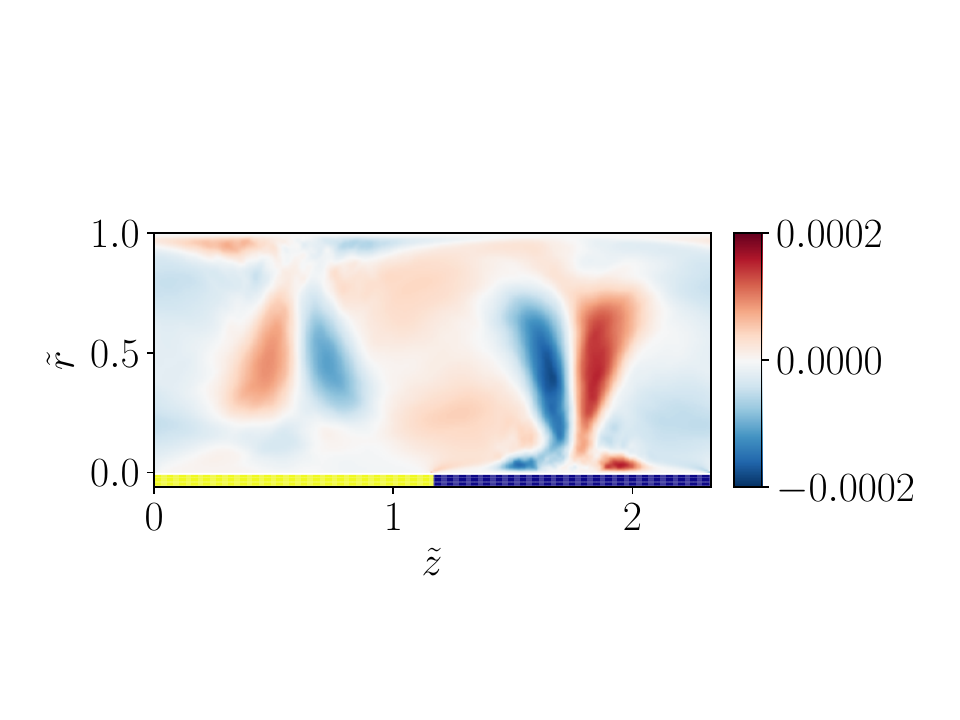}
    \adjincludegraphics[width=0.49\textwidth, trim={{.05\width} {.15\width} {.0\width} {.2\width}},clip]{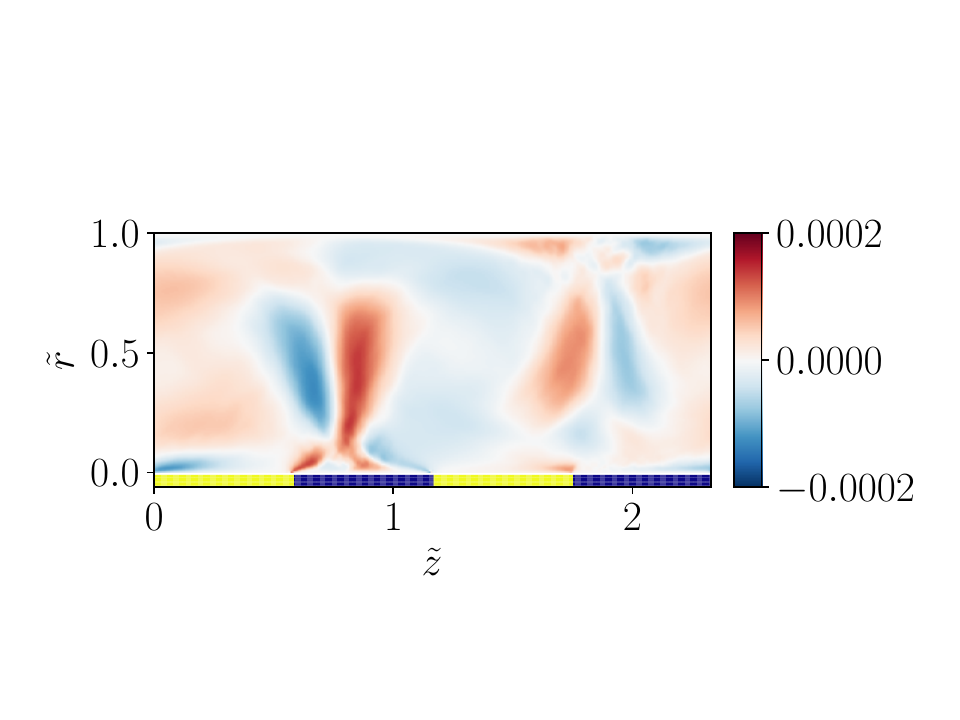}\\
    \adjincludegraphics[width=0.49\textwidth, trim={{.05\width} {.15\width} {.0\width} {.2\width}},clip]{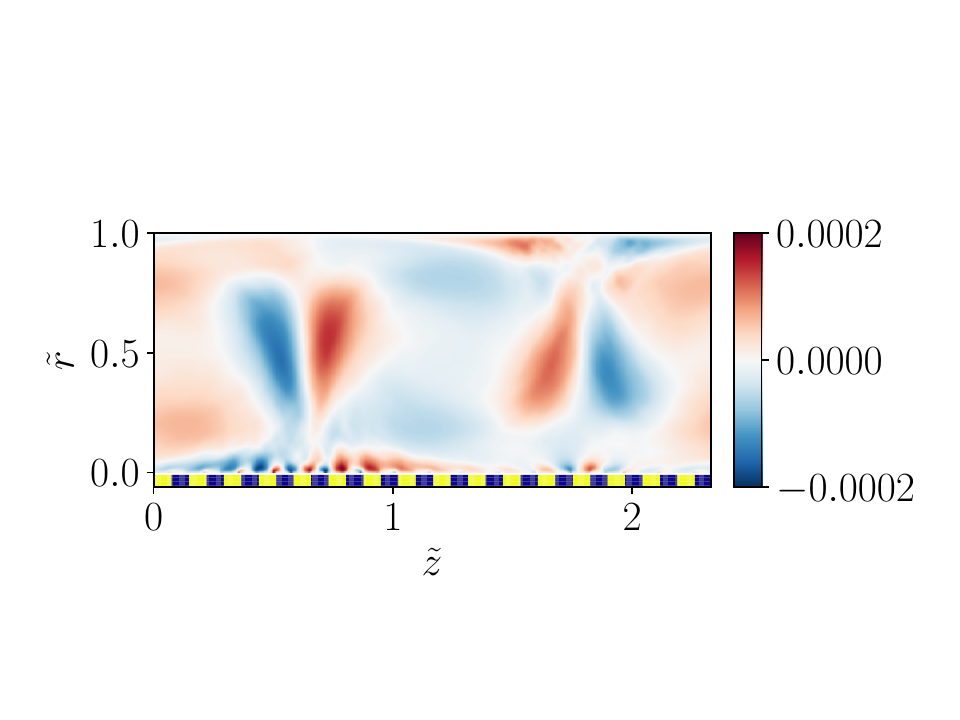}
    \adjincludegraphics[width=0.49\textwidth, trim={{.05\width} {.15\width} {.0\width} {.2\width}},clip]{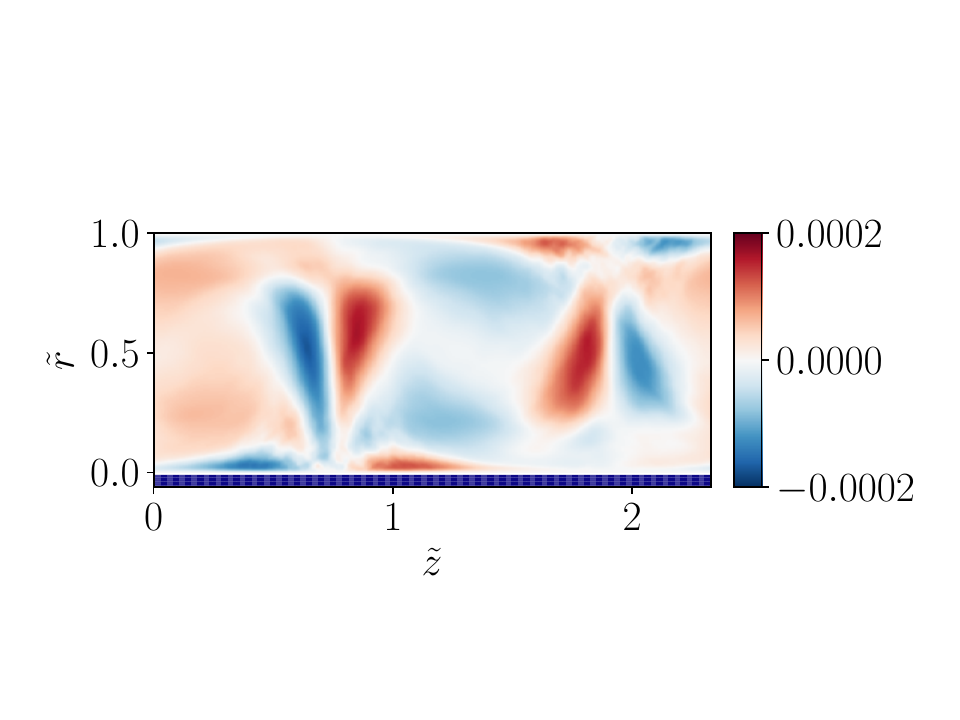}
\centering
\caption{Azimuthally and temporally averaged Reynolds shear stress $\langle u_r u_z \rangle$ for $Re_s =3\times 10^4$ and axial inhomogeneities for (\textit{a}) $f_z=0.43$, (\textit{b}) $f_z=0.86$, (\textit{c}) $f_z=3.43$ and (\textit{d}) No inhomogeneity. Yellow stripes correspond to stress-free zones and blue zones correspond to no-slip zones}
  \label{fig:restresses}
\end{figure}

\begin{figure}
    \adjincludegraphics[width=0.49\textwidth, trim={{.05\width} {.2\width} {.0\width} {.25\width}},clip]{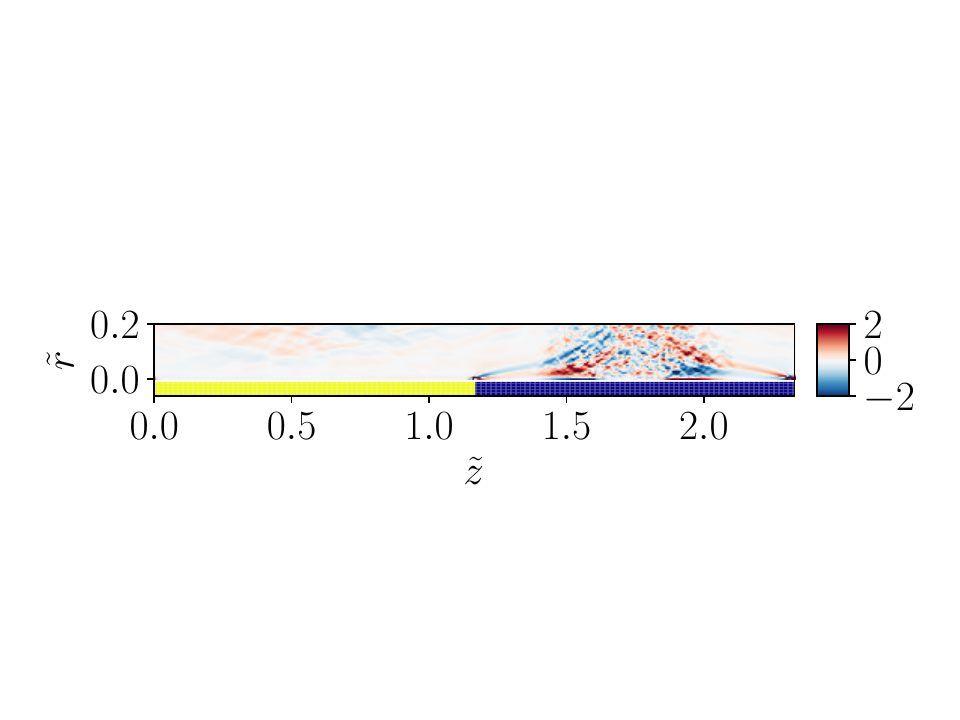}
    \adjincludegraphics[width=0.49\textwidth, trim={{.05\width} {.2\width} {.0\width} {.25\width}},clip]{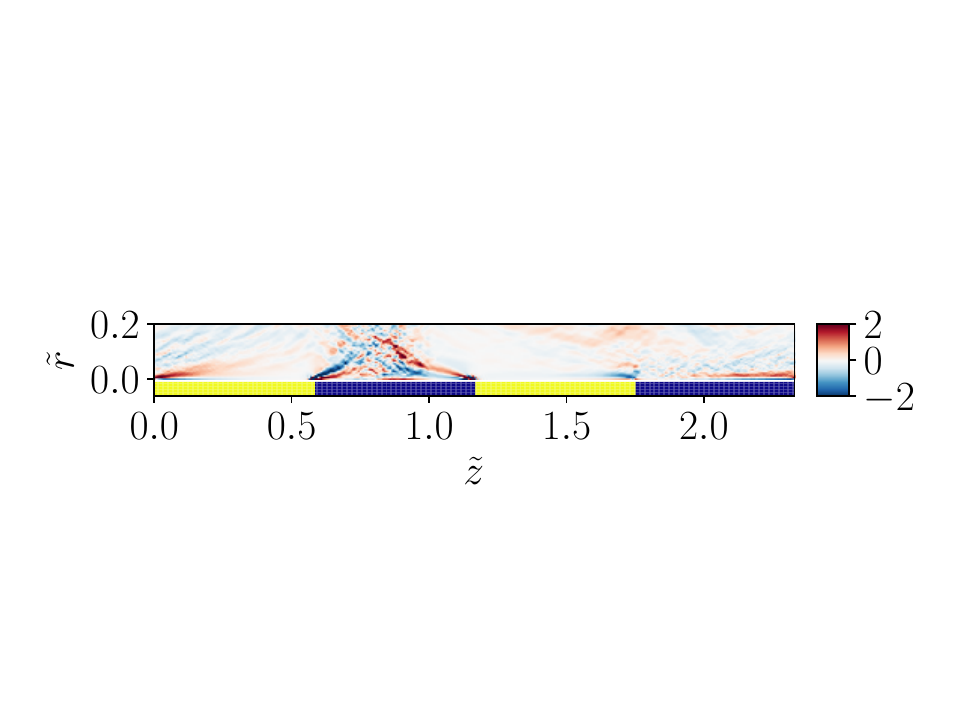}\\
    \adjincludegraphics[width=0.49\textwidth, trim={{.05\width} {.2\width} {.0\width} {.25\width}},clip]{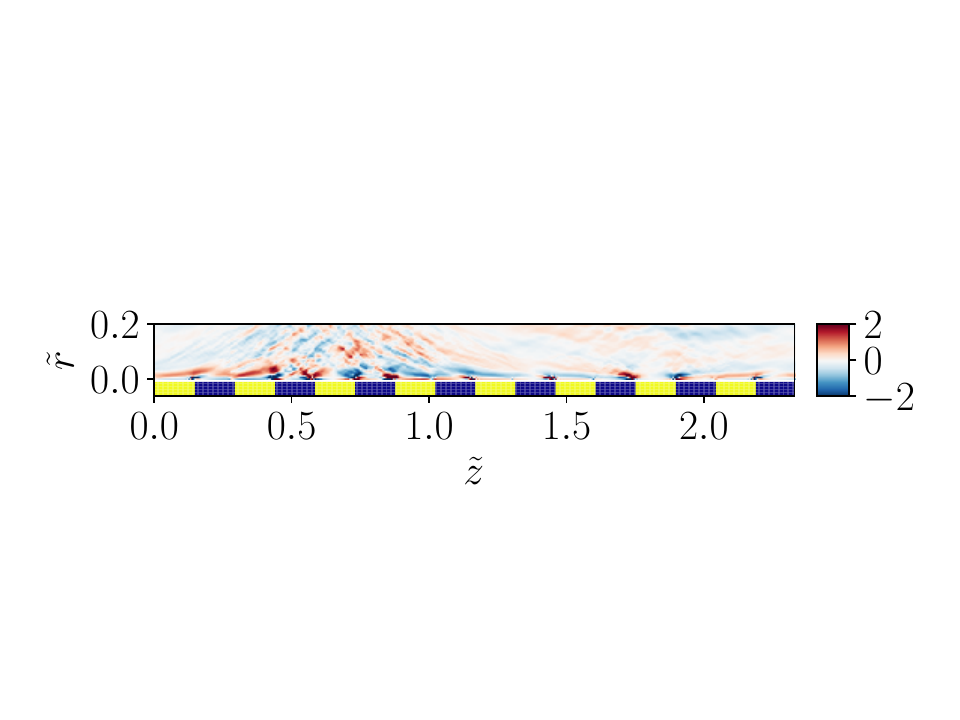}
    \adjincludegraphics[width=0.49\textwidth, trim={{.05\width} {.2\width} {.0\width} {0.25\width}},clip]{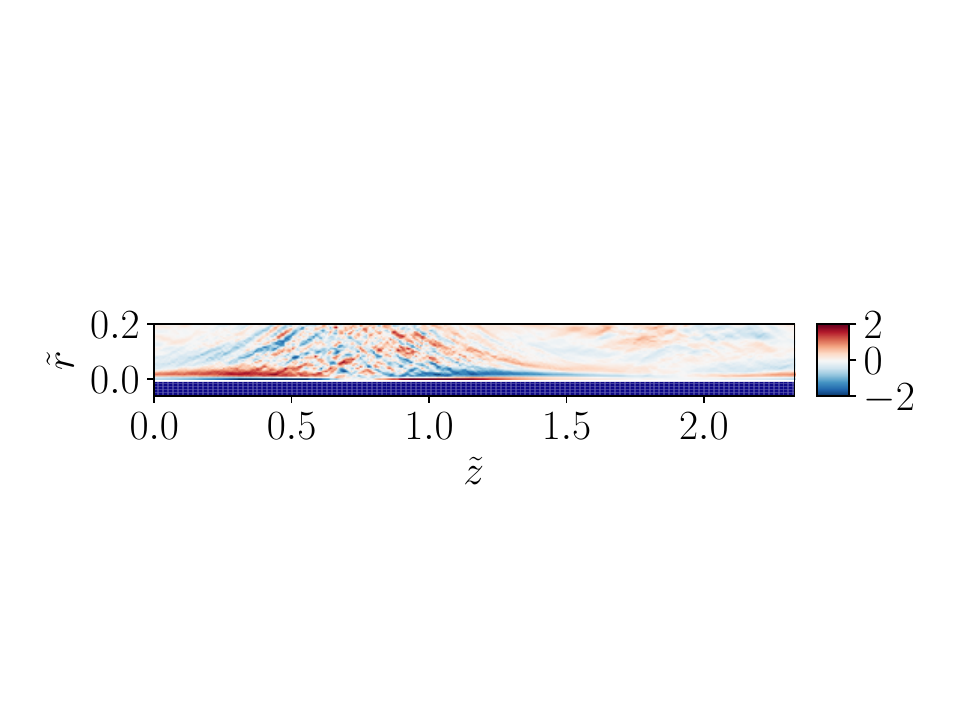}
\centering
\caption{Azimuthally and temporally averaged spanwise Reynolds shear stress imbalance $G_\Omega$ for $\eta=0.909$, $Re_s =3\times 10^4$ and axial inhomogeneities at several frequencies (\textit{a}) $f_z=0.43$, (\textit{b}) $f_z=0.86$, (\textit{c}) $f_z=3.43$ and (\textit{d}) No inhomogeneity. Yellow stripes correspond to stress-free zones and blue zones correspond to no-slip zones.}
  \label{fig:gs-omegax}
\end{figure}

We focus on the generation of $\Omega_x$ through spanwise gradients in shear Reynolds stresses. These types of gradients induce the secondary flows which are known as secondary Prandtl flows of the second type (c.f. $\S$\ref{sec:introduction}). We first show the behaviour of the $\langle u_r u_z \rangle$ Reynolds stress in Figure \ref{fig:restresses}, and in Figure \ref{fig:gs-omegax}, we visualize $G_\Omega=\partial^2_z\langle u_ru_z\rangle$ near the walls. The case with no inhomogeneity shows a concentration of $G_\Omega$ near the walls. $G_\Omega$ shows a spanwise variation which is periodic with a period equal to the rolls. This is hardly surprising as this imbalance keeps the roll stable. For the $f_z=0.43$ case, the pattern cannot impose a different period of spanwise variation of stresses, which are now increased on the no-slip surface. For the $f_z=0.86$ case is where we see a significant interference between the period imposed by the pattern, and that which stabilizes the roll- and this causes the disruption seen above. Finally, while some disruption in the $G$ pattern can be seen in the $f_z=3.43$ case, the oscillations become averaged out due to their higher frequency. This shows how the maximum roll disruption takes place: it is coupled to a disruption of the ``natural'' period of the Reynolds stresses.

\section{Mixed spanwise and streamwise variations}

\subsection{Spiral patterns}

In this subsection, we modify the one-dimensional patterns studied above by placing them at an arbitrary oblique angle to the flow. This results in spiral or oblique patterns, which resemble a barber-pole and introduce chirality into the system. While in a statistical sense, axial symmetry is recovered because every point is equally likely to be no-slip or stress-free as the pattern moves through the domain, axial symmetry is persistently broken in an instantaneous sense and this is known to induce spanwise (axial) velocities \citep{hasegawa2011effects,watanabe2017drag}. This induced velocity makes spiral patterns another good candidate for affecting the rolls, as \cite{ostilla2016near} observed that an axial pressure gradient that generated an axial velocity was capable of moving these structures. 

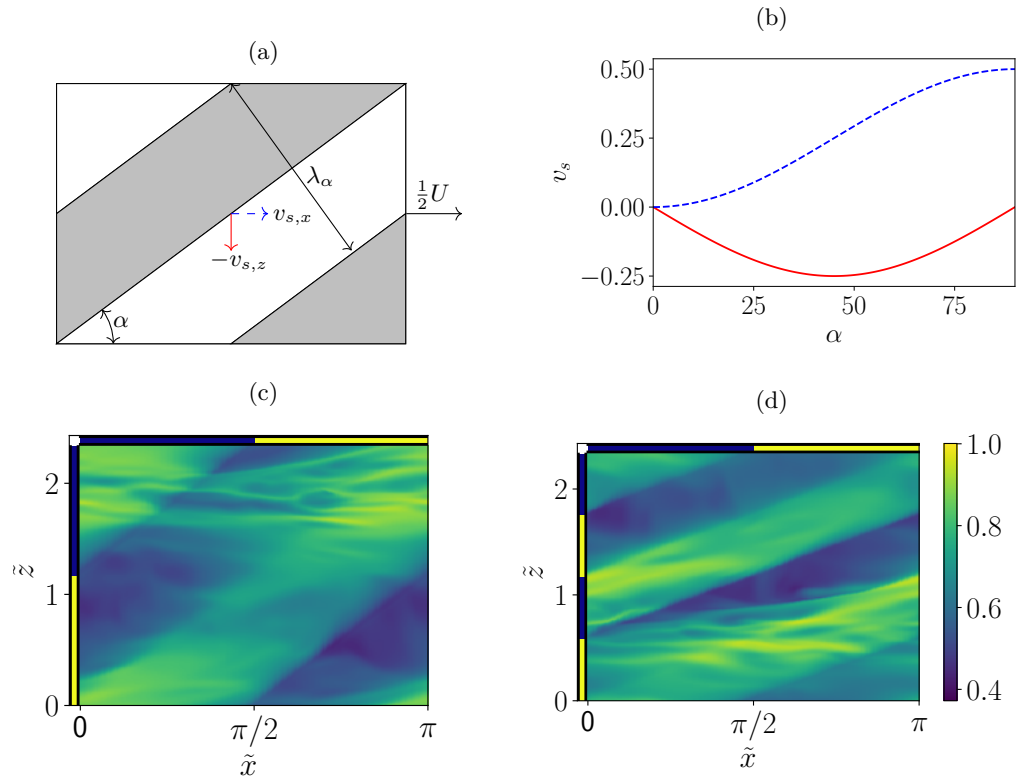
\begin{figure}
\centering
\begin{subfigure}{0.49\textwidth}
\begin{center}
  \caption{}
  \begin{tikzpicture}[scale=0.49]

\draw [black](0,0) -- (9.42,0) -- (9.42,6.99) -- (0,6.99) -- (0,0);

\draw[fill=white]  ((0,6.99*1/2) -- (9.42*1/2,6.99) -- (0,6.99) -- cycle;
\draw[fill=gray!50]  (0,0) -- (9.42,6.99) -- (9.42*1/2,6.99) -- (0,6.99*1/2) -- cycle;
\draw[fill=white]  (9.42*1/2,0) -- (9.42,6.99*1/2) -- (9.42,6.99) -- (0,0) -- cycle;
\draw[fill=gray!50]  (9.42,0) -- (9.42,6.99*1/2) -- (9.42*1/2,0) -- cycle;

  \draw
    (9.42,6.99) coordinate (c)
    (0,0) coordinate (b)
    (9.42,0) coordinate (a)
    pic["$\alpha$", draw=black, <->, angle eccentricity=1.2, angle radius=0.75cm]
    {angle=a--b--c};

\draw[->][red,thin](9.42/2,6.99/2)--+ (0,-1);
\draw[dashed,->][blue,thin](9.42/2,6.99/2)--+ (1,0);
\draw[->][black,thin](9.44,6.99/2)--+ (1.5,0);
\draw[<->, thin][black](9.42/2,6.99) to +(3.25,-1.38*3.25);

\fill[black](9.42/2-0.8,6.99/2-1.3) node [scale=1,anchor=west]{$-v_{s,z}$};
\fill[black](9.42/2+0.9,6.99/2-0.1) node [scale=1,anchor=west]{$v_{s,x}$};
\fill[black](9.42,6.99/2+0.5) node [scale=1,anchor=west]{$\frac{1}{2}U$};
\fill[black](6.5,4.5) node [scale=1,anchor=west]{$\lambda_{\alpha}$};

\end{tikzpicture}
\label{fig:istr}
\end{center}
\end{subfigure}
\begin{subfigure}{0.49\textwidth}
\begin{center}
  \caption{}
  \adjincludegraphics[width=\textwidth,trim={{0\width} {.05\width} {0.05\width} {0.05\width}},clip]{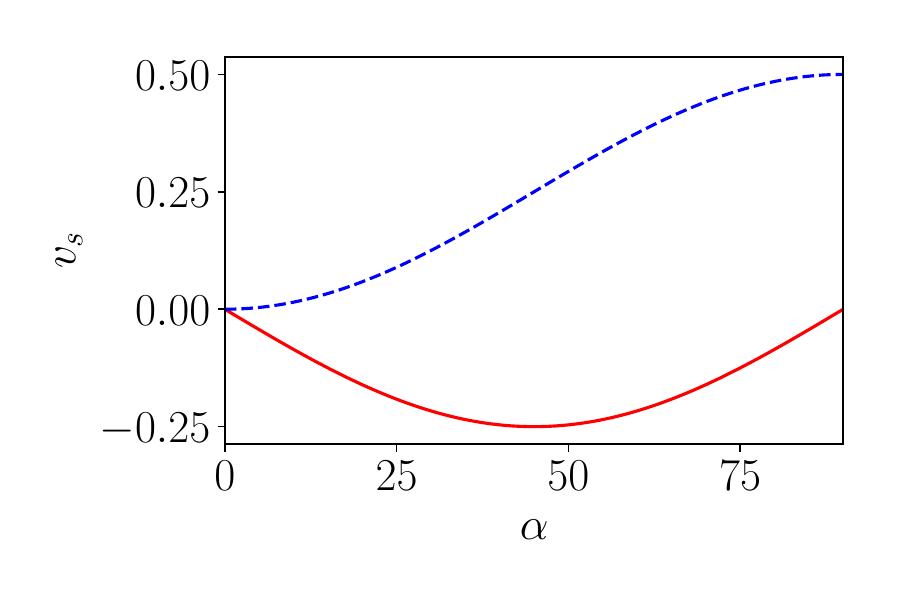}
    \label{fig:vs}
\end{center}
\end{subfigure}
\begin{subfigure}{0.49\textwidth}
\begin{center}
  \caption{}
  \adjincludegraphics[width=\textwidth,trim={{0.14\width} {.05\width} {0.05\width} {0.04\width}},clip]{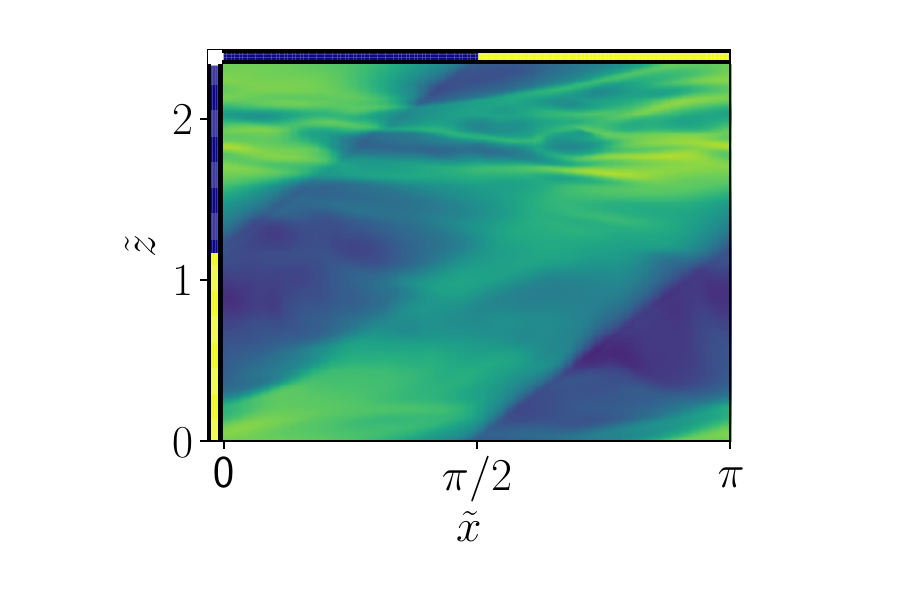}
    \label{fig:iistr36}
\end{center}
\end{subfigure}
\begin{subfigure}{0.49\textwidth}
\begin{center}
  \caption{}
  \adjincludegraphics[width=\textwidth,trim={{0.1\width} {.05\width} {0.05\width} {0.04\width}},clip]{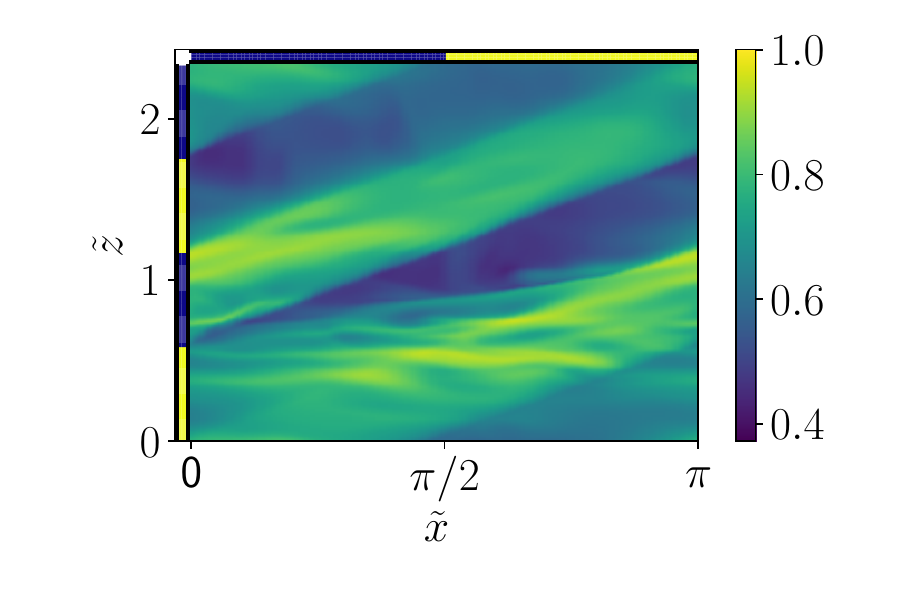}
    \label{fig:iistr20}
\end{center}
\end{subfigure}
\centering
\caption{(\textit{a}) Inclined pattern parameters; (\textit{b}) velocity of the pattern in spanwise $v_{s,z}$ (solid red) and streamwise $v_{s,x}$ (dashed blue) for various angles of inclination; (\textit{c,d}) near-wall instantaneous azimuthal velocity of inclined inhomogeneity for $Re_s=10^4$ (\textit{c}) $\alpha=36.6^{\circ}$, $f_\alpha=0.53$ and (\textit{d}) $\alpha=20.4^{\circ}$, $f_\alpha=0.92$.}
\label{fig:strvel}
\end{figure}

We simulate oblique patterns for three (more) angles between $\theta=0^o$ (purely axial inhomogeneity) and $\theta=90^o$ (purely azimuthal inhomogeneity). A sketch that quantifies the wavelength of the underlying pattern, and the speed at which it moves through the domain is provided in figures \ref{fig:istr} and \ref{fig:vs} respectively. In particular, in figure \ref{fig:vs} we plot the relative spanwise ($v_{s,z}=-U\sin(\alpha)\cos(\alpha)/2$) and streamwise ($v_{s,x}=-U\sin^2(\alpha)/2$) pattern velocities with respect to the domain in figure \ref{fig:vs}, for all possible angles of inclination. The maximum pattern velocity in the $z$ direction is attained when $\alpha=45^\circ$, and is equal to a half of the cylinder speed $U/2$.

Intuitively, we may think that the pattern moves as the stripes do, and that by following the movement of the stripes we are following the pattern. However, what is relevant here is how the boundaries between stripes move, i.e.~tracking the way the inhomogeneous direction moves, because it is the direction across which imbalances are. 

\begin{figure}
\centering
\begin{subfigure}{0.49\textwidth}
\begin{center}
  \caption{}
  \adjincludegraphics[width=\textwidth,trim={{0\width} {.05\width} {0\width} {0\width}},clip]{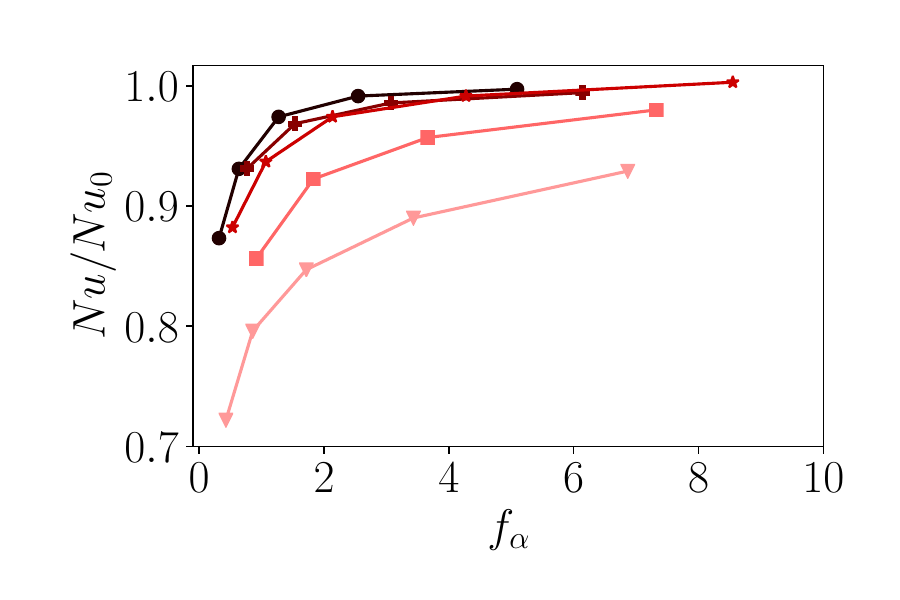}
  \label{fig:itnl}
\end{center}
\end{subfigure}
\begin{subfigure}{0.49\textwidth}
\begin{center}
  \caption{}
  \adjincludegraphics[width=\textwidth,trim={{0\width} {.05\width} {0\width} {0\width}},clip]{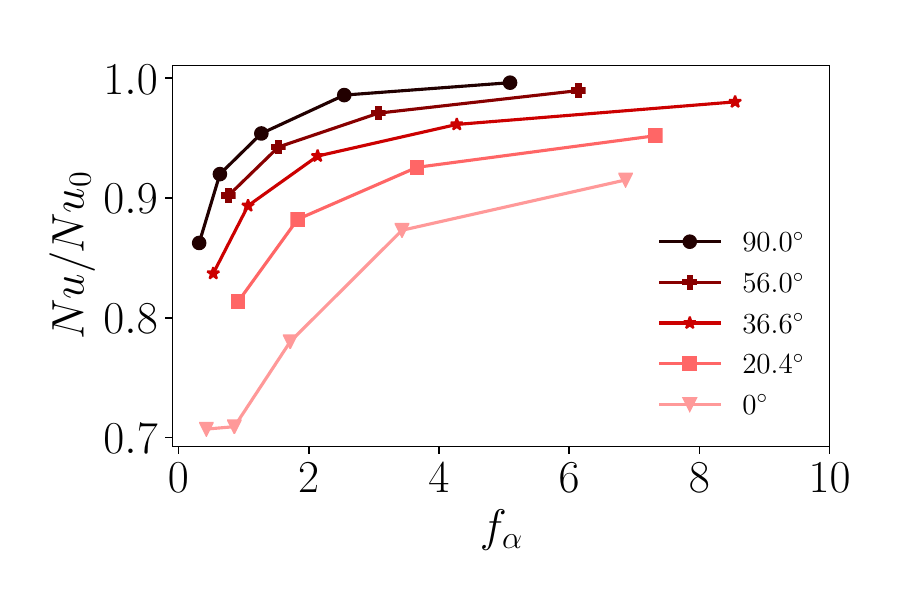}
    \label{fig:itnh}
\end{center}
\end{subfigure}
\centering
\caption{ Torque of $\eta =0.909$ for various $\alpha$ (\textit{a}) $Re_s=7.22\times 10^2$ and (\textit{b}) $Re_s=10^4$.}
    \label{fig:torquespiral}
\end{figure}

We start by visualizing the near-wall azimuthal velocities for $\alpha=36.6^{\circ}$ and $\alpha=20.4^{\circ}$ in figures \ref{fig:iistr36} and \ref{fig:iistr20} respectively to show the effect of inclined pattern on the flow. While the signatures are stronger than for purely azimuthal variations, they appear weaker than those coming from purely axial variations as the elongated streaks are able to bridge the gap between patterns. We continue by plotting the normalized torque for two Reynolds numbers considered in figure \ref{fig:torquespiral}, in the modulated Taylor vortex regime and in the turbulent Taylor vortex regime. The large differences between the resulting torque when using axially- and azimuthally-oriented pattern are apparent in this representation, as they are the two limit cases of spiral patterns. Unsurprisingly, we find that for the most part, the torques of different angles of pattern lie between that of the azimuthal and axial pattern in a predictable, monotonic order, and again consistent with \cite{watanabe2017drag} and \cite{naim2019turbulent}. 

\begin{figure}
\centering
\includegraphics[width=0.48\textwidth]{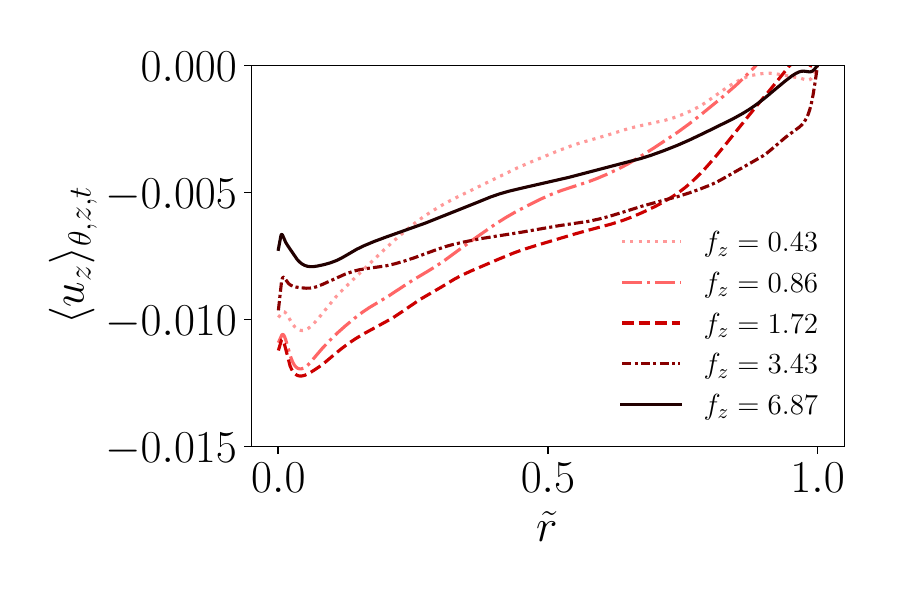}
\includegraphics[width=0.48\textwidth]{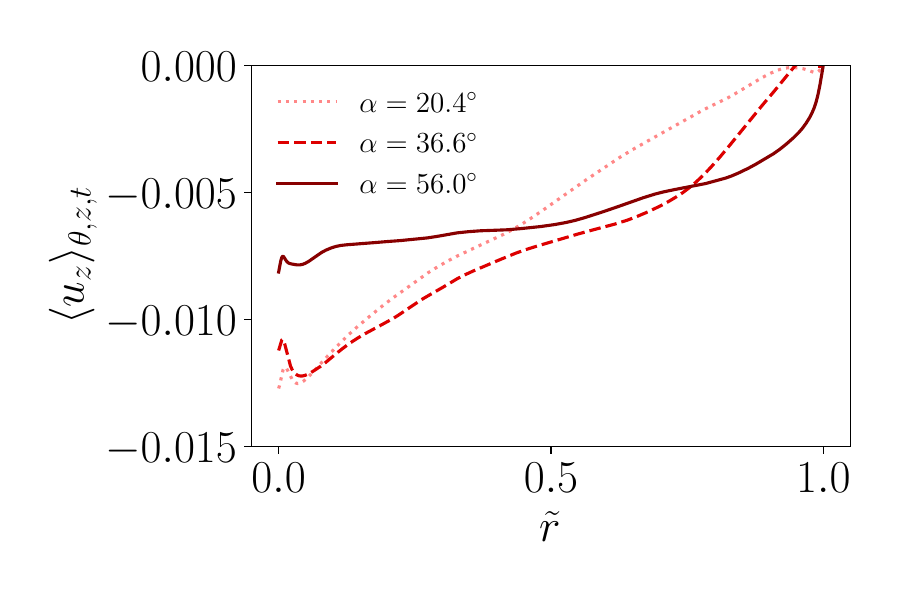}
\includegraphics[width=0.48\textwidth]{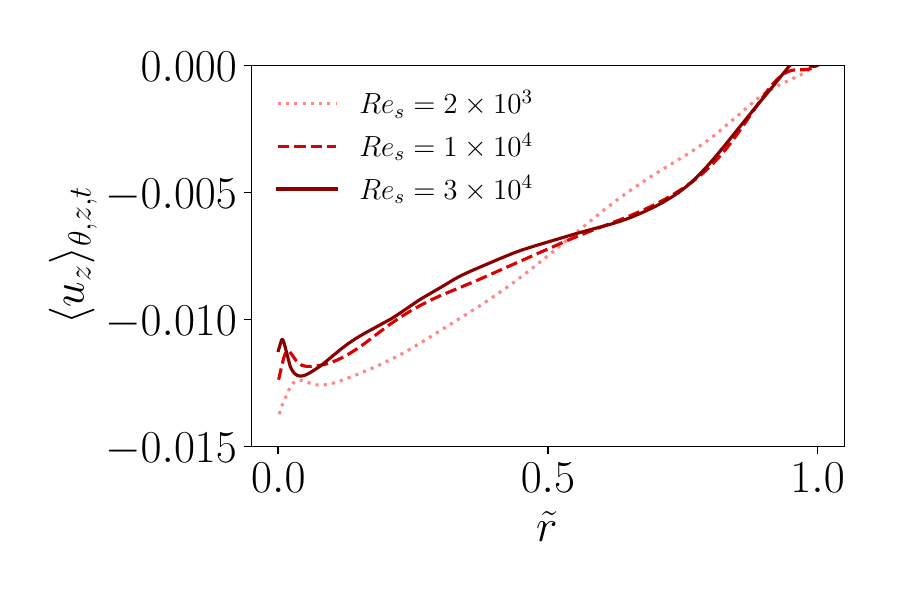}
\includegraphics[width=0.48\textwidth]{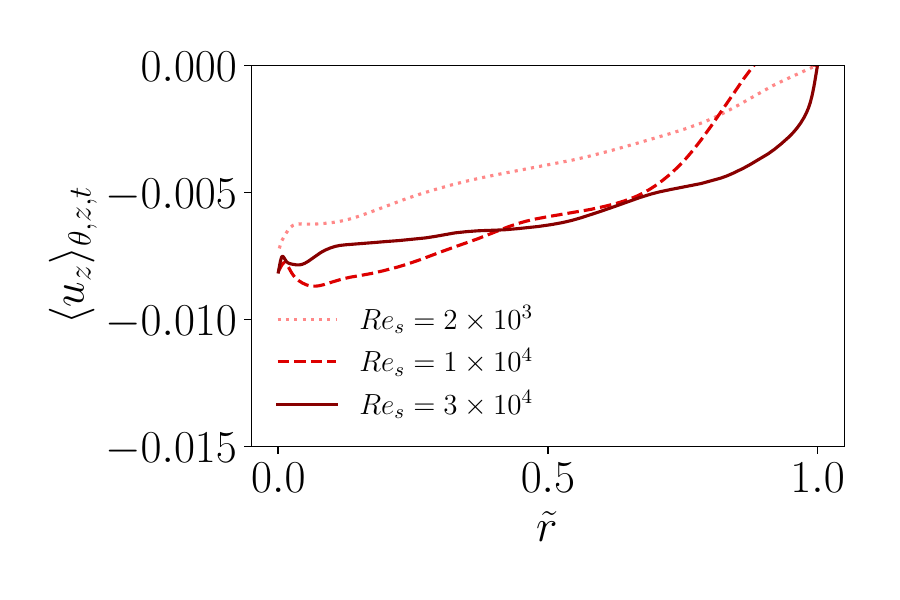}
\caption{ Mean induced axial velocity as a function of radius for: 
(\textit{a}) Varying pattern frequency $f_z$ with constant angle $\alpha=36.6^\circ$ and constant $Re_s=3\times10^4$. (\textit{b}) Varying pattern angle $f_z$ with constant frequency $f_z=1.72$ and constant $Re_s=3\times10^4$.
(\textit{c}) Varying shear Reynolds number $Re_s$ with constant $\alpha=36.6^\circ$ and constant $f_z=1.72$. (\textit{d}) Varying shear Reynolds number $Re_s$ with constant $\alpha=56.0^\circ$ and constant $f_z=1.72$.}
\label{fig:meanvz}
\end{figure}

Before moving to the secondary flows themselves, we quantify the induced axial (spanwise) velocities by showing $\langle v_z \rangle_{\theta,z,t}$ as a function of radius for several patterns and several Reynolds numbers in Figure \ref{fig:meanvz}. As can be seen from the figure, it is hard to discern a simple behaviour from the plots, aside that some indications that as $f_z\to\infty$, the induced velocities go to zero. For $\alpha=56.0^\circ$ and $f_z=1.72$ the induced velocity does not change significantly with $Re_s$, while for $\alpha=36.6^\circ$, the $Re_s$ dependence is much stronger, with $\sim 50\%$ variations. The only real conclusion we can make is that the patterns will induce a velocity, which is of the order of $1-2\%$ of the cylinder velocity. This coincides with the direction of the movement of pattern inhomogeneity, and not the pattern itself.

\begin{figure}
\centering
\begin{subfigure}{0.35\textwidth}
\begin{center}
  \caption{}
  \adjincludegraphics[width=\textwidth,trim={{0.4\width} {.05\width} {0\width} {0\width}},clip]{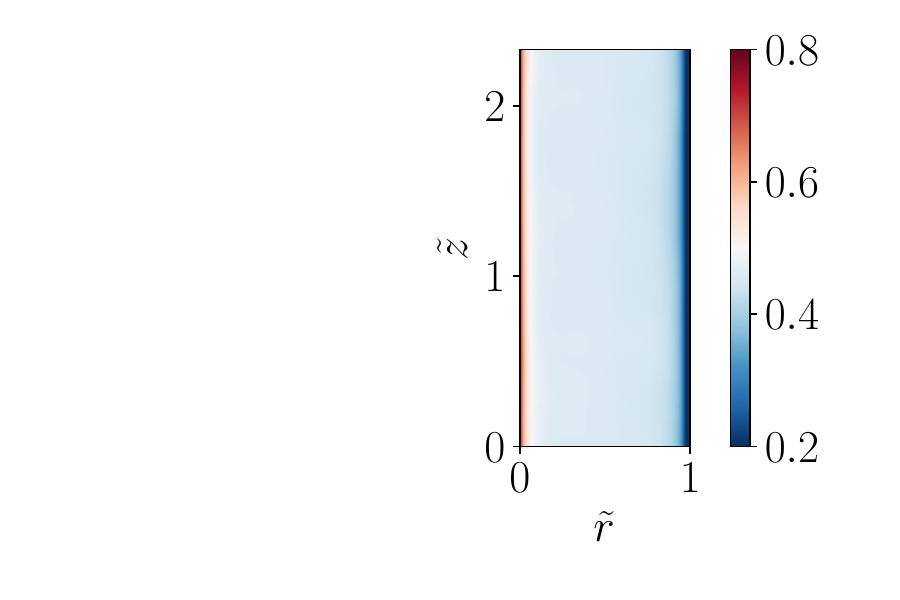}
  \label{fig:azic}
\end{center}
\end{subfigure}
\begin{subfigure}{0.49\textwidth}
\begin{center}
  \caption{}
  \adjincludegraphics[width=\textwidth,trim={{0\width} {0\width} {0\width} {0\width}},clip]{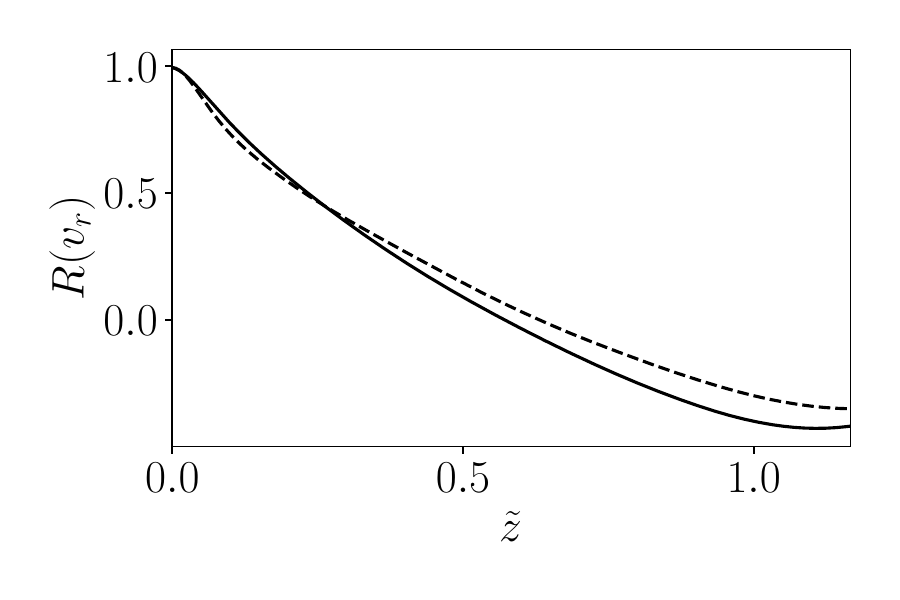}
    \label{fig:corr}
\end{center}
\end{subfigure}
\begin{subfigure}{0.49\textwidth}
\begin{center}
  \caption{}
  \adjincludegraphics[width=\textwidth,trim={{0\width} {.05\width} {0\width} {0\width}},clip]{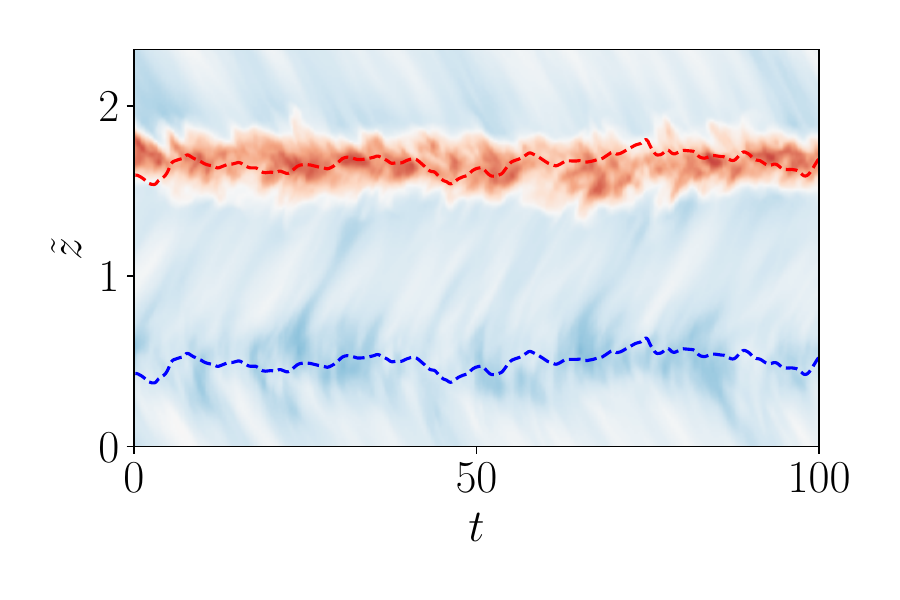}
    \label{fig:spt}
\end{center}
\end{subfigure}
\begin{subfigure}{0.49\textwidth}
\begin{center}
  \caption{}
  \adjincludegraphics[width=\textwidth,trim={{0\width} {.05\width} {0\width} {0\width}},clip]{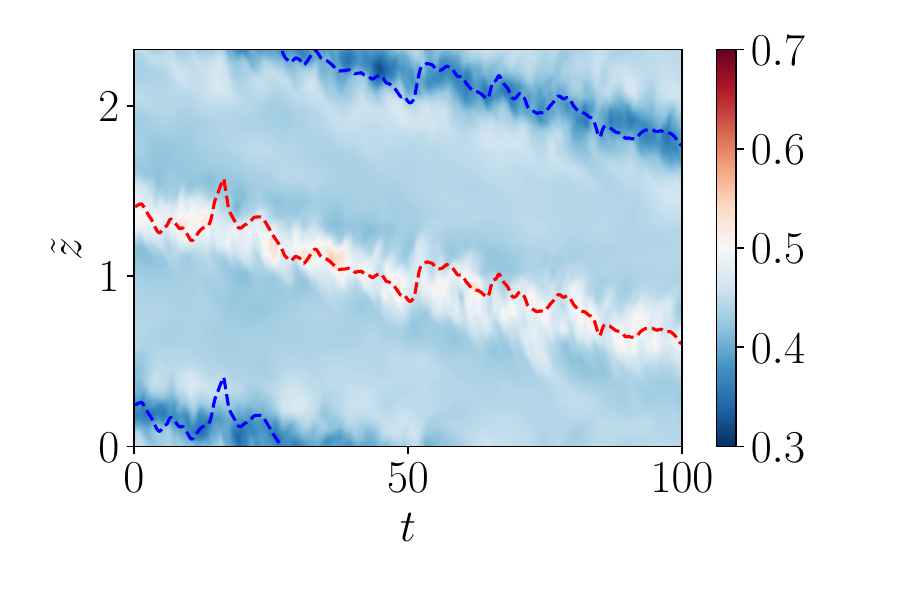}
    \label{fig:spt2}
\end{center}
\end{subfigure}
\centering
\caption{ (\textit{a}) Average azimuthal velocity of inclined pair of pattern $Re_s=10^4$, $\alpha=36.6^{\circ}$ and $f_z=1.72$; (\textit{b}) homogeneous TC (dashed) and $\alpha=36.6^{\circ}$, $f_z=1.72$ (solid) auto-correlation of radial velocity in the mid-gap along the axial direction; the spacetime diagram of the azimuthal velocity in the midgap for (\textit{c}) homogeneous TC; and (\textit{d}) $\alpha=36.6^{\circ}$ and $f_z=1.72$. Red and blue dashed lines correspond to the maximum and the minimum mid-gap average azimuthal velocities in time respectively predicted from the phase angle of the fundamental mode of the Fourier transform.}
\end{figure}

Even if the induced velocities are small they are enough to nudge the large-scale structures. Indeed, when we examine $\langle u_\theta \rangle_{\theta,t}$, we find that the clear signature of the rolls is absent. One such case is presented in figure \ref{fig:azic} which shows $\langle u_\theta \rangle_{\theta,t}$ for the spiral pattern $f_z=1.72$, $\alpha=36.6^{\circ}$ and $Re_s=10^4$. However, strong rolls are still present in an instantaneous velocity field. We quantify this statement by showing the auto-correlation of radial velocity in the mid-gap along the axial direction in figure \ref{fig:corr}. The large scale structures still exist as the radial velocity at the mid-point of the span is negatively correlated at the wavelength of a Taylor roll. 

The next step is to quantify the way the rolls are moving around. To quantify the axial velocity of the rolls, we follow \cite{sacco2019dynamics}, and take the Fourier transform of the azimuthal velocity in the mid-gap. We use the phase $\phi$ of the Fourier mode which is axisymmetric and fundamental in the axial direction to find the axial displacement of the rolls, and hence their velocity $v_{roll,z} = \dot{\phi} \Gamma/(2\pi)$. To show the effectiveness of this approach, we depict space time diagrams of the azimuthally-averaged azimuthal velocity in the mid-gap in figures \ref{fig:spt}-\ref{fig:spt2}, with the predicted location of the roll superimposed. For the first case, shown in \ref{fig:spt} the rolls remain relatively steady, fluctuating around a fixed position. For the second case in \ref{fig:spt2}, we can see that the rolls follow the induced velocity, and are moving \emph{downwards} in time at a certain velocity, instead of moving upwards following the motion of the spiral. To explain this, we return to the diagram in figure \ref{fig:vs}. The no-slip region is swept downwards by the incoming free-slip region, and the Reynolds stress imbalance follows this direction of movement. We postulate that the change in inhomogeneities in the downwards direction tends to move the roll downwards along the direction of change of inhomogeneities. 

\begin{figure}
\centering
\begin{subfigure}{0.49\textwidth}
\begin{center}
  \caption{}
  \adjincludegraphics[width=\textwidth,trim={{0\width} {.05\width} {0\width} {0\width}},clip]{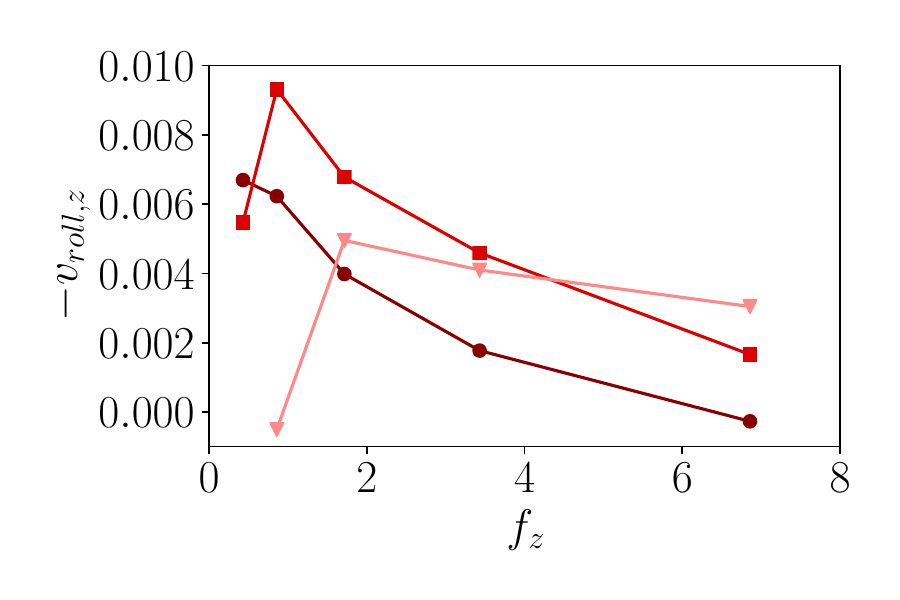}
      \label{fig:rv}
\end{center}
\end{subfigure}
\begin{subfigure}{0.49\textwidth}
\begin{center}
  \caption{}
  \adjincludegraphics[width=\textwidth,trim={{0\width} {.05\width} {0\width} {0\width}},clip]{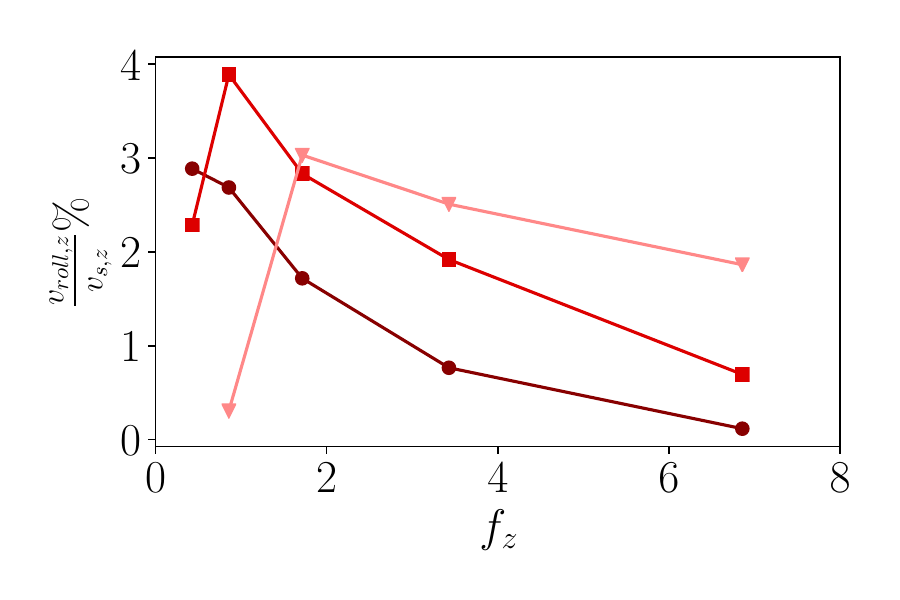}
      \label{fig:rvp}
\end{center}
\end{subfigure}
\begin{subfigure}{0.49\textwidth}
\begin{center}
  \caption{}
  \adjincludegraphics[width=\textwidth,trim={{0\width} {.05\width} {0\width} {0\width}},clip]{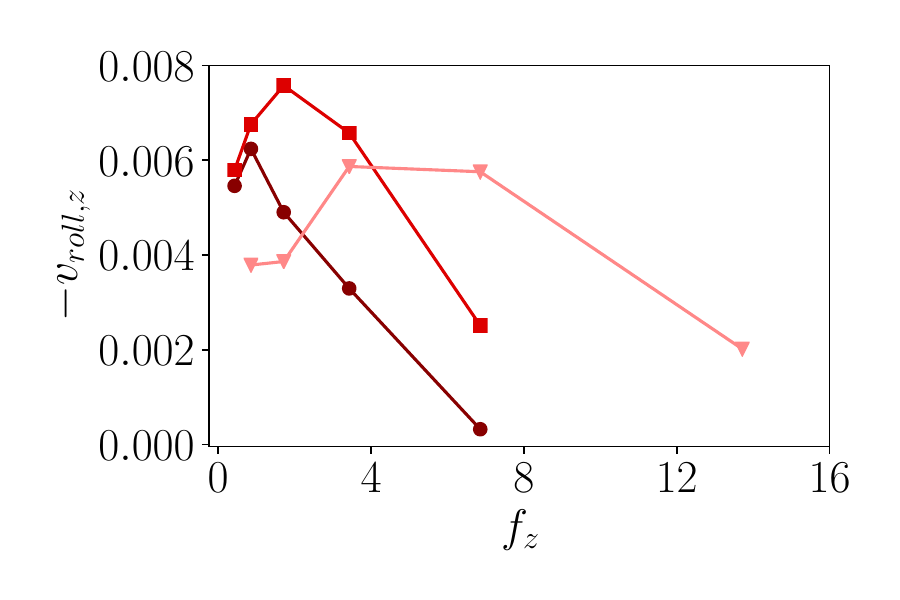}
      \label{fig:rv1}
\end{center}
\end{subfigure}
\begin{subfigure}{0.49\textwidth}
\begin{center}
  \caption{}
  \adjincludegraphics[width=\textwidth,trim={{0\width} {.05\width} {0\width} {0\width}},clip]{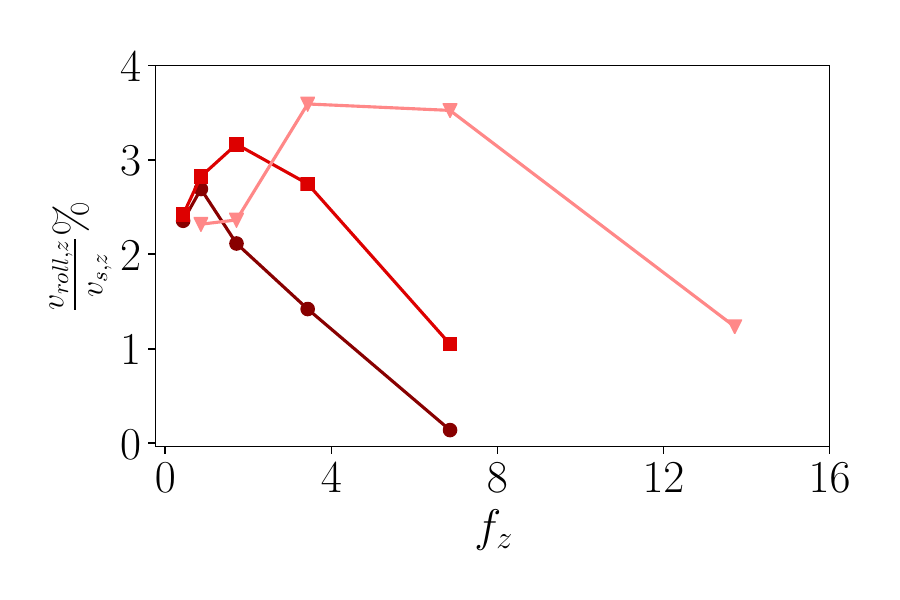}
      \label{fig:rvp1}
\end{center}
\end{subfigure}
\begin{subfigure}{0.49\textwidth}
\begin{center}
  \caption{}
  \adjincludegraphics[width=\textwidth,trim={{0\width} {.05\width} {0\width} {0\width}},clip]{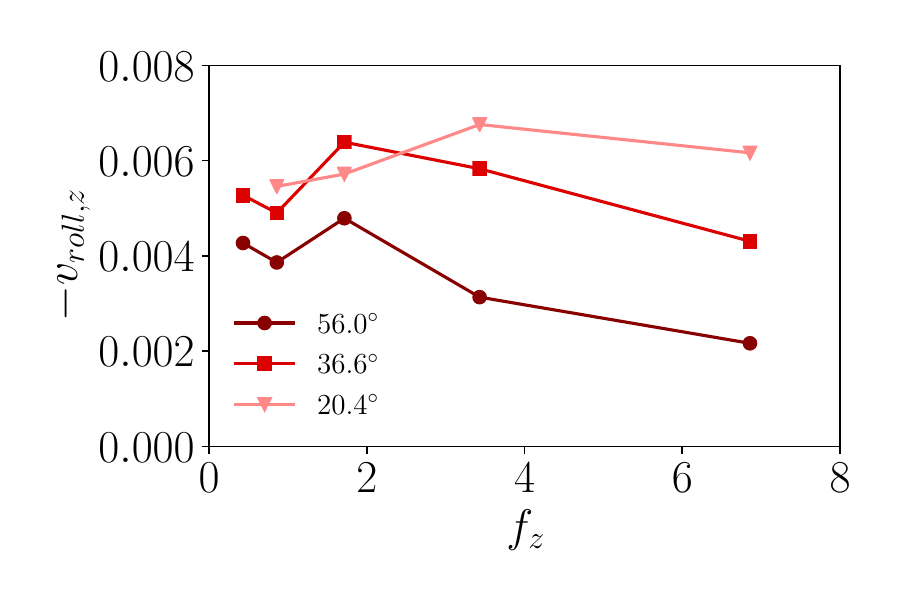}
      \label{fig:rv3}
\end{center}
\end{subfigure}
\begin{subfigure}{0.49\textwidth}
\begin{center}
  \caption{}
  \adjincludegraphics[width=\textwidth,trim={{0\width} {.05\width} {0\width} {0\width}},clip]{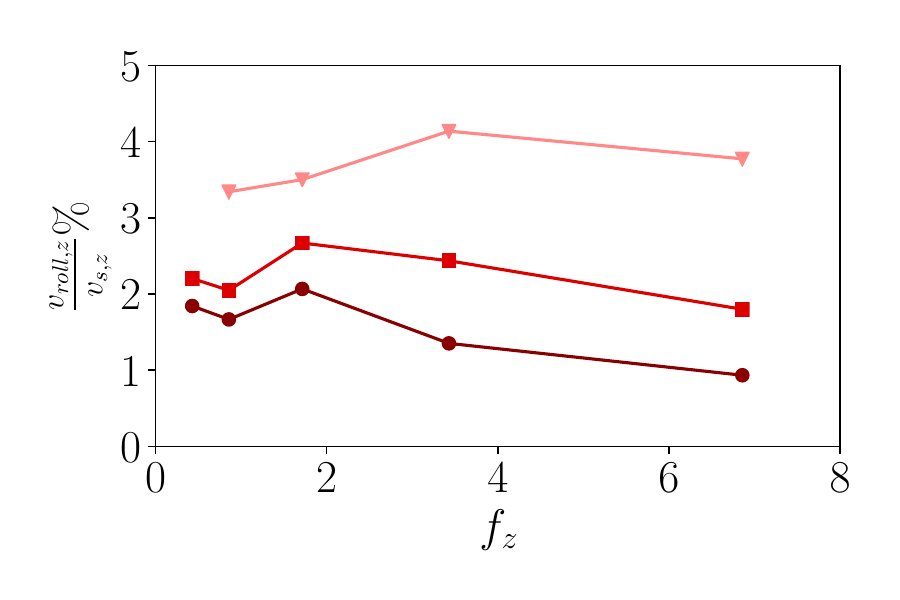}
      \label{fig:rvp3}
\end{center}
\end{subfigure}
\centering
\caption{ Inclined pair of pattern (\textit{a,c,e}) axial roll velocity for $Re_s=2.28 \times 10^3,\ 10^4,\ 3.00 \times 10^4$. (\textit{b,d,f}) axial roll velocity as the percentage of the spanwise pattern velocity for different angles and axial frequencies of the pattern for $Re_s=2.28 \times 10^3,\ 10^4,\ 3.00 \times 10^4$.}
    \label{fig:roll_vel}
\end{figure}

In the left panels of figure \ref{fig:roll_vel}, we show the roll velocity $v_{roll,z}$ for various pattern frequencies and varying Reynolds number. We highlight again that the negative sign of the axial roll velocity denotes that the rolls are moving downwards in the axial direction, and not upwards, following the stripe. For all the angles of inclination of the pattern, the roll velocity is non-monotonic: it increases with pattern frequency until a local maximum is reached, and then decreases as the pattern becomes smaller and its effect becomes confined to the boundary layers. 

The rolls appear to have a certain resistance to moving because they achieve small velocities relative to those of the pattern. Indeed, the roll velocity is of the order of the induced axial velocities. It is very hard to elucidate what the functional form of the $v_{roll,z}(\alpha,f_z,Re_s)$ is. For the $Re_s$ shown, some patterns seem to be present. First, increasing the angle of inclination from $\alpha=20.4^{\circ}$ to $\alpha=36.6^{\circ}$, increases the roll-velocity peak. However, if the angle of inclination further increases to $\alpha=56.0^{\circ}$, a decrease in the roll-velocity peak is observed. To account for this, we show in the right panels of figure \ref{fig:roll_vel} the roll-velocity as the percentage of the pattern velocity in the axial direction is plotted for various pattern frequencies and angles of inclination. It can be clearly seen that for smaller values of $\alpha$, the rolls move with a higher fraction of the geometrical velocity, even if the absolute velocities are smaller. The second noticeable thing is that there is a $f_z$ frequency for which the velocity is maximal, and further increasing or decreasing the pattern size will slow down the roll movement. While due to the limitations of our computer simulations it is impossible to do a proper $f_z$ sweep, this underlying pattern is present for almost all curves. Finally, the dependence on $Re_s$ cannot be properly elucidated. It is clear that as $Re_s$ increases, the rolls keep on moving, in line with the presence of induced velocities for all values of $Re_s$ seen in figure \ref{fig:meanvz}. However, the underlying pattern is not clear. So to summarize, these oblique patterns will generate induced velocities that move the rolls at velocities much slower than the cylinder speed, but how to optimize this velocity is unclear. Further supporting material that shows that oblique patterns induce spanwise velocities and move the rolls is provided in the Appendix \ref{sec:appa} for different roll wavelengths and numbers of rolls.

\begin{figure}
\begin{subfigure}{0.45\textwidth}
\begin{center}
  \caption{}
  \begin{tikzpicture}[scale=0.45]

\draw [black](0,0) -- (9.42,0) -- (9.42,6.99) -- (0,6.99) -- (0,0);

\draw[fill=white]  ((0,6.99*1/2) -- (9.42*1/2,6.99/2) -- (9.42/2,6.99) -- (0,6.99) -- cycle;
\draw[fill=gray!50]  (0,0) -- (9.42/2,0) -- (9.42*1/2,6.99/2) -- (0,6.99*1/2) -- cycle;
\draw[fill=white]  (9.42*1/2,0) -- (9.42,0) -- (9.42,6.99/2) -- (9.42/2,6.99/2) -- cycle;
\draw[fill=gray!50]  (9.42/2,6.99/2) -- (9.42,6.99*1/2) -- (9.42,6.99) -- (9.42/2,6.99) -- cycle;

  \draw
    (9.42,6.99/2) coordinate (c)
    (9.42/2,0) coordinate (b)
    (9.42,0) coordinate (a)
    pic["$\alpha$", draw=black, <->, angle eccentricity=1.2, angle radius=0.75cm]
    {angle=a--b--c};

\draw[black,dashed](0,6.99/2)-- (9.42/2,6.99);
\draw[black,dashed](9.42/2,0)-- (9.42,6.99/2);
\draw[<->, thin][black](9.42*3/8,6.99*7/8) to +(3.25,-1.38*3.25);

\fill[black](5.1,4.1) node [scale=1,anchor=west]{$\lambda_{\alpha}$};

\end{tikzpicture}
  \label{fig:cbstr}
\end{center}
\end{subfigure}
\centering
\begin{subfigure}{0.45\textwidth}
\begin{center}
  \caption{}
  \adjincludegraphics[width=0.9\textwidth,trim={{0.23\width} {.05\width} {0.05\width} {0.\width}},clip]{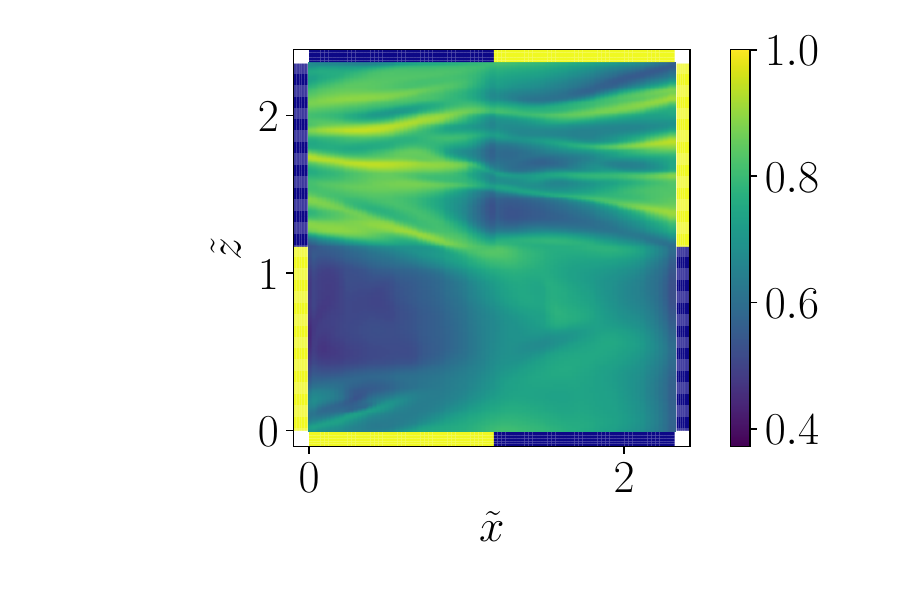}
  \label{fig:cbi}
\end{center}
\end{subfigure}
\centering
\caption{(\textit{a}) Checkerboard pattern parameters and (\textit{b}) near-wall instantaneous azimuthal velocity of $\alpha=45^{\circ}$ square checkerboard pattern.}
\label{fig:checkdef}
\end{figure}

\begin{figure}
\centering
\begin{subfigure}{0.49\textwidth}
\begin{center}
  \caption{}
  \adjincludegraphics[width=\textwidth,trim={{0\width} {.05\width} {0\width} {0\width}},clip]{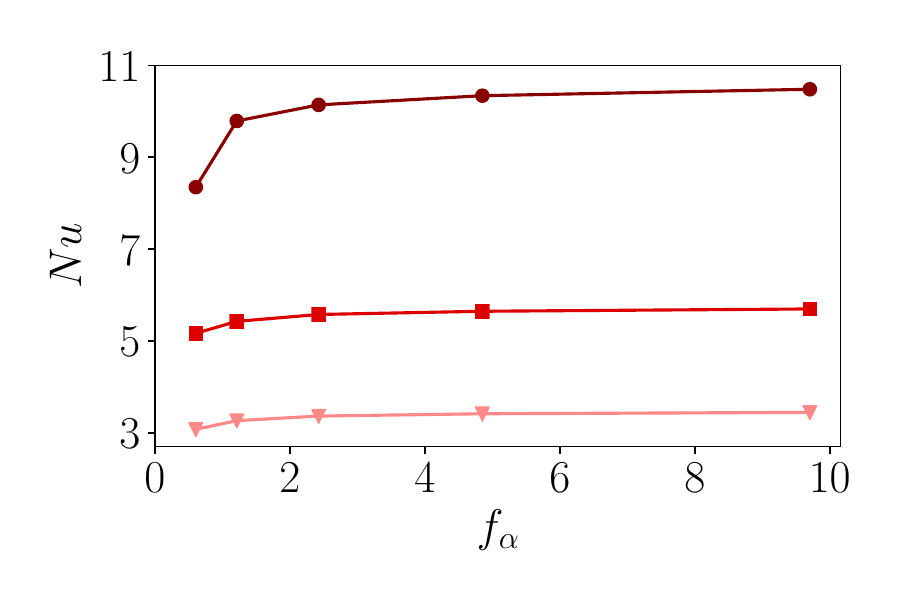}
  \label{fig:cb_f}
\end{center}
\end{subfigure}
\begin{subfigure}{0.49\textwidth}
\begin{center}
  \caption{}
  \adjincludegraphics[width=\textwidth,trim={{0\width} {.05\width} {0\width} {0\width}},clip]{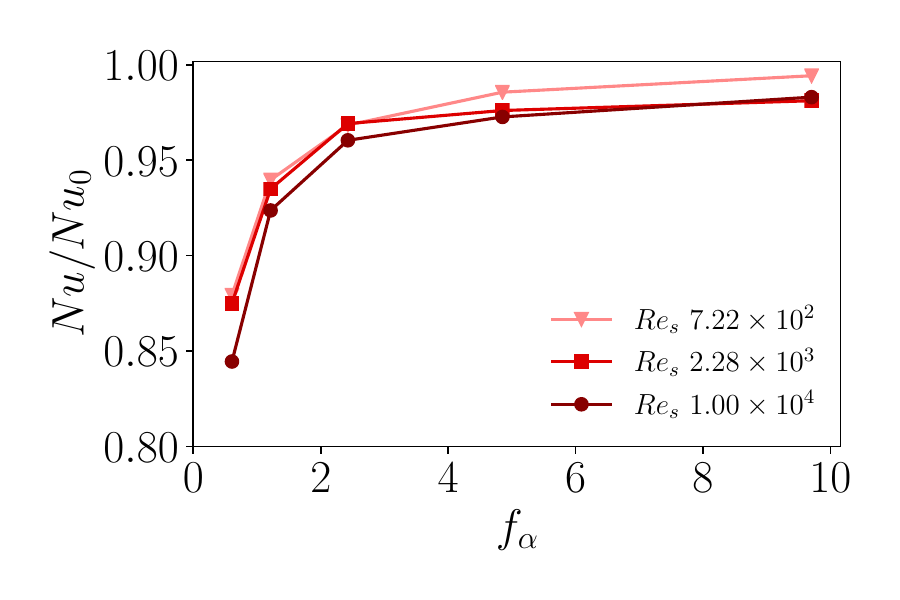}
  \label{fig:cbc_f}
\end{center}
\end{subfigure}
\centering
\caption{(\textit{a}) Non-dimensionalized and (\textit{b}) normalized torque at various square checkerboard frequency and Reynolds number.}
\label{fig:checktorq_f}

\end{figure}

\subsection{Checkerboard patterns}

The final pattern we have studied is checkerboard patterns, as shown in figure \ref{fig:cbstr}. A schematic that defines their parameters is shown in Figure \ref{fig:cbstr}. Similar to spiral patterns, these patterns are inhomogeneous in both axial (spanwise) and azimuthal (streamwise) directions, and will generate spanwise imbalances in the Reynolds stresses which can induce secondary flows. However, checkerboard patterns do not break the axial symmetry, so in principle we would not expect that a persistent axial velocity will result from their application, and this is indeed corroborated when looking at the statistics. Furthermore, as they move with the cylinder, the imbalances change and a symmetry is restored. Thus, their effect on the rolls is hard to guess a priori because the intermittent imbalances could modify the rolls, or the effect could be rapidly absorbed within the boundary layers.

To show the effect of the pattern on the boundary layer, we present the near-wall azimuthal velocity in figure \ref{fig:cbi}. Unlike the spiral pattern, the footprint on the flow appears to be much smaller. Indeed, the flow appears to be very similar to that generated by azimuthal inhomogeneities, where the streak elongation is able to wash away the inhomogeneities very fast. To corroborate this, we show the torque of a square checkerboard pattern (i.e.~$\alpha=45^{\circ}$) for various frequencies as shown in figure \ref{fig:checktorq_f}. The curve is monotonic, and the torque increases with increasing pattern frequency. The checkerboard case is thus not as special as the axial or spiral patterns. 

\begin{figure}
\centering
\begin{subfigure}{0.49\textwidth}
\begin{center}
  \caption{}
  \adjincludegraphics[width=\textwidth,trim={{0\width} {.05\width} {0\width} {0\width}},clip]{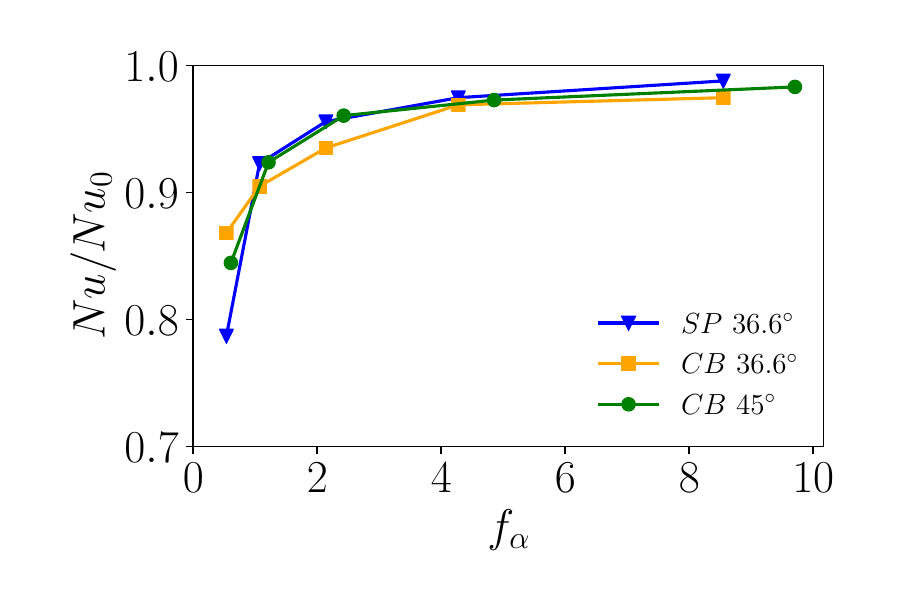}
  \label{fig:cb_sp_torque}
\end{center}
\end{subfigure}
\begin{subfigure}{0.49\textwidth}
\begin{center}
  \caption{}
  \adjincludegraphics[width=\textwidth,trim={{0\width} {.05\width} {0\width} {0\width}},clip]{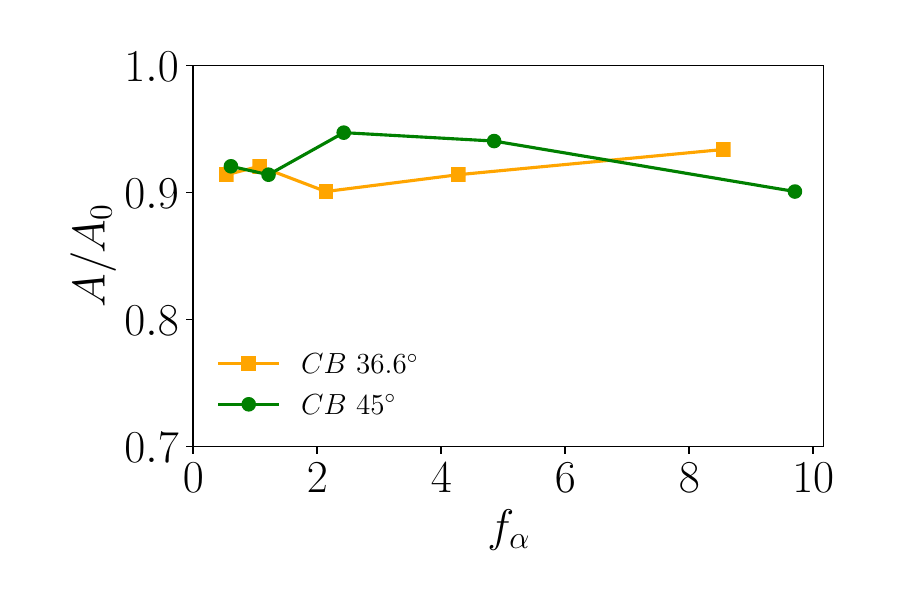}
  \label{fig:cb_sp_amp}
\end{center}
\end{subfigure}
\centering
\caption{(\textit{a}) Normalized torque for different checkerboard (CB) frequencies $\alpha=36.6^{\circ}$ spiral pattern (SP), $\alpha=36.6^{\circ}$ rectangle checkerboard and $\alpha=45^{\circ}$ square checkerboard for $Re_s=10^4$. (\textit{b}) Normalized roll amplitude for different pattern frequencies and two types of checkerboard pattern at $Re_s=10^4$ }
\label{fig:checktorq}
\end{figure}

To further analyze the square checkerboard pattern, we can compare its effect on the torque with the spiral pattern at $\alpha=36.6^{\circ}$. To do this, we also simulated a similar rectangular checkerboard pattern at $\alpha=36.6^{\circ}$. In figure \ref{fig:cb_sp_torque}, we show the normalized torques for the two checkerboard patterns considered, and the spiral case, for $Re_s=10^4$. For the lower frequencies, the spiral pattern has a much lower torque than the checkerboard patterns, whereas for the higher frequencies there is no significant differences in the torque.

The absence of a large dip in the torque for small frequencies is another indication that the checkerboard pattern barely disrupts the rolls. This is further corroborated by Figure \ref{fig:cb_sp_amp}, which shows the normalized roll amplitude for the two checkerboard cases studied. The values of $A/A_0$ seen for both checkerboard geometries are in line with those observed for azimuthal variations. Both azimuthal and checkerboard patterns do not significantly disrupt the rolls and induce no meaningful secondary flows, making them of low interest to our future research. Therefore, our analysis on this case is brief.

\section{Summary and conclusion}
\label{sec:summary_and_conclusion}
 
Four pattern geometries of stress-free and no-slip boundary conditions have been applied on the inner cylinder of a Taylor-Couette system to study the effect of boundary inhomogeneities on existing pinned secondary flows (known as Taylor rolls) as well as other flow statistics. We found that the azimuthal variations do not significantly alter the rolls and reduce the torque by moderate amounts (as low as $10\%$ for most of the Reynolds numbers simulated) even if $50\%$ of the surface is patterned and not able to transmit shear.  The natural elongation of streaks in the streamwise direction counters the effect of azimuthal patterns effectively. In addition, because the azimuthal pattern is moving with the cylinder, the flow will see intermittent stress-free and no-slip conditions which restore symmetry on an average sense. 

The picture is different for the axial variations. A certain portion of the cylinder is always stress-free at any time, and this generates a persistent imbalance. Axial variations reduce the torque to values as low as $65\%$ of the homogeneous TC torque value, consistent with the low Reynolds number results found by \cite{watanabe2017drag} and \cite{naim2019turbulent}. In addition, the $f_z=0.86$ pattern was found to generate a Reynolds stress imbalances that induce a secondary flow that destructively interferes with the rolls, and this reduces the torque beyond what is expected from just boundary layer effects analyzed through effective slip lengths and velocities. 

We also applied spiral variations which break axial reflection symmetries of the problem and introduce chirality. This induces a persistent axial velocity, which does not heavily modify the roll but it is able to move it around along the direction of relative pattern velocity with respect to the cylinder. For the three angles of inclinations studied, there is an optimum frequency at each inclination angle for which the roll movement becomes maximum, which does not always coincide with the frequency that generated the maximum induced velocity. When normalizing the roll velocity by the relative pattern velocity, we found that the patterns with smaller angles of inclination tend to move the rolls with a higher fraction of this velocity. 

The last pattern we studied was the checkerboard pattern which neither generates an induced velocity, nor a persistent Reynolds stress imbalance. Because of this, it ends up not having a significant effect on the rolls, while modifying the torque in a similar manner to the azimuthal patterns.
 
This study has served as confirmation that the Taylor rolls can be affected by careful patterning of the inner cylinder boundary. Limitations of this study are primarily due to the fact that the parameter space to be explored is very large, and that we are only dealing with a narrow-gap TC system with radius ratio $\eta=0.909$ under pure inner cylinder rotation. While this selection was  advantageous to focus on Taylor rolls, as they are present even at higher $Re_s$, in order to fully comprehend the effect of boundary heterogeneity, we must further simulate radius ratios or combinations of different cylinder rotation where the rolls are not present at higher $Re_s$. Further work will also include the analysis of more realistic finite-slip length patterns, which are obtainable in the laboratory.

\textbf{Acknowledgments:} We thank the Research Computing Data Core (RCDC) at the University of Houston for providing computing resources. We acknowledge funding from the National Science Foundation through grant NSF-CBET-1934121.

\textbf{Declaration of Interests}. The authors report no conflict of interest.

\pagebreak 

\appendix
\section{Nusselt number results}\label{sec:appb}

\begin{table}
    \centering
\begin{tabular}{ccc|ccc|ccc|ccc|ccc}
\hline
$\alpha$ & $f_{\alpha}$ & $Nu$ & $\alpha$ & $f_{\alpha}$ & $Nu$ & $\alpha$ & $f_{\alpha}$ & $Nu$ & $\alpha$ & $f_{\alpha}$ & $Nu$ & $\alpha$ & $f_{\alpha}$ & $Nu$ \\
\hline
 \multicolumn{15}{c}{$Re_s=7.22\times10^2$, $Nu_0=3.47$} \\
 \hline 
$0^\circ$~ & 6.87 & 3.23& $20.4^\circ$   & 7.32          & 3.41 & $36.6^\circ$   & 8.55 & 3.49&$56.0^\circ$   & 6.14          & 3.46 & $90^\circ$   & 5.09  & 3.46\\
~ & 3.43 & 3.11 & ~ & 3.66 & 3.33  & ~ & 4.27 & 3.45 &               & 3.07          & 3.43  & ~ & 2.55          & 3.44\\
~ & 1.72 & 2.98 & ~ & 1.83 & 3.21 & ~ & 2.14 & 3.39 &  ~ &1.54          & 3.37& ~ & 1.27          & 3.38\\
~ & 0.86 & 2.75 & ~ & 0.92 & 2.98 & ~ & 1.07 & 3.26 &  ~ &0.77          & 3.24 & ~ & 0.64          & 3.23\\
~ & 0.43 & 2.52 & ~ & ~    & ~    & ~ & 0.53  & 3.07 &  ~ &~          & ~ & ~ & 0.32          & 3.03\\
\hline
 \multicolumn{15}{c}{$Re_s=2.28\times10^3$, $Nu_0=5.77$} \\
\hline 
$0^\circ$ & 6.87 & 5.18 & $20.4^\circ$   & 7.32          & 5.65 & $36.6^\circ$   & 8.55          & 5.75 &$56.0^\circ$   & 6.14          & 5.78&$90.0^\circ$   & 5.09          & 5.85\\
~ & 3.43 & 4.93 & ~ & 3.66 & 5.53 & ~ & 4.27          & 5.70&               & 3.07          & 5.73   & ~ & 2.55          & 5.83\\
~ & 1.72 & 4.75&~ & 1.83 & 5.38 & ~  & 2.14          & 5.60&               & 1.54          & 5.64 & ~ &1.27          & 5.62\\
~ & 0.86 & 4.45& ~ & 0.92 & 4.52 & ~ & 1.07          & 5.41&               & 0.77          & 5.48 & ~ & 0.64          & 5.38\\
~ & 0.43 & 3.83& ~ & ~ & ~ & ~  & 0.53          & 4.80 & ~ & ~ & ~ & ~ & 0.32          & 5.08\\
\hline
 \multicolumn{15}{c}{$Re_s=1.0\times10^4$, $Nu_0=10.6$} \\
\hline 
~$0^\circ$~    & 6.87          & 9.73 & $20.4^\circ$   & 7.32          & 10.1 &$36.6^\circ$   & 8.55          & 10.4 & $56.0^\circ$   & 6.14          & 10.5 & $90.0^\circ$   & 5.09          & 10.6\\
~ & 3.43 & 9.20 & ~ & 3.66          & 9.82&~& 4.27          & 10.2&~ & 3.07          & 10.3 & ~  & 2.55          & 10.5\\
~ & 1.72 & 8.38& ~  & 1.83          & 9.36&~& 2.14          & 9.92&~& 1.54          & 10.0 & ~& 1.27          & 10.1\\
~ & 0.86 & 7.70& ~& 0.92          & 8.63 &~& 1.07          & 9.48&~& 0.77          & 9.57 & ~  & 0.64          & 9.76\\
~ & 0.43 & 7.50& ~ & ~ & ~ & ~& 0.53          & 8.88 & ~ & ~ & ~ & ~ & 0.32          & 9.15\\
\hline
 \multicolumn{15}{c}{$Re_s=3.0\times10^4$, $Nu_0=21.1$} \\
\hline 
$0^\circ$ & 6.87 & 18.1  &$20.4^\circ$ & 7.32 & 19.1&$36.6^\circ$ & 8.55 & 19.7 &$56.0^\circ$ & 12.3 & 20.7 &$90.0^\circ$ & 5.09 & 20.3 \\
~ & 3.43 & 16.4 & ~ & 3.66 & 17.8 & ~ & 4.27 & 19.0 & ~ & 6.14 & 20.1 & ~ & 2.55 & 19.2\\
~ & 1.72 & 15.5 & ~ & 1.83 & 16.4 & ~ & 2.14 & 18.0 & ~ & 3.07 & 18.0 & ~ & 1.27 & 18.4\\
~ & 0.86 & 14.7 & ~ & 0.92 & 15.3 & ~ & 1.07 & 17.0 & ~ & 1.54 & 18.3 & ~ & 0.64 & 17.1\\
~ & 0.43 & 14.3 & ~ & ~    & ~    & ~ & 0.53 & 15.9 & ~ & 0.77 & 16.8 & ~ & 0.32 & 15.7\\
\hline 
\end{tabular}
    \caption{Nusselt number for cases with one-dimensional stripes and $\lambda_{TR}=2.33$ and one roll}
    \label{tab:nusseltnormal}
\end{table}

\begin{table}
    \centering
\begin{tabular}{ccc|ccc|ccc|ccc}
\hline
$Re_s$ & $f_{\alpha}$ & $Nu$ & $Re_s$ & $f_{\alpha}$ & $Nu$ &$Re_s$ & $f_{\alpha}$ & $Nu$ &$Re_s$ & $f_{\alpha}$ & $Nu$ \\
\hline
$7.22\times10^2$ & 5.09 & 3.46 & $2.28\times10^3$ & 5.09  & 5.73 & $1.0\times10^3$ &  5.09 & 10.4 & $1.0\times10^3$ &5.09 & 10.3\\
$\alpha=45^\circ$ & 2.55 & 3.43 & $\alpha=45^\circ$ & 2.55 & 5.70 & $\alpha=45^\circ$ &  2.55          & 10.3 & $36.6^\circ$& 2.55 & 10.3\\
~ & 1.27 & 3.37& ~ & 1.27 & 5.66 & ~ & 1.27 & 10.2 & ~ & 1.27          & 9.92\\
~ & 0.64 & 3.27& ~ & 0.64 & 5.46 & ~ & 0.64 & 9.80 & ~ & 0.64          & 9.60\\
~ & 0.32 & 3.06& ~ & 0.32 & 5.11 & ~ & 0.32 & 8.96 & ~ & 0.32          & 9.21\\
\hline 
\end{tabular}
    \caption{Nusselt number for cases with checkerboard patterns and $\lambda_{TR}=2.33$ and one roll}
    \label{tab:nusseltnormalcb}
\end{table}

\section{Box size dependence}
\label{sec:appa}

In this Appendix we show additional results which corroborate the independence of our conclusions from the some geometrical and numerical control parameters chosen. In particular, we vary the number of roll pairs $n_r$ and the wavelength of the rolls $\lambda_{TR}$. The main study was conducted using $\Gamma=\lambda_{TR}=2.33$, and $n_r=\Gamma/\lambda_{TR}=1$. In this Appendix we independently vary $\Gamma$ and $n_r$, and we also vary the order of rotational symmetry.

\subsection{Axial patterns}

\begin{figure}
\centering
\begin{subfigure}{0.49\textwidth}
\begin{center}
  \caption{} 
  \adjincludegraphics[width=\textwidth,trim={{0\width} {.05\width} {0\width} {0\width}},clip]{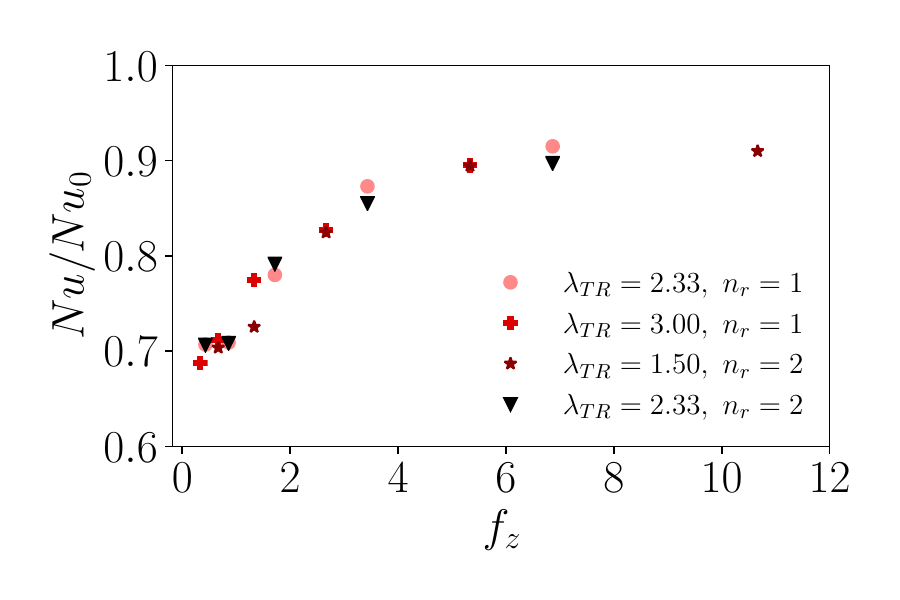}
\label{fig:torque_roll}
\end{center}
\end{subfigure}
\begin{subfigure}{0.49\textwidth}
\caption{} 
\begin{center}
  \adjincludegraphics[width=\textwidth,trim={{0\width} {.05\width} {0\width} {0\width}},clip]{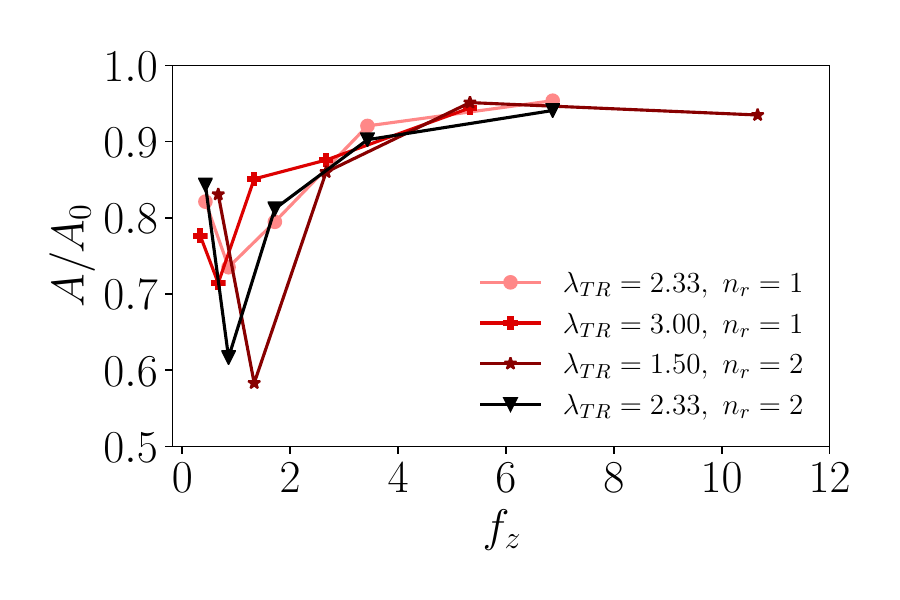}
\end{center}
\label{fig:torque_amp}
\end{subfigure}
\centering
\caption{(\textit{a}) Normalized torque and (\textit{b}) Roll amplitudes at various wavelengths and different types of Taylor rolls for $Re_s=10^4$. }
\label{fig:roll_types}
\end{figure}

In figure \ref{fig:torque_roll} we show the $Nu(f_z)/Nu_0$ curves at $Re_s=10^4$ for different roll wavelengths and numbers. All curves can be seen to have approximately the same shape. Figure \ref{fig:torque_amp} shows the normalized roll amplitude as a function of $f_z$ for different cases. All curves show a large dip in the roll amplitude for patterns at half the frequency of the roll, consistent with what was observed in the figure \ref{fig:rollamp0}.

\subsection{Spiral patterns}

\begin{figure}
\centering
\begin{subfigure}{0.49\textwidth}
\begin{center}
  \caption{} 
  \adjincludegraphics[width=\textwidth,trim={{0\width} {.05\width} {0\width} {0\width}},clip]{Plots/19a.pdf}
\label{fig:torque_1x_roll}
\end{center}
\end{subfigure}
\begin{subfigure}{0.49\textwidth}
\caption{} 
\begin{center}
  \adjincludegraphics[width=\textwidth,trim={{0\width} {.05\width} {0\width} {0\width}},clip]{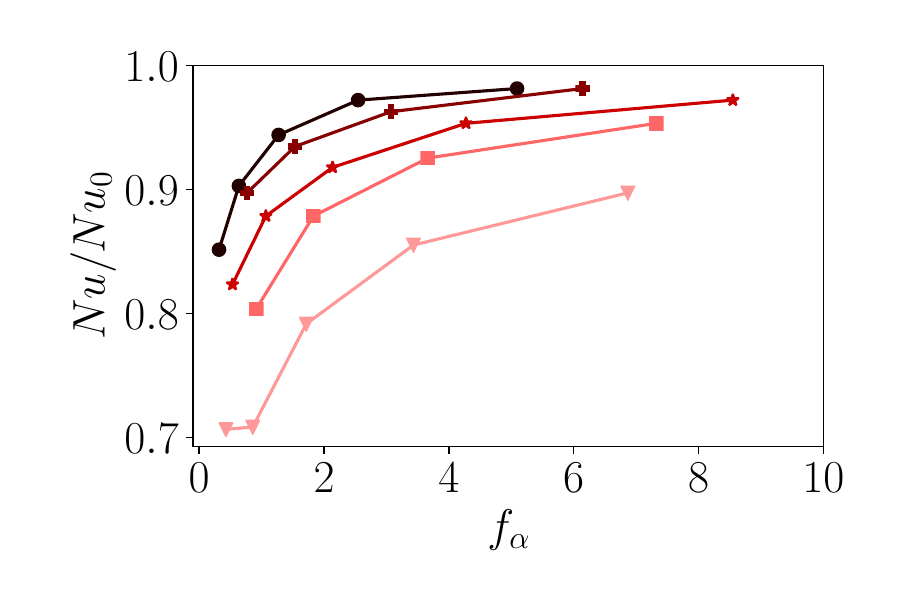}
\end{center}
\label{fig:torque_2x_roll}
\end{subfigure}
\centering
\caption{Normalized torque for $Re_s=10^4$ of (\textit{a}) $\lambda_{TR}=2.33,\ n_r=1$ and (\textit{b}) $\lambda_{TR}=2.33,\ n_r=2$. }
\label{fig:domain_comp}
\end{figure}

Figure \ref{fig:domain_comp} shows the normalized torque for the spiral patterns at $Re_s=10^4$ for the domain studied above ($\Gamma=2.33$ and $n_{sym}=20$), and a computational box with twice the domain size ($\Gamma=4.66$ and $n_{sym}=10$) such that two rolls are present. We can see that the graphs show very similar behaviour, with the same ordering of the curves according to pattern angle.

\begin{figure}
\centering
\begin{subfigure}{0.54\textwidth}
\begin{center}
  \caption{}
  \adjincludegraphics[width=\textwidth,trim={{0\width} {0\width} {0\width} {0\width}},clip]{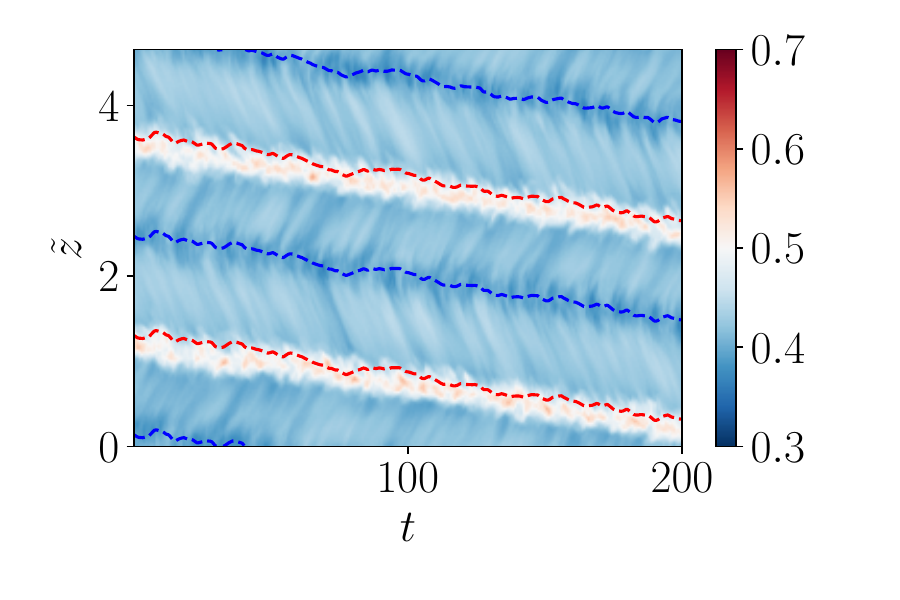}
    \label{fig:spt2x}

\end{center}
\end{subfigure}
\begin{subfigure}{0.49\textwidth}
\begin{center}
  \caption{}
  \adjincludegraphics[width=\textwidth,trim={{0\width} {.05\width} {0\width} {0\width}},clip]{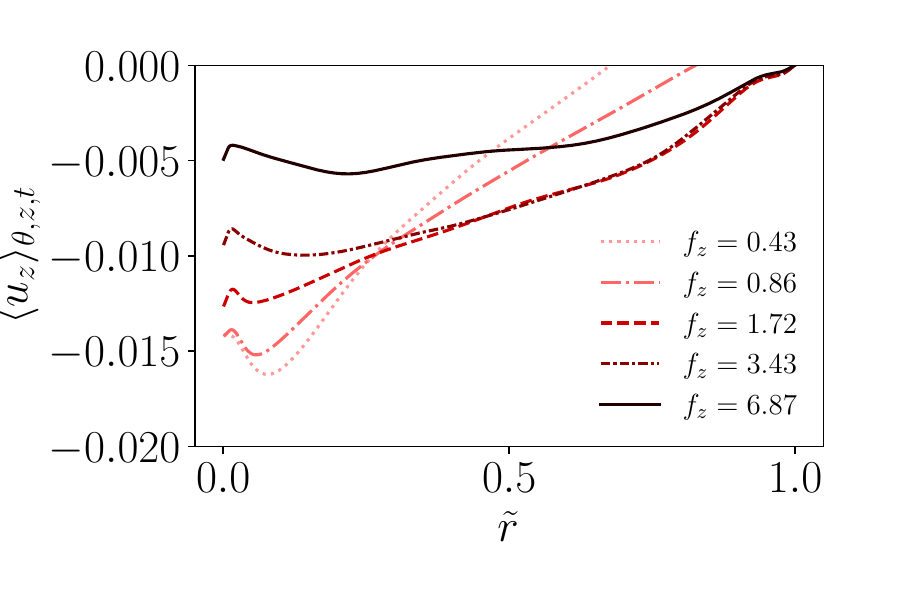}
    \label{fig:induced2x}
\end{center}
\end{subfigure}
\begin{subfigure}{0.49\textwidth}
\begin{center}
  \caption{}
  \adjincludegraphics[width=\textwidth,trim={{0\width} {.05\width} {0\width} {0\width}},clip]{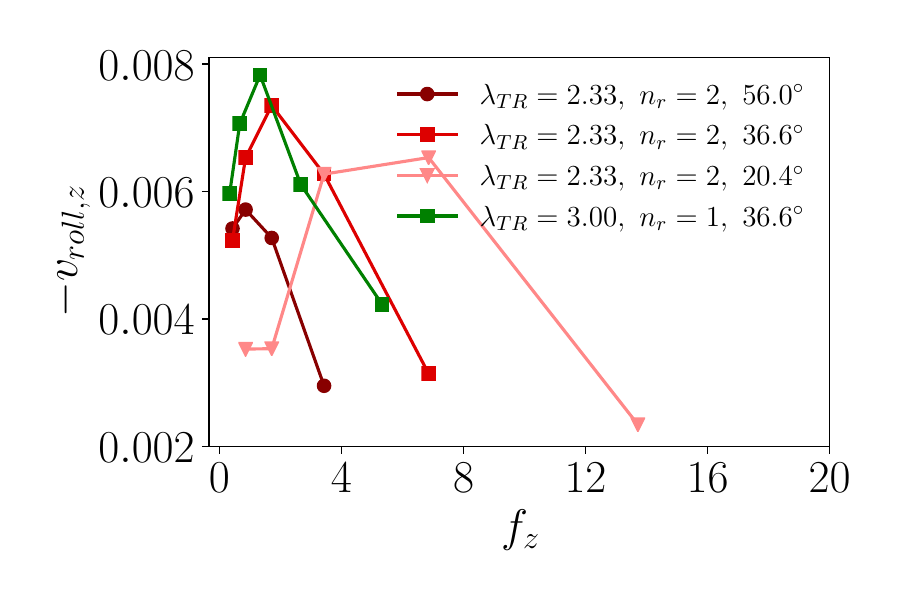}
    \label{fig:rollvel2x}
\end{center}
\end{subfigure}
\centering
\caption{ (\textit{a}) Roll movement with two pairs of rolls: space-time plot of the averaged azimuthal velocity at the mid-gap for $\lambda_{TR}=2.33$, $n_{r}=2$, $\alpha=36.6^{\circ}$ and $f_z=1.72$. Red and blue dashed lines correspond to the maximum and the minimum mid-gap average azimuthal velocities in time respectively predicted from the phase angle of second mode of the Fourier transform; (\textit{b}) Induced axial velocity as a function of pattern frequency $f_z$ with constant angle $\alpha=36.6^\circ$ for $\lambda_{TR}=2.33$, $n_r=2$ and $Re_s=10^4$. (\textit{c}) Axial roll velocity of inclined pattern pair for $Re_s= 10^4$.}
\label{fig:spiralrollvelocity}
\end{figure}

The roll motion is analyzed in figure \ref{fig:spiralrollvelocity}. The first panel shows how we apply the Fourier method to determine roll velocity in the case of two rolls: by taking the second fundamental axial Fourier mode. The second panel shows the induced velocities as a function of $f_z$. Again, velocities are more or less of the order of $1-2\%$ of the inner cylinder velocity, are in the negative direction and are highest close to the inner cylinder. Finally, the third panel shows how the roll velocity has a maximum for intermediate values of $f_z$, consistent with what was seen in the study (figure \ref{fig:roll_vel}), and the velocities are also of the order of less than $1\%$ the roll velocity. This is true both for rolls of different wavelengths, and for cases with two rolls.

\subsection{Nusselt number data}

We present the Nusselt number data for this section in Tables \ref{tab:nusselttworolls} and \ref{tab:nusseltgamma}.

\begin{table}
    \centering
\begin{tabular}{ccc|ccc|ccc|ccc|ccc}
\hline
$\alpha$ & $f_{\alpha}$ & $Nu$ & $\alpha$ & $f_{\alpha}$ & $Nu$ & $\alpha$ & $f_{\alpha}$ & $Nu$ & $\alpha$ & $f_{\alpha}$ & $Nu$ & $\alpha$ & $f_{\alpha}$ & $Nu$ \\
\hline
~$0^\circ$~ & 6.87 & 9.60 & $20.4^\circ$   & 7.32 & 10.2 &$36.6^\circ$   & 8.55 & 10.4 & $56.0^\circ$ & 6.14 & 10.5 & $90.0^\circ$ & 5.09 & 10.5\\
~ & 3.43 & 9.15 & ~ & 3.66 & 9.90 & ~ & 4.27 & 10.2 & ~ & 3.07 & 10.3 & ~ & 2.55 & 10.4\\
~ & 1.72 & 8.47 & ~ & 1.83 & 9.40 & ~ & 2.14 & 9.82 & ~ & 1.54 & 10.0 & ~ & 1.27 & 10.1\\
~ & 0.86 & 7.58 & ~ & 0.92 & 8.60 & ~ & 1.07 & 9.40 & ~ & 0.77 & 9.57 & ~ & 0.64 & 9.66\\
~ & 0.43 & 7.56 & ~ & ~    & ~    & ~ & 0.53 & 8.81 & ~ & ~    & ~    & ~ & 0.32 & 9.11\\

\hline 
\end{tabular}
    \caption{Roll number dependence: $Nu$ for cases with one-dimensional stripes, $Re_s=10^4$, $\lambda_{TR}=2.33$ and two rolls. The homogeneous Nusselt number is $Nu_0=10.7$.}
    \label{tab:nusselttworolls}
\end{table}

\begin{table}
    \centering
\begin{tabular}{ccc|ccc|ccc|ccc}
\hline
$\alpha$ & $f_{\alpha}$ & $Nu$ & $\alpha$ & $f_{\alpha}$ & $Nu$ & $\alpha$ & $f_{\alpha}$ & $Nu$ & $\alpha$ & $f_{\alpha}$ & $Nu$ \\
\hline
\multicolumn{9}{c|}{$\lambda_{TR}=3$, $Nu_0=10.3$} & 
\multicolumn{3}{c}{$\lambda_{TR}=1.5$, $Nu_0=10.3$} \\
\hline
~$0^\circ$~ & 6.87 & 9.22 & $36.6^\circ$ & 8.55 & 9.99 & $90.0^\circ$ & 5.09 & 10.1 & ~$0^\circ$ & 6.87 & 9.22\\
~ & 3.43 & 8.52 & ~ & 4.27 & 9.75 & ~ & 2.55 & 9.90 & ~ & 3.43 & 8.52\\
~ & 1.72 & 7.99 & ~ & 2.14 & 9.38 & ~ & 1.27 & 9.55 & ~ & 1.72 & 7.98\\
~ & 0.86 & 7.42 & ~ & 1.07 & 8.87 & ~ & 0.64 & 9.04 & ~ & 0.86 & 7.33\\
~ & 0.43 & 7.11 & ~ & 0.53 & 8.28 & ~ & 0.32 & 8.43 & ~ & 0.43 & 7.08\\
\hline 
\end{tabular}
    \caption{Roll wavelength dependence: $Nu$ for cases with one-dimensional stripes, $Re_s=10^4$ and changing $\lambda_{TR}$.  }
    \label{tab:nusseltgamma}
\end{table}

\FloatBarrier
\bibliographystyle{jfm}

\bibliography{literature}

\end{document}